\documentclass[sn-basic]{sn-jnl}

\usepackage{graphicx}%
\usepackage{multirow}%
\usepackage{amsmath,amssymb,amsfonts}%
\usepackage{amsthm}%
\usepackage{mathrsfs}%
\usepackage[title]{appendix}%
\usepackage{xcolor}%
\usepackage{textcomp}%
\usepackage{manyfoot}%
\usepackage{booktabs}%
\usepackage{algorithm}%
\usepackage{algorithmicx}%
\usepackage{algpseudocode}%
\usepackage{listings}%
\usepackage{geometry}%

\usepackage{bm}
\usepackage[thinc, italic]{esdiff}
\usepackage{siunitx}
\usepackage{ragged2e}  
\usepackage{tabularx}
\usepackage{pdflscape} 
\usepackage{xspace}

\newcommand{\teff}{\ensuremath{T_{\mathrm{eff}}}}
\newcommand{\logg}{\ensuremath{\log g}}

\newcommand{\kms}{km s$^{-1}$}

\newcommand{\Magritte}{\texttt{MAGRITTE}\xspace}
\newcommand{\Stagger}{\texttt{STAGGER}\xspace}
\newcommand{\Bifrost}{\texttt{BIFROST}\xspace}
\newcommand{\Dispatch}{\texttt{DISPATCH}\xspace}
\newcommand{\MultiD}{\texttt{MULTI3D}\xspace}
\newcommand{\Balder}{\texttt{BALDER}\xspace}
\newcommand{\Scate}{\texttt{SCATE}\xspace}
\newcommand{\Porta}{\texttt{PORTA}\xspace}
\newcommand{\RH}{\texttt{RH}\xspace}
\newcommand{\Asset}{\texttt{ASS$\epsilon$T}\xspace}
\newcommand{\PhoenixD}{\texttt{PHOENIX/3D}\xspace}
\newcommand{\OptimD}{\texttt{OPTIM3D}\xspace}
\newcommand{\Iris}{\texttt{IRIS}\xspace}
\newcommand{\LinforD}{\texttt{Linfor3D}\xspace}
\newcommand{\RHD}{\texttt{RH1.5D}\xspace}
\newcommand{\Nicole}{\texttt{Nicole}\xspace}
\newcommand{\NLTED}{\texttt{NLTE3D}\xspace}
\newcommand{\Cloudy}{\texttt{Cloudy}\xspace}
\newcommand{\Detail}{\texttt{DETAIL}\xspace}
\newcommand{\Multi}{\texttt{Multi}\xspace}
\newcommand{\Tlusty}{\texttt{Tlusty}\xspace}

\makeatletter
\input{aas_macros.sty}
\def\ref@jnl#1{{\jnl@style#1\ }}
\makeatother

\graphicspath{ {./figs/} }

\begin{document}

\title[3D NLTE radiation transfer]{3D NLTE radiation transfer: theory and applications to stars, exoplanets, and kilonovae}

\author*[1]{\fnm{Maria} \sur{Bergemann}}\email{bergemann@mpia.de}
\equalcont{These authors contributed equally to this work.}

\author*[1,2]{\fnm{Richard} \sur{Hoppe}}\email{hoppe@mpia.de}
\equalcont{These authors contributed equally to this work.}

\affil*[1]{\orgname{Max-Planck Institute for Astronomy}, \orgaddress{\street{K\"{o}nigstuhl 17}, \city{Heidelberg}, \postcode{69117}, \country{Germany}}}

\affil[2]{\orgname{Ruprecht-Karls-Universit\"{a}t}, \orgaddress{\street{Grabengasse 1}, \city{Heidelberg}, \postcode{69117}, \country{Germany}}}

\abstract{
Most of the physical information about astrophysical objects is obtained via the analysis of their electromagnetic spectra. Observed data coupled with radiation transfer models in physical conditions representative of stars, planets, kilonovae, and ISM, yield constrains on their physical structure,  gas flow dynamics at the surface, mass loss, and detailed chemical composition of the systems. All these core astrophysical parameters are only as reliable as the physical quality of the models that are employed for simulations of radiation transfer. Recent advances in multi-D transfer modelling with Non-Local Thermodynamic Equilibrium (NLTE) in inhomogeneous time-dependent systems revealed systematic shortcomings of canonical models. Owing to major complexities of solving coupled multi-frequency RT equations in 3D geometry, a number of approximations have been introduced. This review presents an overview of the physical problem, standard solutions, and recent methodological advances. We also provide an overview of main results in the area of 3D NLTE radiation transfer and its applications to modelling diverse astrophysical environments, including FGKM type- and OBA-type stars, multi-epoch spectra of kilonovae, and atmospheres of rocky and gaseous exoplanets.
}

\keywords{Radiative transfer, Numerical methods, Stellar abundances, Stellar atmospheres}

\maketitle

\setcounter{tocdepth}{3} 
\tableofcontents

%
%
%
%

\section{Introduction}
Physical modeling of the emergent radiation field of stars\footnote{Throughout this paper, for convenience we follow the spectral type classification presented by Eric Mamajec at \url{https://www.pas.rochester.edu/~emamajek/EEM_dwarf_UBVIJHK_colors_Teff.txt}}, their remnants, and their companions is the backbone of modern astrophysics. These calculations produce synthetic observables, including energy-dependent models of emergent fluxes and spatially-resolved surface intensities. These quantities are used in nearly every area of astronomy research, to address scientific questions ranging from exoplanets to stellar structure, and Galaxy evolution.

\begin{figure}[ht]
	\includegraphics[width=\columnwidth]{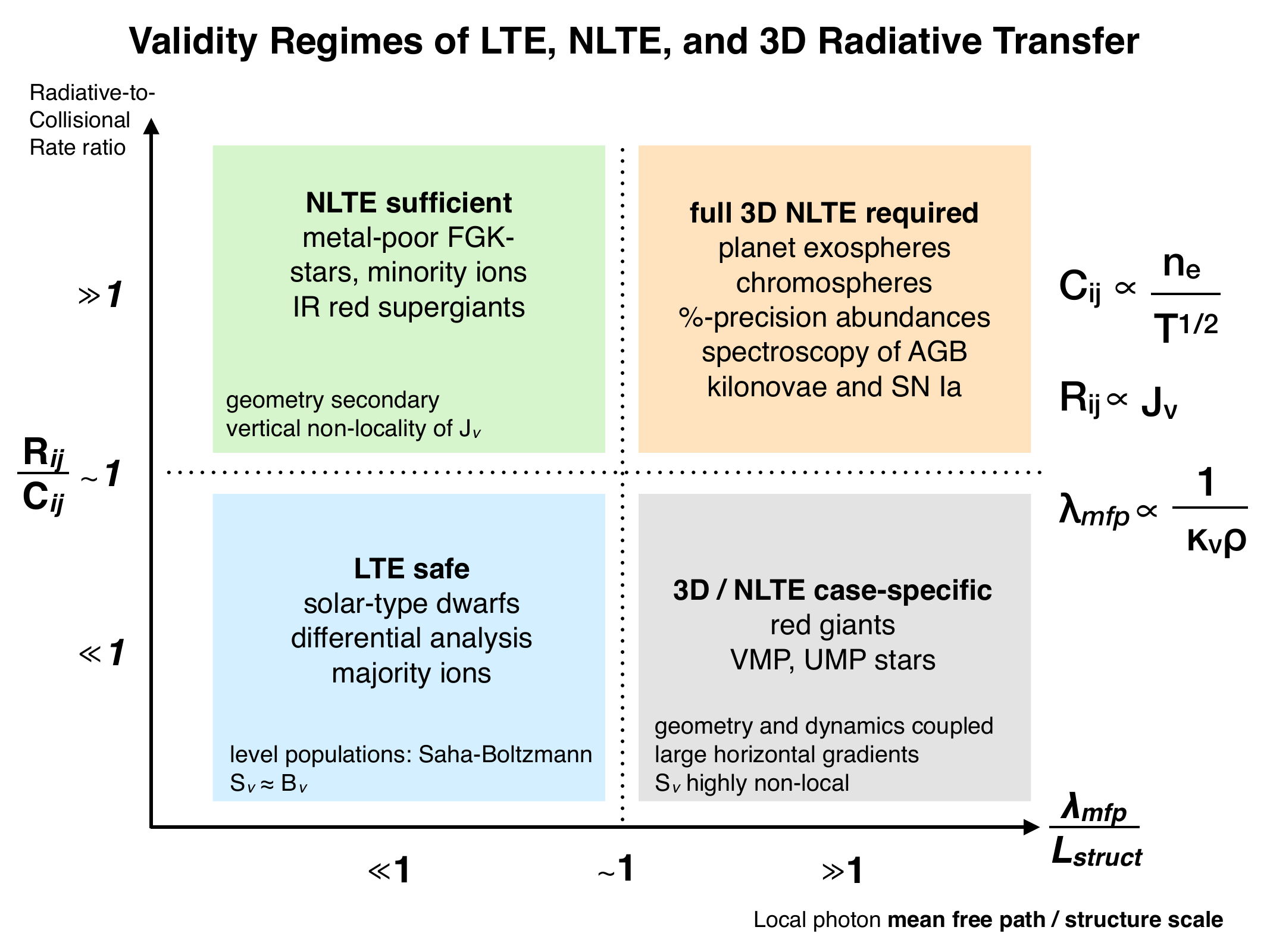}
    \caption{Validity regimes of multi-D and NLTE radiation transfer. We define the boundaries by the  physical origin (see text).}
    \label{fig:regimes}
\end{figure}

The goal of this paper is to review the physical and numerical approaches used in modern 3D NLTE calculations of emergent electromagnetic radiation of stars. The definition of 3D NLTE is as follows: three-dimensional (3D) radiation transfer (RT) simulations with level populations computed in Non Local Thermodynamic Equilibrium (NLTE)\footnote{The word shall not be hyphenated, as per recommendation of the anonymous referee. Indeed, the conventional form with a hyphen distorts the meaning of the term. The hyphenated form would imply that the systems in NLTE are those that \textit{are} in Thermodynamic Equilibrium, just that the TE is "non-local", which physically is not meaningful. The prefix "non" applies to the full expression "LTE" and not only to the first adjective ("local") (Mats Carlsson, priv. comm).}. To make the point more clear, we would like to re-cite \citet{Mihalas1973} directly: \textit{"Departures from LTE occur simply because stars have a boundary through which photons escape into space"} which obviously applies to any other astronomic system, such as atmospheres of exoplanets or kilonovae. Furthermore, sub-surface convection in stars is driven by entropy losses caused by radiative cooling at the surface \citep{Nordlund2009}. This implies that a major part of physics of stellar interior is controlled by the properties of radiation propagation and loss from stellar atmospheres. Multi-dimensionality is generally required whenever the physical structure - as compared to the scales of the photon - shows substantial spatial variability or non-monotonicity along the path of the photon. NLTE (or NLTE) is required whenever the radiation field is so intense that it outweighs the relevance of thermalising collisions, leading to level populations of atoms or molecules set by radiation. These two conditions are generally met in most atmospheres, but they particularly occur in the systems shown in Fig. \ref{fig:regimes}: stars with strong sub-surface convection (FGKM-type stars), massive stars (OB-type) with extreme radiation fields, and expanding shells of exploding and/or post-merger systems, such as kilonovae. 

\begin{figure*}[ht!]
	\includegraphics[width=\columnwidth]{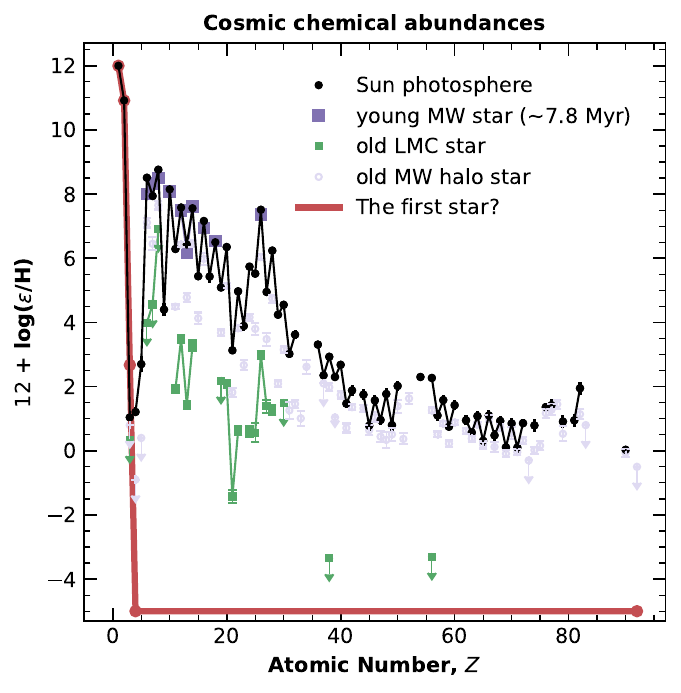}
    \caption{Stellar abundances obtained using NLTE RT spectroscopy methods for the Sun \citep{Lodders2025, Bergemann2025}, a massive B-type supergiant HD 14818 in the Milky Way disc \citep{Wessmayer2022}, a very old ultra-metal-poor star in the halo of the Large Magellanic Cloud \citep{Ji2026}, an r-process enhanced red horizontal branch star HD 222925 in Galactic halo \citep[][with NLTE corrections applied to selected elements]{Roederer2022} and the expected abundances for a hypothetical primordial star formed from the ISM with a BBN composition (values for $Y_{p}$ and $^{7}$Li are from \citealt[][their eqs. 5.3, 5.6]{Fields2020}). Here we represent with $-5$ all elements that were absent in the primordial composition. The primordial value of A($^{4}$He) is based on $Y_p = 0.24691 \pm 0.00018$ and the proto-solar value is based on $Y_p = 0.2752 \pm 0.0077$ (based on the Standard Solar Model, see \citealt[][]{Bergemann2025}.)}
    \label{fig:cosmicab}
\end{figure*}

As a consequence, 3D NLTE is essential for precision studies of the solar chemical composition \citep[e.g.]{Asplund2021,Caffau2011,Magg2022}, chemical abundances of metal-poor stars \citep[e.g.]{Jofre2019,Nissen2018,Lind2024}, probes of stellar convection, granulation, and sub-surface dynamics \citep[e.g.][]{Hoefner2018, Kravchenko2021, Goldberg2022, Chiavassa2024}, analyses of stellar variability in the context of planetary transits \citep[][]{Maxted2018, Dravins2021, Maxted2023}, constraints on stellar magnetism \citep{Ludwig2023, Kostogryz2024}, studies of atmospheric inhomogeneities and mass loss of massive stars \citep{Sundqvist2018, DelbroekSundqvist2025}. In Fig. \ref{fig:cosmicab}, we demonstrate, in particular, some applications of modern NLTE RT methods to analyses of chemical elements in stars, highlighting the abundances obtained using NLTE methods for the Sun \citep{Lodders2025, Bergemann2025}, a massive B-type star HD 14818 in the Milky Way disc \citep{Wessmayer2022}, a very old ultra-metal-poor star in the halo of the Large Magellanic Cloud \citep{Ji2026}, and - for the context - the expected abundances for a primordial star formed from the ISM with a BBN composition (values for $Y_{p}$ and $^{7}Li$ adopted from \citealt[][their eqs. 5.3, 5.6]{Fields2020}). Finally, we note that 3D and/or NLTE RT methods are becoming progressively adapted to modern analyses of exoplanet atmospheres \citep{Koskinen2013, Fossati2021} and expanding shells of kilonovae \citep{Tarumi2023}.

In this review, we primarily focus only on those studies, in which advances in 3D and/or NLTE RT methods were reported, or quantitative estimates of astrophysical parameters of \textit{stars, planets, or kilonovae} obtained using 3D and/or NLTE RT were presented. For completeness, we point out that aspects of 3D and NLTE RT are relevant in other fields of astrophysics, including modeling and diagnostics of ISM \citep[e.g.][]{Juvela1997,Vandertak2007}, polarisation of circumstellar discs \citep[][]{Bjorkman2005}, dusty emission from galaxy discs \citep[][]{Baes2011}, solar spectro-polarimetry \citep{SocasNavarro2015,TrujilloBueno2018, CruzRodriguez2019, delPino2021} and solar prominences \citep{Labrosse2010}, accretion discs in AGN \citep{Hubeny2001}, and Type Ia supernovae \citep[e.g.][]{Lentz2000,Kromer2009, Botyanszki2017}. These topics, however, were subjects of a comprehensive review in other contributions focussing on Monte-Carlo RT and 3D RHD models \citep{Hoefner2018, Noebauer2019, Chiavassa2024}, therefore we do not address these topics here. 

For academic clarity, we begin with the definitions (Sect.\ref{sec:definitions}) and observational motivation in Sect. \ref{sec:obs}. In Sects. \ref{sec:basicinputs} respectively we devote a substantial space to the physical data that are required as input to 3D and NLTE RT, including the model atmospheres and NLTE models. Sec. \ref{sec:num3DRT} provides a conceptual discussion of RT and we progressively move to more complex geometric and physical setups throughout the text. The NLTE concepts and numerical solution methods are reviewed in Sect. \ref{sec:nlte}. We then proceed with a summary of the main-stream codes used in the community in Sect. \ref{sec:codes}. Finally, the main astrophysical results based on recent 3D and/or NLTE calculations are presented in Sect. \ref{sec:results}. The review is closed with conclusions, written by AI (Sect. \ref{sec:conAI} and by ourselves (Sect.\ref{sec:conHuman}.) Throughout the text we devote attention to the present limitations of the models and highlight the aspects of the field where progress still needs to be made from the numerical, physical, and observational perspective.
\section{Definitions and basic considerations}\label{sec:definitions}
Radiation transfer is a physical process that enables determination of the emitted radiation field from the surface of an object. Numerical modeling of this process represents a highly complex boundary condition problem \citep{Auer1969, Mihalas1988}, which requires inputs from atomic and molecular physics, as well as fluid dynamics, statistical mechanics, and energy transport. Typically the problem is split into three steps:

\begin{itemize}

\item Given the upper and lower boundary values of physical parameters, such as radial size (set by the assumed mass of the system), entropy, and the incoming flux, the equations the describe the underlying physics (mass, charge, momentum, energy conservation) are solved. The result of these calculations is the prediction for the physical state of matter and of the approximate emergent radiation field. The latter is coarse and it is primarily useful as a statistical representation of the total energy density of the radiation field, that is, of the bolometric radiative flux.

\item In the vast majority of scientific problems, this first step is not sufficient, and the second step is to compute radiation transfer again, using more realistic (and thus sophisticated) micro-physics, a finer and wider frequency grid, more sources of radiative opacities, flexible chemical abundances of different elements and isotopic structure, which yields a more realistic spectral energy distribution or spectrum of an object. 

\item In the third step, this radiation field is subject to iteration, because the physical interaction between the matter itself and the radiation field (which is typically handled in a very coarse manner in step 1), by allowing the radiation field to act upon (that is, change) the energy states of particles, such as atoms and molecules. A classical example of this process is the photo-ionization process (as in electron-hole pair production in CCDs, also known as photo-excitation through the band gap) or the photo-dissociation process in plants (as in photolysis, involving dissociation of H$_{2}$O by photons into an H atom and OH, which further are converted to O$_2$). As a result of this change, the radiation field changes, but also the particles involved change their internal energy states according to quantum-mechanics selection rules. These particles, in turn, change the radiation field, and so until some form of equilibrium or steady-state is established. 

\end{itemize}

In this review, the focus is on the formal solution of the RT equation in time-dependent systems with a 3D geometry and in NLTE. Numerically, 3-dimensional (3D) radiation transfer (RT) is not different from its solution in a 1D or 2D system. The same 1D RT equation is solved in both cases, and one may adopt either a plane-parallel geometry or spherical symmetry. However, physically there is a major difference, because the solution is done in qualitatively different types of physical systems. These systems are typically defined as a 1D (radial column), 2D (a slice), or 3D (a cube). It has become common in astronomy to associate 3D RT with formal solutions of the 1D radiation transfer equation in non-stationary 3D models of atmospheres. Here, an \textit{atmosphere} is the surface region of the object, from which the radiation escapes into the ISM. For stars, these are 3D Radiation-HydroDynamics (RHD) simulations of sub-surface convection \citep{Nordlund2009, Kupka2017, Leenaarts2020, Chiavassa2024}, which may - depending on the size of the convection envelope - cover either a small patch of the surface (box-in-a-star) or a representative fraction of the star (star-in-a-box). For exoplanets, these are externally-irradiated 3D Magneto-HydroDynamics (MHD) models with global circulation \citep[e.g.][]{Parmentier2021, Christie2021}. These may including prescriptions for formation of dust and clouds \citep[e.g.][]{Helling2013, Lines2018} and they resemble the approaches used in modeling very cool Asymptotic Giant Branch (AGB) stars \citep{Hoefner2018}. 
Kilonovae are typically represented as 1D spherically-symmetric expanding systems \citep[e.g.][]{Jerkstrand2011, Kerzendorf2014} similar to massive stars with winds. These atmospheric models are vastly different in complexity and computational burden, and may include magnetic fields and different types of scattering and dust formation, or mass loss.

NLTE refers to the description of the physical state of matter, in which the RT is solved. In LTE, which is a special case of NLTE (Mats Carlsson, priv. comm.), it is assumed -- often without justification -- that collisional rates between particles dominate over radiative transition rates caused by the radiation field. Hence, the number densities of particles and, in particular, their distribution with respect to the internal energy states (ionisation, excitation, chemical equilibrium) can be described by the Saha--Boltzmann equilibrium statistics. The other option is not to assume this equilibrium, which is known as NLTE. In NLTE the full rate equations are solved. In the time-independent case and when the advection term is ignored, these equations simplify to the equations of Statistical Equilibrium (SE), which are the ones of most widespread use in the NLTE community. Therefore in this review, we reserve the term NLTE to represent the \textit{absence of LTE} and maintain that NLTE is not an assumption, but the state of matter, in which LTE is not assumed. We advise the users and practitioners of NLTE methods to refer to NLTE SE, when SE equations and not the full rate equations are solved.
\section{Observational evidence}\label{sec:obs}

\subsection{Stellar spectra and line bisectors}

Convection leaves a unique imprint in the shapes of spectral lines. However,  until recently studies of convective blue- and red-shifts in stellar spectra were fundamentally limited by the capabilities of the astronomical instrumentation and the wavelength calibration. For the Sun, resolving power in excess of $500\,000$ is possible with Fourier Transform Spectroscopy (FTS) facilities. However, only very few instruments, like ESPRESSO at the Very Large Telescope (VLT), yield the data quality that can be used to study convection in other stars.
%
%
%

\begin{figure}[ht]
    \centering
    \includegraphics[width=0.5\textwidth]{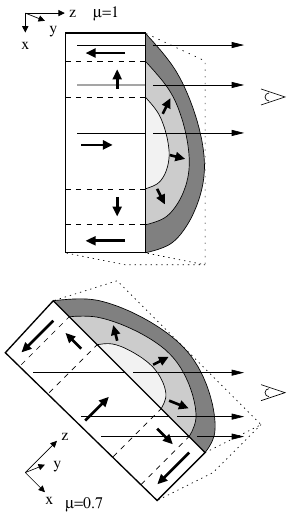}
    \caption{Figure illustrates the critical effect of line-of-sight geometry in the presence of granulation, that is upflows (granules, which occupy most of the stellar surface) and downflows (inter-granular lanes). Here one granule is schematically shown at $\mu=1$ and 0.7. The consequence of this effect is that as the disc centre mostly vertical (longitudinal) flows contribute to intensities, whereas closer to the stellar limb the contribution of horizontal (transveral) motions dominates the emitted radiation field. Image reproduced with permission from \cite{Frutiger2005}, copyright by ESO.}
    \label{fig:velshift0}
\end{figure}

\begin{figure}[ht]
    \centering
    \includegraphics[width=1\linewidth]{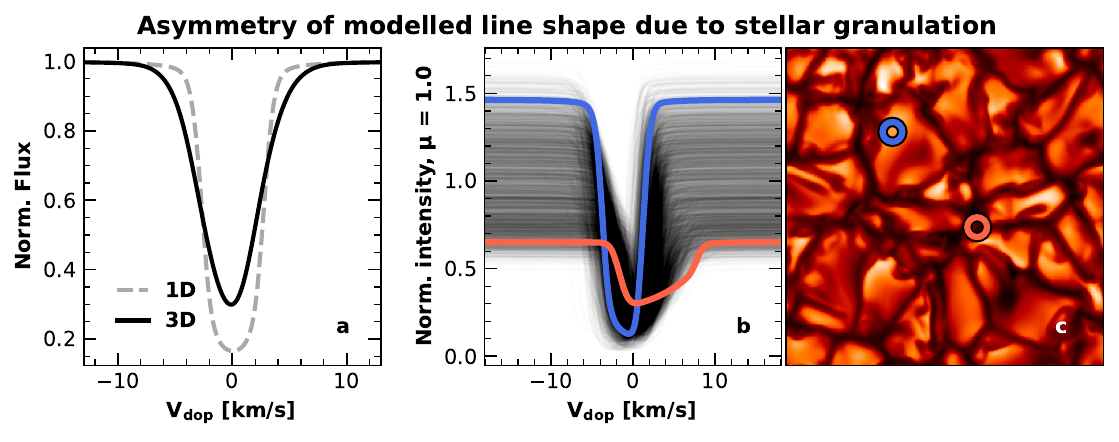}
    \caption{Left panel: Line shape of the 4502.2 \AA\ Mn I line. 1D line computed using a solar \texttt{MARCS} model atmosphere vs. 3D line shape from solar \texttt{STAGGER} snapshot. Middle panel: Variation of 3D line shape across the solar granulation pattern. Wavelengths on the x-axis have been converted to corresponding Doppler shift velocities. Right panel: Continuum intensity at 5000~\AA\ as a function of surface position. The overplotted circles mark the corresponding position of the highlighted profiles in the middle panel. Convection cells produce blue shifted line profiles. Intergranular lanes produce redshifted line profiles.}
    \label{fig:velshift1}
\end{figure}

\begin{figure}[ht]
    \centering
    \includegraphics[width=1\linewidth]{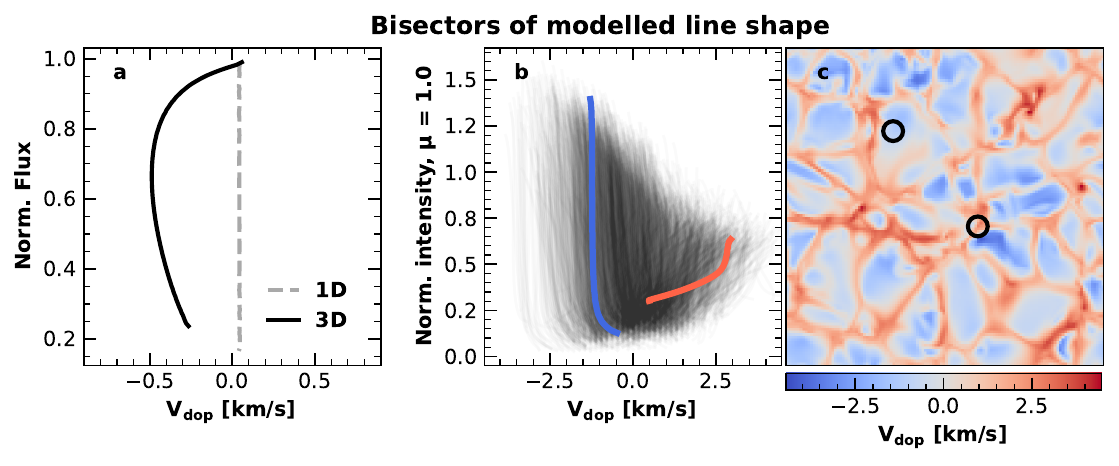}
    \caption{Left and middle panel: Same as Fig.~\ref{fig:velshift1} but showing only line-bisectors instead of line profile. Right most panel: Doppler shift at half maximum of line-bisectors across solar granulation pattern.}
    \label{fig:velshift2}
\end{figure}

\begin{figure}[ht]
    \centering
    \includegraphics[width=1\linewidth]{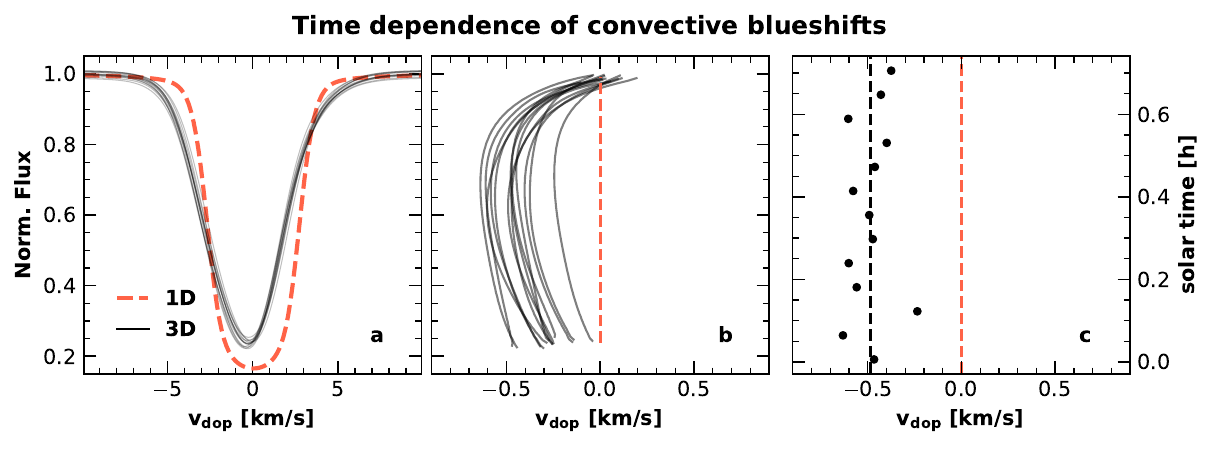}
    \caption{Left panel: Line profiles of the 4502.2 \AA\ Mn I line computed from 13 different \texttt{STAGGER} snapshots. Middle panel: corresponding line bisectors. Right panel: convective blueshift variation over simulation time. Each dot represents one snapshot. The black dashed line shows the average convective blueshift of all 13 snapshots. }
    \label{fig:velshift3} 
\end{figure}

As firmly established from studies of FGK-type stars, convective motions lead to a preferential type of asymmetries in the spectroscopic line shapes, which are known as $C$-shaped profiles \citep[e.g.][]{Dravins1981, Dravins1982, Dravins1990a, Dravins1990b, Asplund2000, Dravins2008, AllendePrieto2002a, Bergemann2019, Dravins2021}, although also inverse $C$-shape have been reported. The asymmetries, and consequently the C-shape bisectors, are due to the line formation dominated by the up-flows \citep{Beeck2013b}, which occupy about $2/3$ of the area across the stellar disc. Figures~\ref{fig:velshift0}, \ref{fig:velshift1}, \ref{fig:velshift2}, and \ref{fig:velshift3} demonstrate how the line asymmetry arises owing to granulation and correlated motions in stellar atmospheres. 

\begin{figure}[ht]
    \centering
    \includegraphics[width=0.7\textwidth]{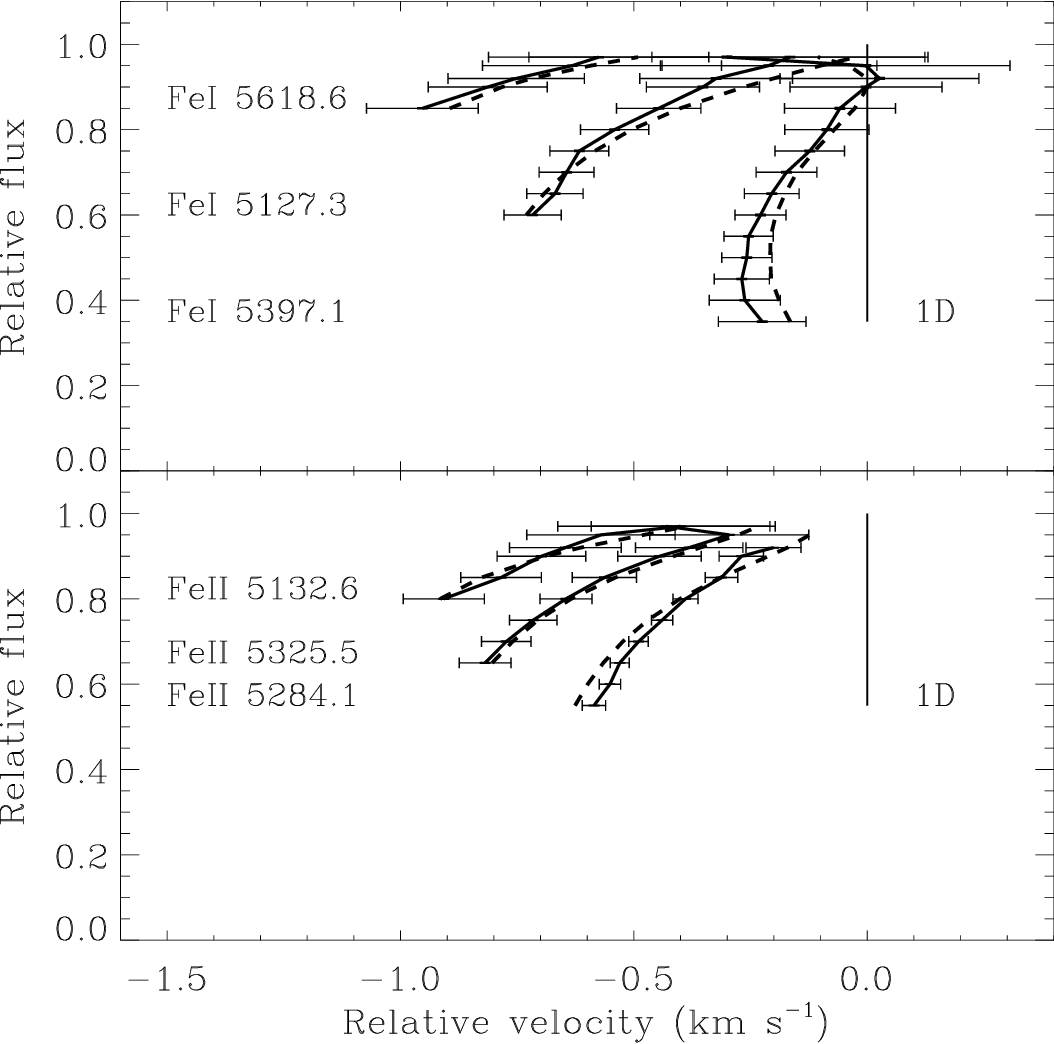}
    \caption{Convective bisectors observed in high-resolution solar (top) and stellar (bottom) spectra in comparison with bisectors obtained from the 3D LTE radiation transfer models. Here the data are shown with solid lines and error bars. The 3D synthetic spectral models are shown with dashed lines. Image reproduced with permission from \cite{AllendePrieto2002b}, copyright by AAS.}
    \label{fig:bisec}
\end{figure}

Among the earlier studies, \citet{AllendePrieto1998} and \citet{AllendePrieto2002b} investigated the effects of convection in the Fe I lines in the very high-resolution spectrum of the Sun and of an F-type star Procyon, see Fig.~\ref{fig:bisec}. They showed that 3D RHD models yield a much improved description of observed line bisectors compared to hydrostatic 1D model atmosphere models. 3D models predict the bisectors to a precision of 50 m/s, except the strongest lines with the EWs in excess of 100 m\AA, for which a small systematic offset at the level of a few 100 m/s between models and data was detected. The origin of the offset is presently unclear and it could possibly be caused by NLTE or effect of magnetic fields. 

\begin{figure}[ht]
    \centering
    \includegraphics[width=0.9\textwidth]{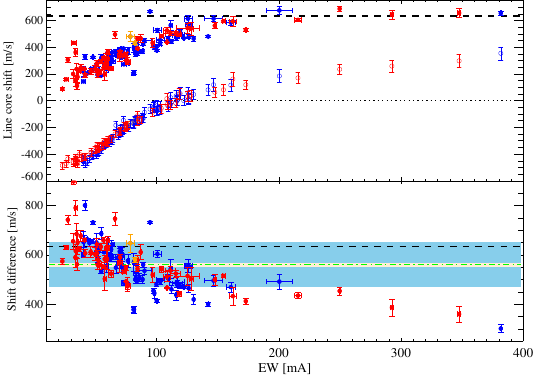}
    \caption{Observed radial velocity shifts obtained from the cores of over 140 Fe I lines in the solar spectrum (here the spectrum of the Moon), compared to the RV shifts computed in LTE using the solar 3D RHD CO$^5$BOLD model. The theoretical solar gravitational redshift 633 m s$^{-1}$ (Einstein) is shown with the black line. Image reproduced with permission from  \cite{Gonzalez2020}, copyright by ESO.}
    \label{fig:RVshifts}
\end{figure}

The systematic variation of convective shifts and bisectors in the solar data has also been explored in \citet{Dravins2008} and \citet{EllwarthEhmann2023}. Early papers included a comparison of observed solar line shifts and those from 3D models for a small number of diagnostic features, however, although satisfactory agreement was demonstrated for weak lines \citep{Asplund2000,AllendePrieto2009}, systematically increasing differences between data and solar models were reported for stronger lines with equivalent widths (EWs) above $\sim$ 60--80~m\AA\ \citep[][]{Gonzalez2020}. Figure~\ref{fig:RVshifts} compares the RV shifts in the solar spectrum with predictions of a 3D RHD models.

Convective bisectors are also seen in the observed positions of lines in the high-resolution spectra of other main-sequence stars \citep{Frutiger2005, Sheminova2020}, and red giants, such as the F-type Procyon \citep{AllendePrieto2002a, Dravins2008} and K-type metal-poor red giant branch star $\alpha$ Boo \citep{AllendePrieto2002b}. \citet{Meunier2017} found systematically stronger convective blueshifts of Ti and Fe lines in the observed spectra of more metal-poor stars, with an amplitude that is larger compared to the predictions of 3D RHD simulations. Analysis of intensity spectra of exoplanet host stars, obtained through spectral reconstruction during the planet's transit, confirms that the spectral lines are redshifted and generally shallower at the limb compared to the disc centre \citep{Dravins2017a, Dravins2017b, Dravins2021}. This finding is qualitatively consistent with predictions of 3D radiative transfer models.

In summary, most of the knowledge on the convective variability of spectral line profiles is currently of an empirical nature. Spectra based on 3D radiation transfer models are indispensable to further understand the nature and amplitude of the variability. 3D RHD simulations support the large (to $\sim -600$ m/s) blueshifts of weaker lines \citep{AllendePrieto1998, Reiners2016}, and these models also predict substantial average RV shifts of up to 0.3 \kms\ for different types of stars \citep{AllendePrieto2013}. Explaining the RV shifts quantitatively may require more refined 3D convective models of stellar atmospheres possibly taking into account magnetic fields.

\subsection{Centre-to-limb variation in lines}
Another important observational effect associated with 3D NLTE radiation transfer is the behaviour of spectral lines across the stellar disc, also known as the CLV.

\begin{figure}[ht]
    \centering
    \includegraphics[width=0.9\textwidth]{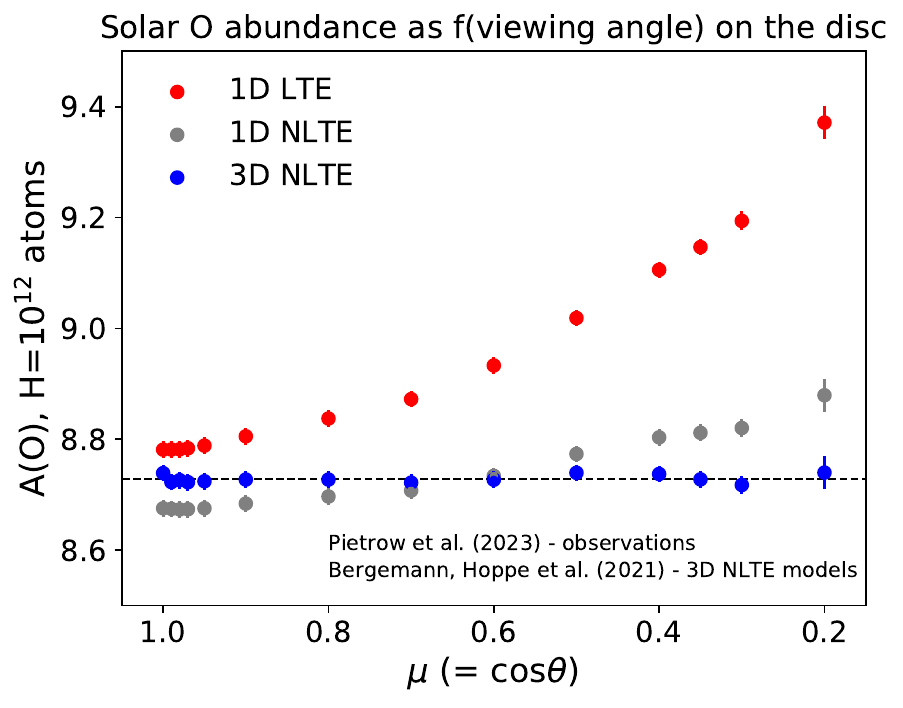}
    \caption{The centre-to-limb variation of the solar oxygen abundance, computed using 1D LTE, 1D NLTE, and 3D NLTE models. The NLTE models for O and the SST data are from \citet{Bergemann2021} and \cite{Pietrow2023b}.}
    \label{fig:CLVO}
\end{figure}

Different studies explored the centre-to-limb variation (CLV) of lines with the primary aim to test the quality of models for abundance retrievals of the solar atmosphere, in particular for iron \citep{Lind2017} and oxygen \citep{Pereira2009,Steffen2015}, the latter being a critical element for the solar structure \citep{Serenelli2009,Vinyoles2017}.
The fact that 1D LTE models of the oxygen triplet fail at the limb has been well known since the 90’s \citep{KiselmanNordlund1995}. In more recent studies, \citet{Bergemann2021} and \citet{Pietrow2023b} combined the new solar data from the SST facility and 3D NLTE radiation transfer to test the ability of the models to describe the variation of line shapes across the solar disc. They found, from the analysis of the line shapes of the O I line at 7772 \AA, that 1D LTE models tend to massively under-predict the strength of the line at the solar limb, thereby over-estimating the chemical abundance of the element by 0.6 dex (a factor 4) compared to the measurements carried out at the disc centre. Fig. \ref{fig:CLVO} demonstrates the applications of the CLV test to the solar O abundance.

Another recent study \citep{Lagae2023} explored in detail the RV shifts of the Ca II triplet lines in 3D NLTE; these are relevant as key metallicity diagnostics in the Gaia RVS spectra.  They found RV shifts of the order $-1$ to $1$ km/s relative to predictions obtained using 1D LTE or 3D LTE models, with differences depending on the type of stars, properties of the instrument (e.g., Gaia versus higher-resolution).

\begin{figure}[ht]
    \centering
    \includegraphics[width=0.9\textwidth]{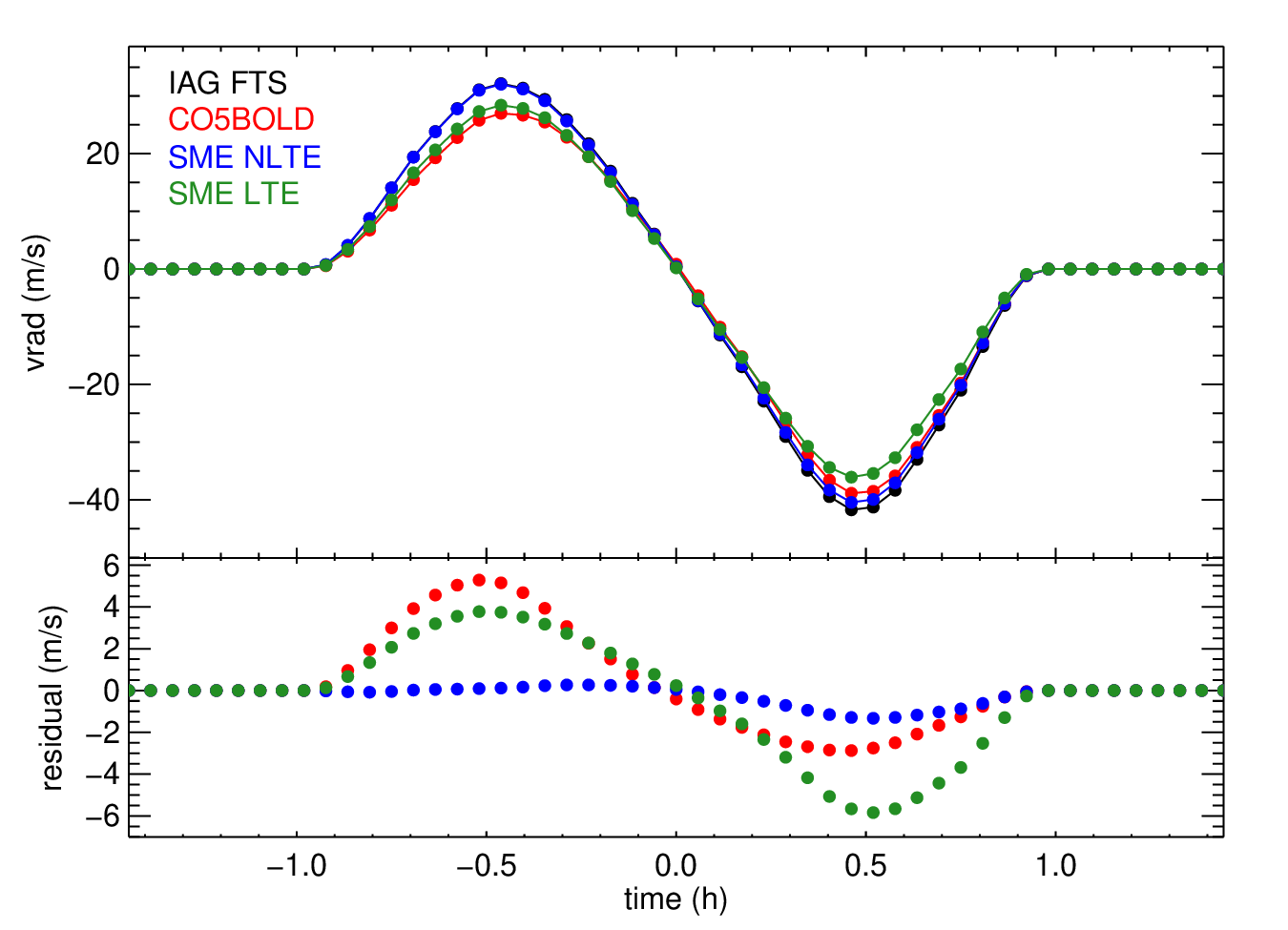}
    \caption{Comparison of the predicted Rossiter--Maclaughlin RV signals in the cores of the Na I D lines as computed using different 1D and 3D LTE / NLTE radiation transfer models. The curves simulate the transit of a hot Jupiter HD 189733b across a star similar to the Sun. Image reproduced with permission from \cite{Reiners2023}, copyright by the author(s).}
    \label{fig:RossMac}
\end{figure}

Other recent investigations address the Rossiter--Maclaughlin effect \citep[see][and references therein]{Ohta2005} with 1D LTE, 1D NLTE, and full 3D RHD models of synthetic solar spectra \citep{Reiners2023}. This effect is the systematic and functionally well-defined distortion of the RV shifts of stellar spectral lines as a function of time, that is caused by the planet transit in the system, where the orbital  axis of the planet and the spin of the host star are approximately aligned \citep{Rossiter1924, McLaughlin1924}. A detailed analysis suggests that solar 3D RHD models (in LTE) do not yield a better description of the Rossiter--Maclaughlin effect in the lines of Na I, Ca I, Mg I, and Fe I compared to 1D models \citep{Reiners2023}. In fact, 1D NLTE models or even 1D LTE models provide a better description of the observed RV signal (Fig.~\ref{fig:RossMac}). A similar study using 3D NLTE \citep{Canocchi2024} indicated an improvement in the Rossiter--Maclaughlin  diagnostics using the solar Na I D and K I lines. Other stellar spectral features, which are relevant for exoplanet diagnostics, however, still remain to be fully explored using 3D NLTE models.

\subsection{Limb darkening of the continuum}

\begin{figure}[ht]
\centering
\includegraphics[width=0.4\linewidth]{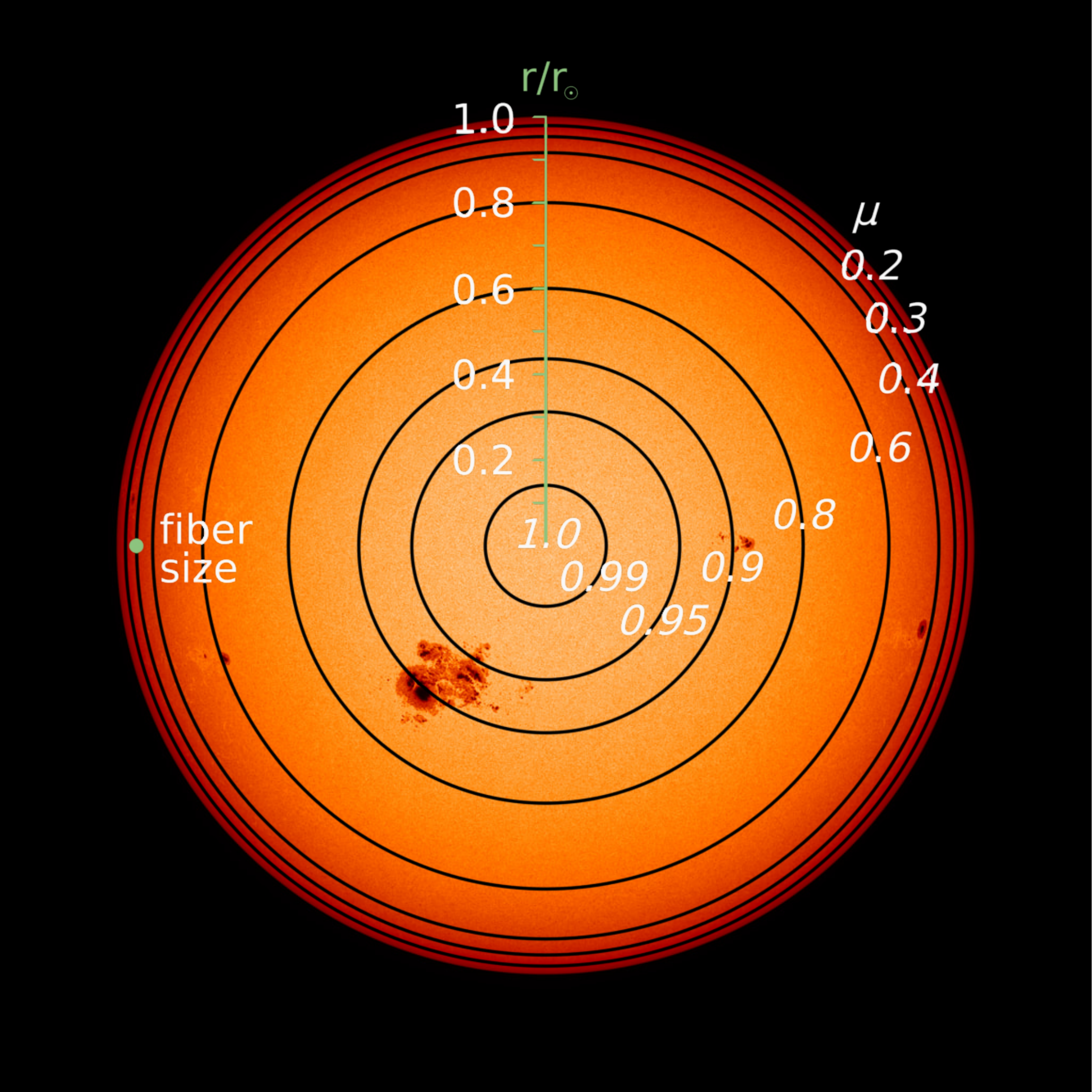}
\includegraphics[width=0.5\linewidth]{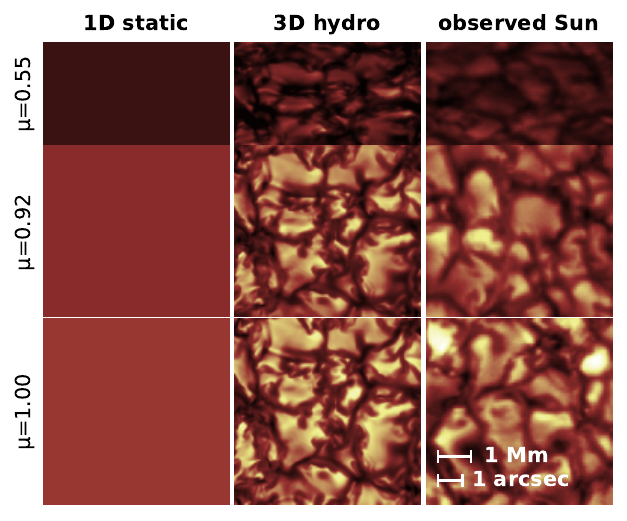}
    \label{fig:clv2}
    \caption{Left: A view of the solar disk by the Helioseismic and Magnetic Imager (copyright by NASA's Scientific Visualization Studio). We overplot the concentric rings to illustrate the viewing positions of constant $\mu$-angle and the corresponding radii. The point shows the fiber size of 32.5 arcsec of the IAG instrument. Right: Comparison observed solar G-band intensity at $\sim 400$ nm with the predictions of 1D hydrostatic and 3D RHD solar model atmospheres at different positions across the solar disc. Both figures from \citet{Hoppe2026}.}
\end{figure}

\begin{figure}[ht]
    \centering
    \includegraphics[width=1\linewidth]{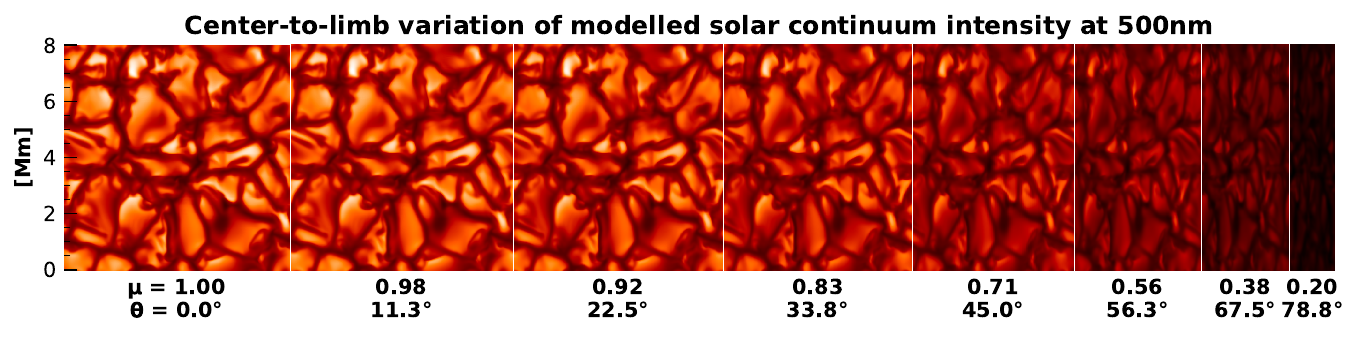}
    \caption{Continuum intensity at \SI{500}{\nm} as a function of surface position and viewing angle. The employed atmosphere model is a solar \texttt{STAGGER} snapshot.}
    \label{fig:CLV500}
\end{figure}

Observations of the solar surface in narrow-band photometric filters can be carried out with different solar telescopes, such as SST. These can be directly compared to predictions of 3D solar RT simulations (Fig.~\ref{fig:clv2}). Figure~\ref{fig:CLV500} shows the surface intensities at 500 nm computed in LTE using the \texttt{Stagger} 3D RHD solar model for different viewing angles. 

\begin{figure}[ht]
    \centering
    \includegraphics[width=\textwidth]{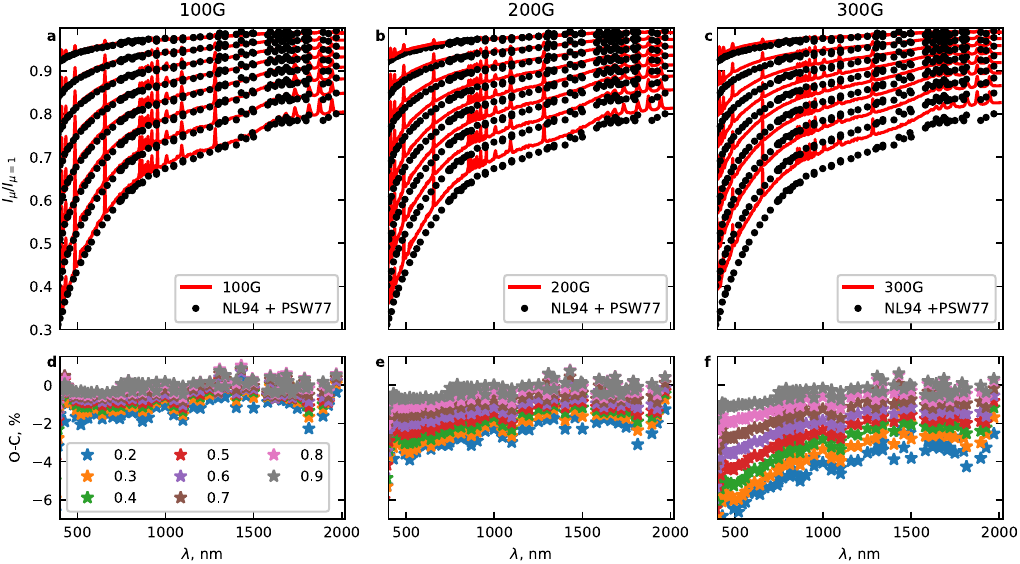}
    \caption{Comparison of the observed solar center-to-limb variation of intensities with the predictions of 3D SSD models. The models assume different levels of the initial vertical magnetic field (see inset). Image reproduced with permission from \cite{Kostogryz2024}, copyright by the author(s).}
    \label{fig:SSD-MHD}
\end{figure}

Comparative theoretical studies of the differences between 1D and 3D radiation transfer effects in the broad-band continuum of the Sun were carried out by \citet{Beeck2012}, \citet{Pereira2013}, and \citet{Eitner2024}. These studies suggest that 3D models yield a good agreement in the optical and near-IR CLV continua with the observed solar data. The good agreement is also seen when comparing intensities computed using the small-scale-dynamo (SSD) 3D RHD solar models \citep{Kostogryz2024}, see Fig.~\ref{fig:SSD-MHD}. An important unresolved question is the observed continuum brightening in the near-UV. 3D RT models under-estimate the continuum intensities at $\lesssim400$ nm, especially at the limb. 

\begin{figure}[ht]
    \centering
    \includegraphics[width=0.8\textwidth]{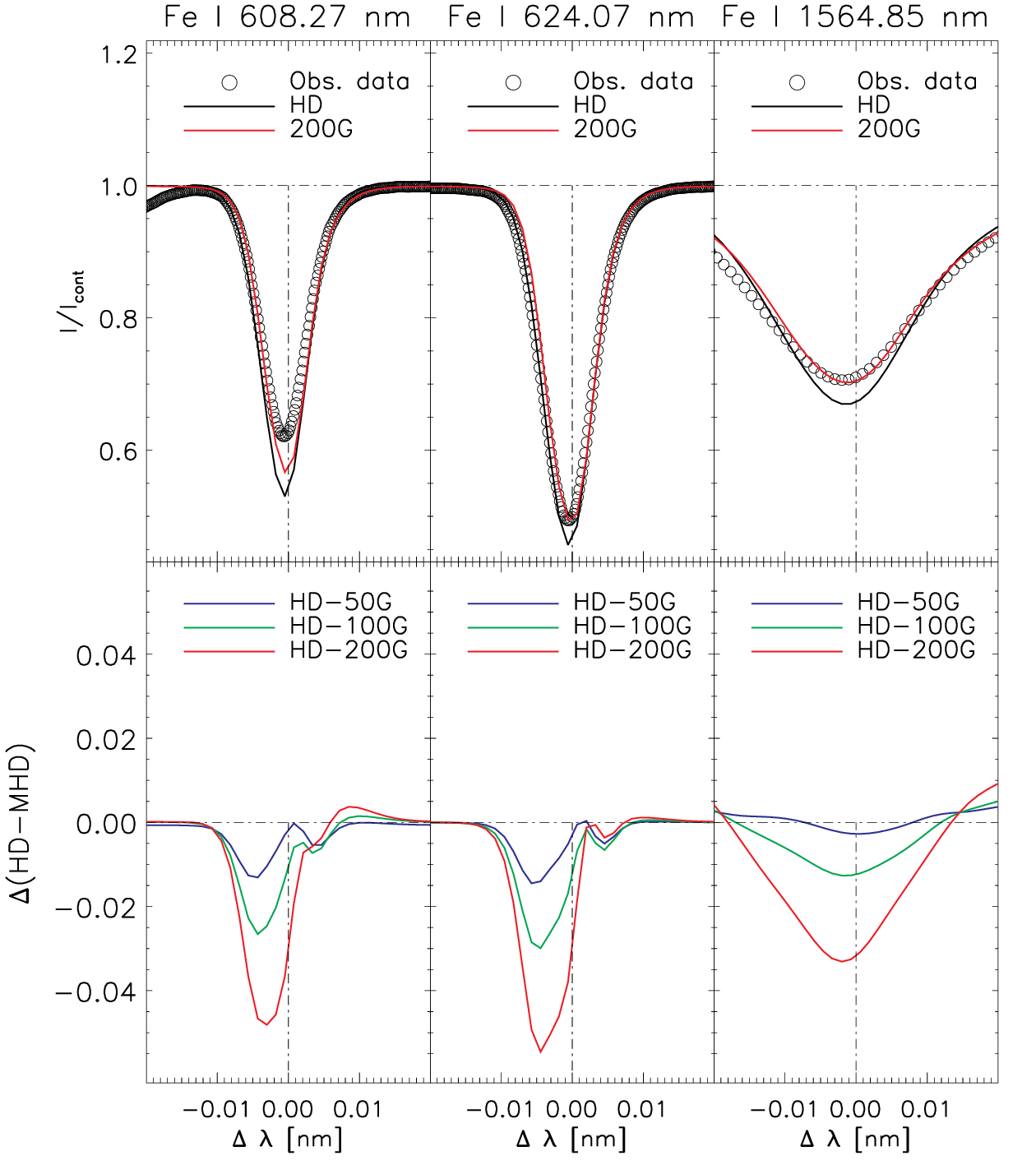}
    \caption{Comparison of the observed solar Fe I lines with the predictions of 3D MHD solar \texttt{stagger} models. The models assume different levels of the initial vertical magnetic field (see inset). Image reproduced with permission from \cite{Fabbian2010}, copyright by AAS.} 
    \label{fig:FeMHD}
\end{figure}

Taking magnetic fields into account \citep[][]{Pereira2013, Kostogryz2024}, yields systematically brighter intensities at all wavelengths and all limb angles. The effect is shown in Fig.~\ref{fig:SSD-MHD}, where the initial uniform  magnetic field of 100 to 300 gauss was applied to the SSD models. The difference with the observed solar data is not a limitation of models, but could also be explained by the observational effect: the solar CLV data were obtained for the quiet Sun, avoiding magnetic concentrations. Indeed, the 10 milli-Tesla (100 gauss) 3D MHD model, which is of the order of the mean magnetic field of the quiet Sun \citep{TrujilloBueno2006}, describes the IR solar observations better compared to a non-magnetic 3D RHD model \citep{Pereira2013}. The magnetized 3D model is slightly hotter in the photospheric line formation regions, by up to $\sim +100$ K, compared to a non-magnetic RHD model \citep[][their Fig.~3]{Fabbian2010}. As seen in Fig.~\ref{fig:FeMHD}, the improvement due to MHD effects is also visible in some of the observed solar Fe I lines  \citep{Fabbian2010}. Hence, it still remains to be explored what is the cause of the puzzling mismatch between the observed solar ultra-violet CLV data and the models.

With the availability of TESS lightcurves, it has become possible to place observational constraints on limb darkening of other stars \citep{Maxted2018, Maxted2023}. Comparison of the data with spatially resolved intensities derived from 1D and 3D model atmospheres suggests that the latter provide an improved agreement with the data for main-sequence stars \citep{Maxted2023} compared to 1D LTE hydrostatic models. However, it is also possible to mimic the effects of stellar sub-surface convection by tuning the free parameters, such as those associated with simplified theories describing convecting overshoot and mixing length, in 1D LTE model atmospheres \citep[e.g.][]{Kostogryz2022}. 

Spectral synthesis using 3D MHD simulations further helps to improve the accuracy of the CLV predictions for Sun-like stars. \citet{Ludwig2023} explored 3D MHD models in the context of reproducing stellar optical observations obtained within the Kepler space mission. The principal physical effect is the relative brightening of the limb due to magnetic fields \citep[see the discussion in][]{Nordlund2009}. A 3D model with 200 to 400 gauss can be up to 200~K warmer at the optical surface compared to the analogous non-magnetic 3D model with the same \teff, \logg, and [Fe/H] \citep{Ludwig2023}. However, at larger values of $\langle B_{\rm z} \rangle~ \gtrapprox 1$ kG a drastic cooling of the entire structure due to suppressed convective motions occurs, leading to the photospheric temperatures drop by almost 1000~K. Also the temperature fluctuations generally increase with increasing $\langle B_{\rm z} \rangle$. 

\begin{figure}[ht]
    \centering
    \includegraphics[width=\textwidth]{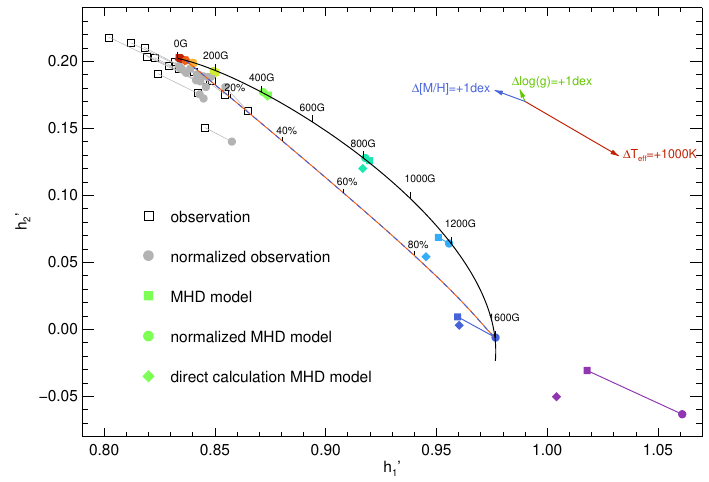}
    \caption{Comparison of intensity parameters (h1, h2) predicted by 3D MHD models with the observed data obtained from Kepler light-curves. See text. Image reproduced with permission from \cite{Ludwig2023}, copyright by the authors(s).}
    \label{fig:CLV-MHD}
\end{figure}

As a consequence, stellar intensities change substantially when 3D MHD models are used in the calculations of stellar limb darkening (Fig.~\ref{fig:CLV-MHD}). Specifically, in the optical wavelength range from 450 nm to approximately 900 nm, the relative continuum intensities at $\mu = \cos(\theta) = 2/3$ ($\theta = 48^\circ$) are about 10\% higher in 3D MHD models with magnetic fields of around 1 kG than at the disc centre (denoted here as h$'_{1}$; see Eq.~5 in \citealt{Ludwig2023}). Moreover, the intensities at the limb ($\mu = \cos (\theta) = 1/3$, $\theta = 70.5^{\circ}$) are another 10\% brighter compared to that at 50 deg. The sensitivity of intensities to magnetic effects at the level of 0.5 kG is comparable to that of the change in \teff~~by nearly 1000 K.

In summary, the relative brightening of the limb due to magnetic fields is well-established \citep[e.g.][]{Ludwig2023,Kostogryz2024}, but it remains to be explored to which extent do these differences influence studies of stellar activity based on light-curves. Current 3D MHD models for late-type stars consistently under-predict the limb brightening compared to Kepler light-curve data \citep{Maxted2018}, which may be attributed to a more complex morphology and distribution of magnetically active regions at the stellar surface \citep{Ludwig2023}.

\subsection{Atmospheric dynamics}

Signatures of atmospheric dynamics are also seen in the spectra of red supergiants. The nearby asymptotic giant branch (AGB) star S Ori and red supergiant Betelgeuse have been, in particular, placed in the focus of spatially-resolved studies of atmospheric dynamics with 3D models \citep{Kravchenko2019, Kravchenko2020, Kravchenko2021}. Photocenter variability of Betelgeuse as seen in Gaia were furthermore attributed to large-scale granulation and giant convective cells \citep{Chiavassa2020, Chiavassa2022}, although other studies \citep{Kochanek2023} find that the instrumental noise is too large to claim the detection of surface sub-structure on RSGs with current facilities like Gaia.

Recently, 3D simulations of red supergiants were carried out in spherical-symmetry using the Athena++ code \citep{Goldberg2022}. Such 3D star-in-a-box models are essential to map the connection between stellar pulsations, atmospheric dynamics, and mass ejection \citep{MacLeod2023}, and these models have paved the way to understand interesting observations such as the 'Great Dimming' event of Betelgeuse \citep{Montarges2021, Dupree2022} and similar extragalactic events reported in the literature \citep[e.g.][]{Jencson2022}.

\subsection{Exoplanets and kilonovae}

Whereas 3D and NLTE RT in the context of modelling of these two type of objects will be a subject of this review, there are currently no direct spatially-resolved astronomical observations that allow to test the relevance of 3D hydrodynamics and NLTE modelling in their photospheres. Here we adopt the definition that the photosphere is the outer region of the object, where the radiation becomes optically thin and is lost from the object into the interstellar matter, which of course depends on the wavelength.

It, however, is obvious that the main reason for including the effects of geometry and appropriate representation of space and time is primarily the fact that all objects are three-dimensional in nature and conditions in the photospheres of these objects favour deviations from hydrostatic equilibrium and local thermodynamic equilibrium. 

\section{Basic inputs}\label{sec:basicinputs}
\subsection{Model atmospheres} \label{subsec:atmospheres}

In order to perform 3D NLTE radiative transfer, pre-computed 3D RHD stellar model atmospheres are typically used. Different codes are available and can be used to compute 3D radiation-hydrodynamics models (RHD) of FGKM type stars, with codes such as \texttt{MURaM} \citep{Vogler2004, Vogler2005}, \texttt{CO$^{5}$BOLD} \citep{Freytag2012, Hoefner2018}, Stagger \citep{Magic2013a} and M3DIS \citep{Eitner2024}. Also, 3D HD models have become available \citep[e.g.][]{DelbroekSundqvist2025}. In this review, we only provide a brief summary of the models and techniques, and we refer to other excellent reviews dedicated to 3D M(RHD) modelling of sub-surface stellar convection \citep{Nordlund2009, Kupka2017, Leenaarts2020, Chiavassa2024}.

All current 3D RHD models of stars assume LTE in the calculations of opacity and source function of lines and continua. To the best of our knowledge, no 3D NLTE modeling of atmospheres has never been done. However, an approximate treatment of scattering in the lines and/or continuum can be included \citep{Hayek2010, Hayek2011}, primarily in the framework of coherent and isotropic scattering, representing Rayleigh scattering in the continuum. 

\begin{figure}[ht]
    \centering
    \includegraphics[width=0.8\linewidth, angle=0]{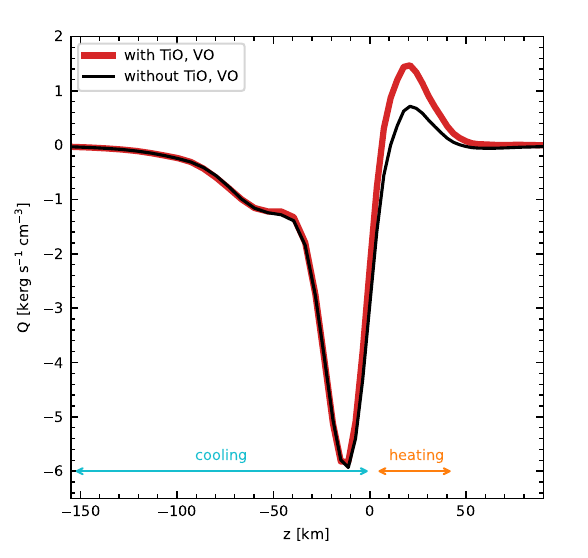}
    \caption{Effect of line opacities in molecular bands of TiO and VO on the radiative energy term $Q$ in the energy equation. The figure shows the data for a 3D RHD simulation of an M dwarf \citep{Perdomo2023}.} 
    \label{fig:surfaceQ3D}
\end{figure}

Comparative analysis of the basic thermodynamic properties of these models suggest excellent agreement in the deeper adiabatic layers \citep{Beeck2012, Eitner2024}, where radiative transfer is computed in the diffusive approximation. But slightly larger deviations between the models are seen in the outer layers, which are sensitive to the detailed micro-physics of line blanketing \citep{Beeck2012}. Figure~\ref{fig:surfaceQ3D} illustrates the effect of line opacities in the molecular bands of TiO and VO on the radiative energy term $Q$ in the energy equation, specifically for a cool M-type dwarf. Inclusion of these molecules in LTE RT, especially for cold stars, leads to a substantial surface heating in the region where these lines contribute to opacity. Of course, radiative cooling would also take place, once the transitions become optically thin. The analysis of the influence of opacity binning on the model structure was the subject of \citet{Ludwig1992} and is also extensively investigated in \citet{Perdomo2023}.

\begin{figure}[htbp]
    \centering
    \includegraphics[width=0.9\textwidth]{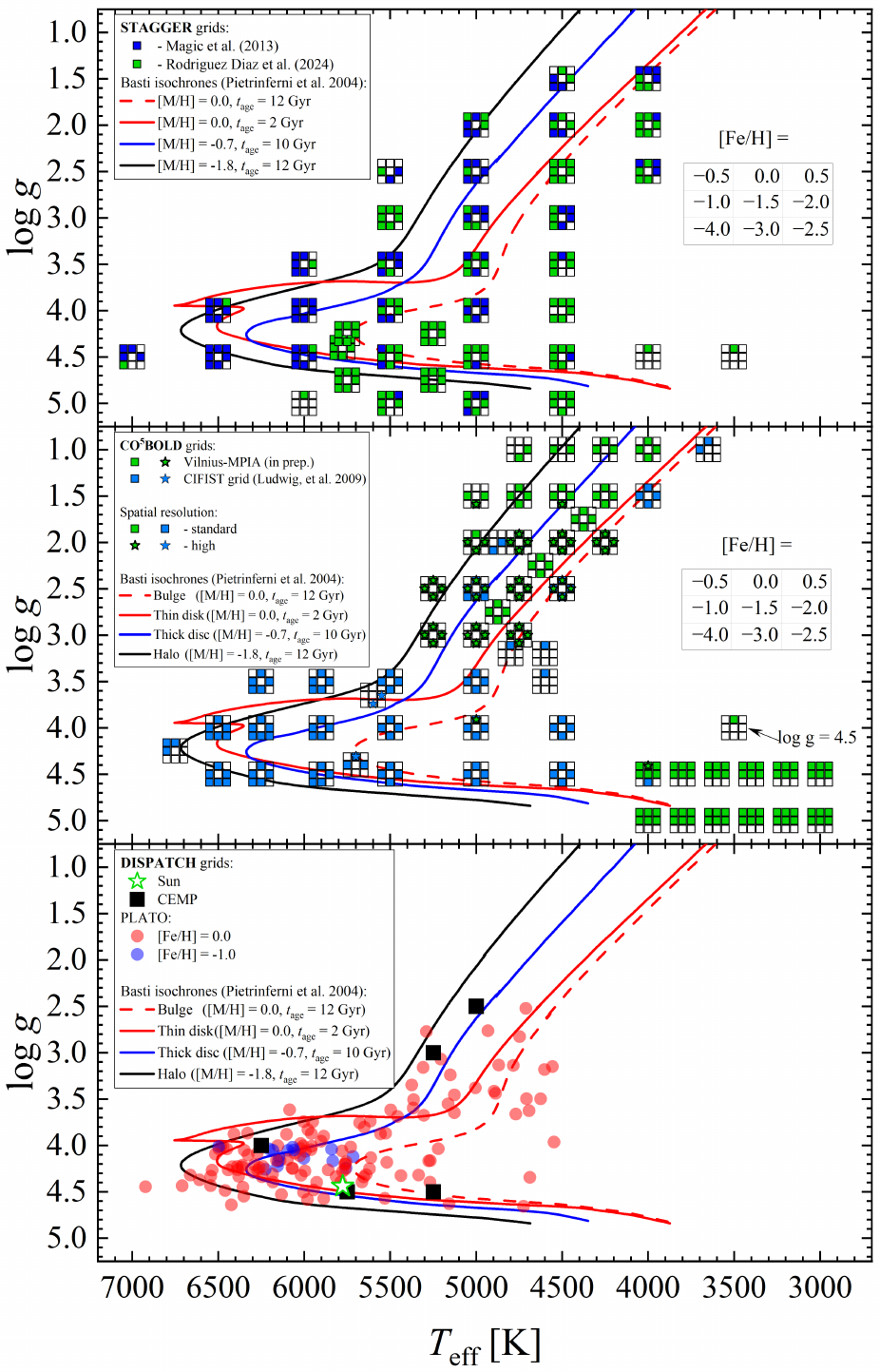}
    \caption{The Stagger, 3D CIFIST (Co5BOLD), MPIA (M3DIS at Dispatch) grid of 3D RHD models. Image courtesy of Jonas Klevas. The MPIA grid is available at \url{nlte.mpia.de}, whereas the \texttt{CO$^{5}$BOLD} MPIA-Vilnius grid for M dwarfs is under development by J. Klevas \citep{Klevas2022}.}
    \label{fig:stellargrid3D}
\end{figure}

Some of the afore-mentioned codes were used successfully to generate grids of 3D models for entire parameter space of FGK-type stars \citep{Ludwig2009b, Magic2013a, Magic2013b, Bonifacio2018, Kucinskas2018}. Each model within the grid typically comprises of selected time-slices, that is 3D snapshots, of stellar atmospheres that tabulate temperature $T$, density $\rho$, and velocities in the x, y, and z directions $v_{x}, v_{y}, v_{z}$ on a spatial grid consisting of (nx$\times$ny$\times$nz) points. For detailed spectroscopy, number densities (or partial pressures) of all species (atoms, molecules) are usually computed from $T$ and $\rho$ within the post-processing NLTE and spectrum synthesis codes. Figure~\ref{fig:stellargrid3D} shows the Stagger, the 3D CIFIST, and MPIA grid of 3D RHD models. The coverage is large enough to enable a complete exploitation of 3D RHD structures over the cool part of the HRD, covering the domain of FGKM type stars.

3D RHD simulations require a certain minimum spatial resolution and a surface area in order produce realistic convective patterns and accurate thermodynamic (temperature, pressure, velocity) structures, and to ensure that the radiative balance is captured accurately at all depths. The surface area has to be large enough to contain a representative number of convection cells. Usually, 10 to 20 convection cells are viewed as sufficient to capture the statistical properties of the atmosphere, and such setup is typically used in spectroscopic studies \citep[e.g.][]{Collet2007,Bergemann2012a, Magic2013a, Amarsi2017}. The vertical resolution typically lies around 10~km between adjacent voxels\footnote{This is a term that comes from the discipline of computer graphics and may not be taught in physics academic curricula. Here we provide the definition of the term as provided by Wikipedia: ``{Voxel is a three-dimensional analogue of a pixel; a volume element representing some numerical quantity of a point in a 3D space}''}. However, the spatial discretization can vary substantially depending on the desired application \citep[][their Tab. 2]{Eitner2024}, and on the types of stars \citep[][their Fig.~3]{Rodriguez2024}. The vertical step size ranges from $\delta z$ of $\sim 10$ km around the optical surface to $\delta z \approx$ 3 Mm at the bottom of the simulation for a typical red giant (total size of the 3D box over 250 Mm). \citet{Eitner2025} furthermore show that a major improvement in the accuracy of the RT solution, and hence a substantial gain in computational efficiency, can be obtained by using a conservative RT solver. That is, in the calculations using short-characteristics solver, the intensities are computed at cell interfaces, not in cell centers. This approach conserves energy and allows to reduce the vertical resolution by a factor of 2 to 3. 

\begin{figure}
\includegraphics[width=0.9\linewidth, angle=0]{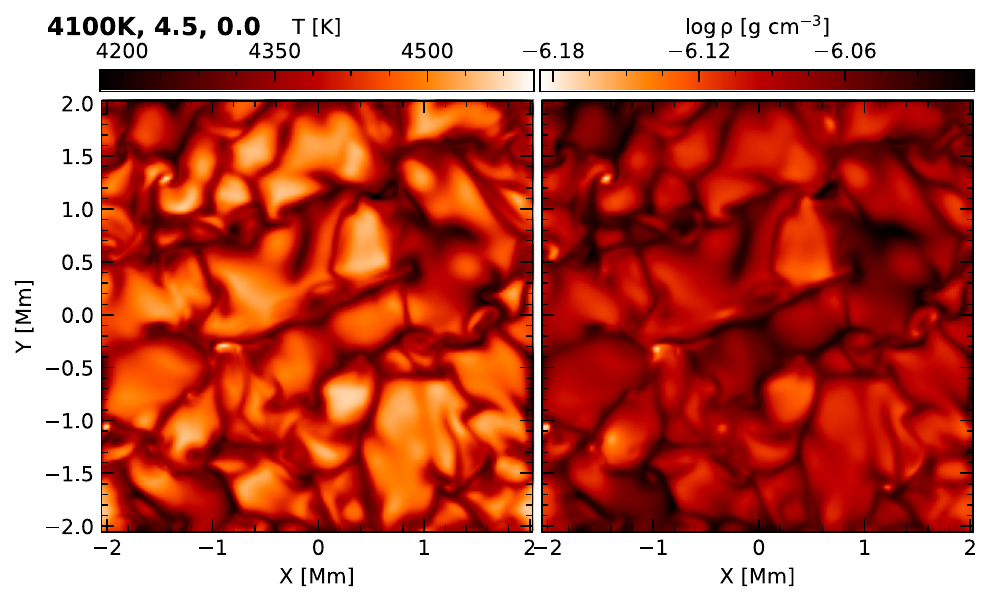}
\includegraphics[width=0.92\linewidth, angle=0]{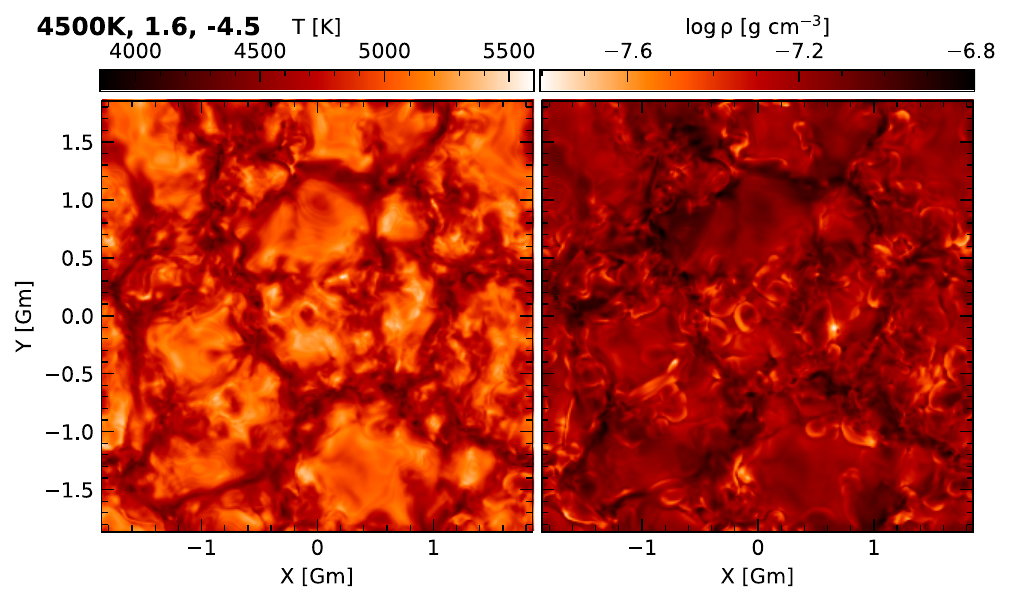}
\caption{Surface temperature structure at optical depth unity for two types of stars, a typical M dwarf and a M-type red giant. Note the difference in spatial scales of the simulation boxes, ranging from a few Mm to Gm for a red giant. The M dwarf 3D model adopts the solar metallicity, whereas the red giant 3D RHD model is a CEMP model computed using [Fe/H]$=-4$. Images courtesy of Philipp Eitner, see \citet{Eitner2025} for details.} 
\label{fig:cubes}
\end{figure}

\begin{figure}
\centering
\includegraphics[width=0.8\linewidth]{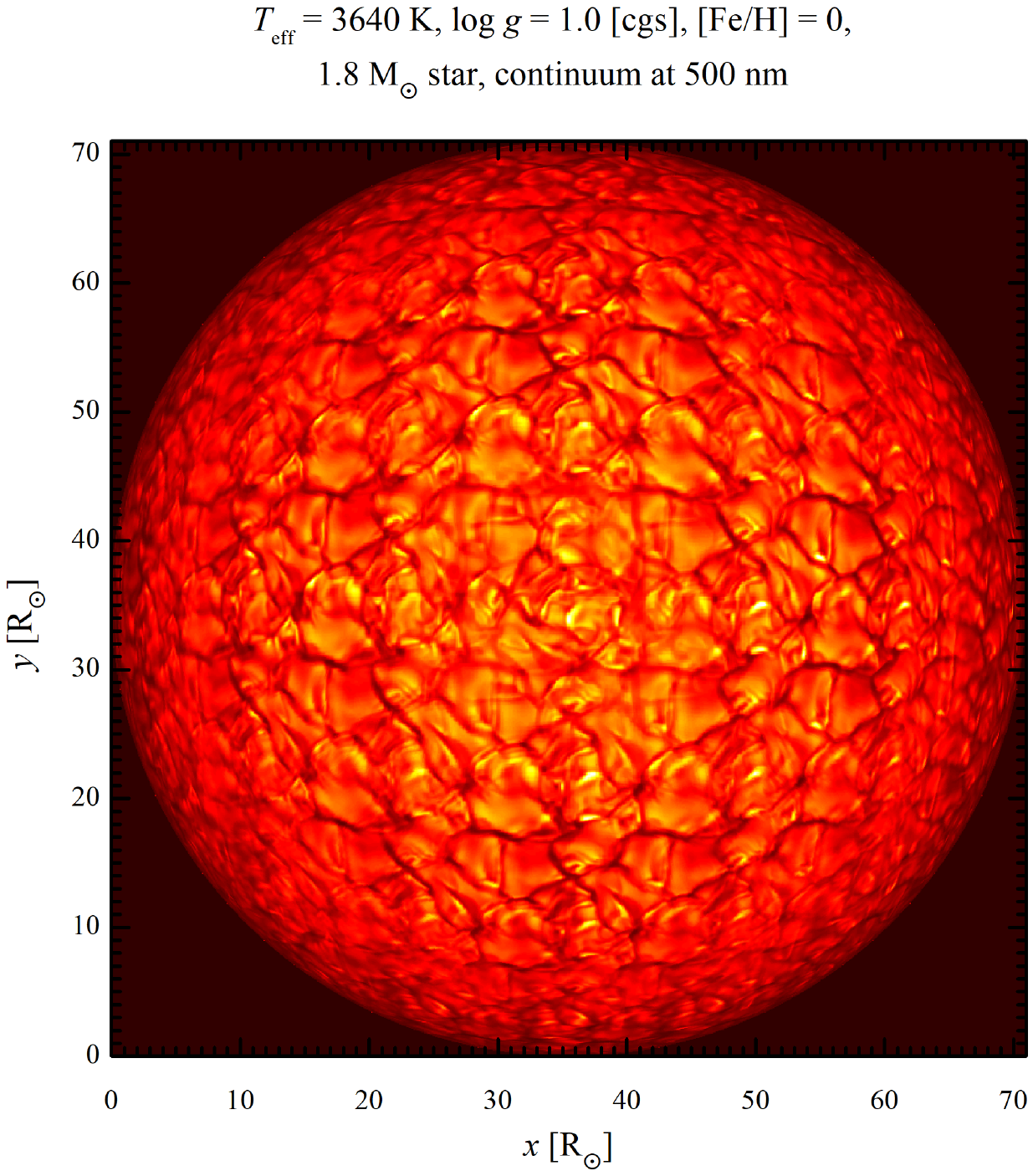}
\caption{Radiation field at 500 nm computed using a 3D RHD model of a red giant with parameters indicated in the figure header. The 3D RHD cube (box-in-a-star) was used to fill the entire sphere using the radius of 70 R$_{\odot}$ for a mass of 1.8 M$_{\odot}$, and tiled to account for the limb darkening. Image courtesy Jonas Klevas.} 
\label{fig:fullsphere}
\end{figure}

\begin{figure}
\includegraphics[width=0.9\linewidth, angle=0]{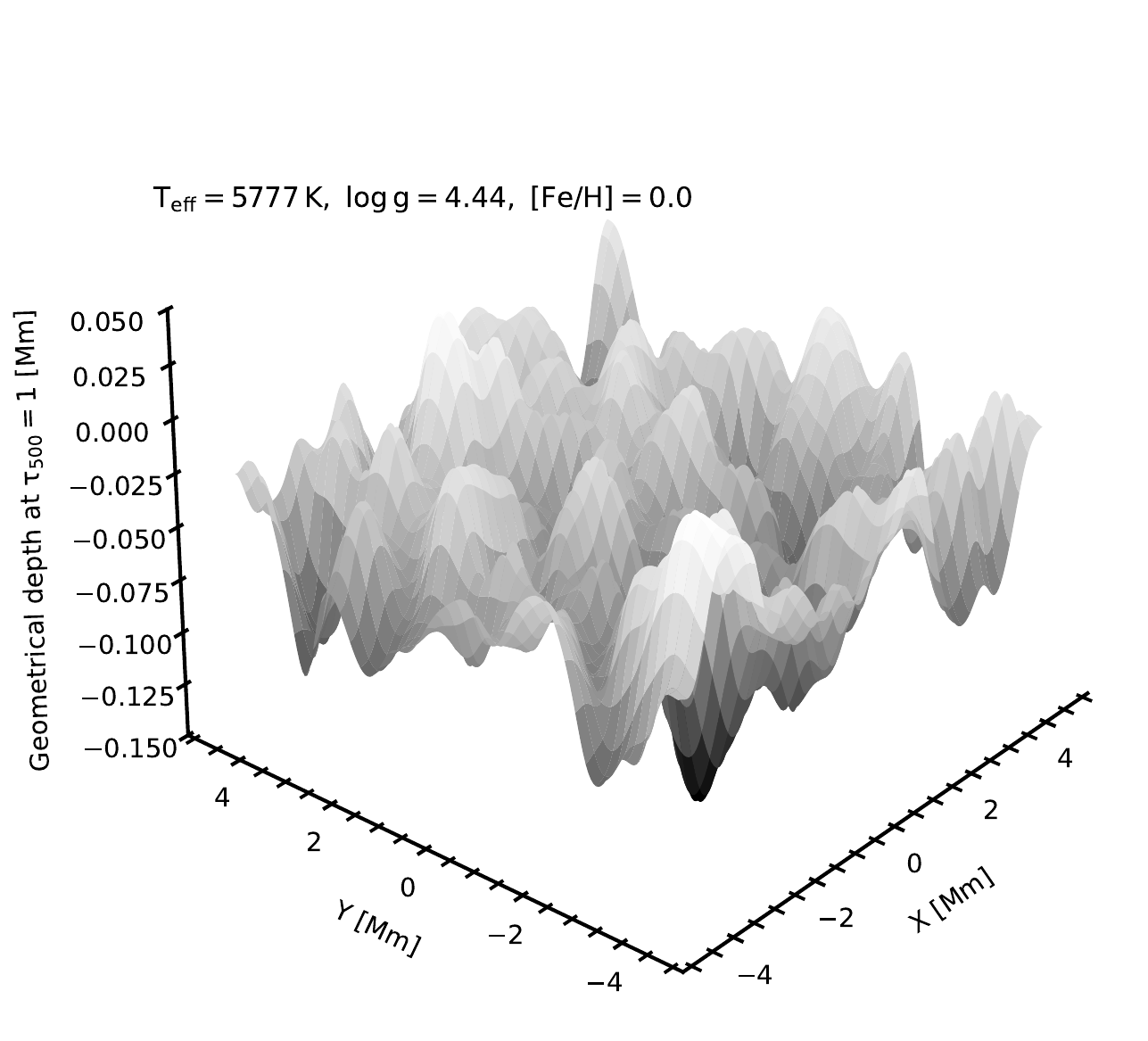}
\includegraphics[width=0.92\linewidth, angle=0]{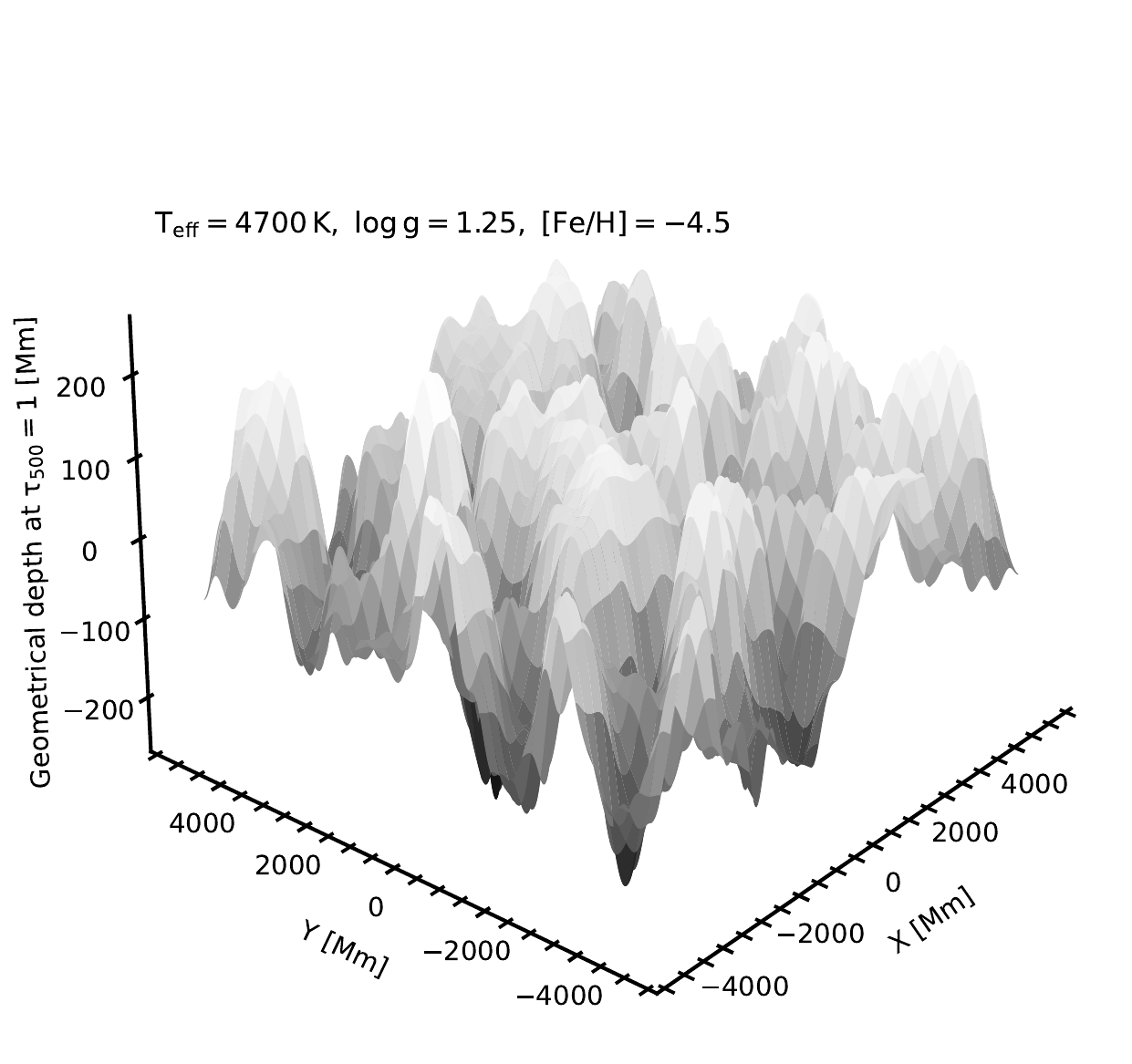}
\caption{Optical depth unity surfaces of two representative stellar model atmospheres computed at different values of $T_{\rm eff}$, log$g$, and metallicity [Fe/H]. Images courtesy of Philipp Eitner, see \citet{Eitner2025} for details.} 
\label{fig:optdepth}
\end{figure}

RT is then carried out in the quasi-stationary setup, using the so-called 3D cubes of physical quantities extracted using regular time steps, typically minutes to hours, depending on whether a main-sequence or an RGB star is analysed. These cubes are commonly referred to as "snapshots". In Fig.~\ref{fig:cubes}, we show the surface temperature structure at optical depth unity for two types of stars, a typical M dwarf and a red giant branch star. We note that the boxes are very small compared to the total size of the star, and for a more intuitive understanding we additionally show in Fig.~\ref{fig:fullsphere} how the 3D RHD box-in-a-star cubes fit into the entire stellar sphere (here a red giant) in physical units. Additionally, in Fig. \ref{fig:optdepth} we show the optical depth surfaces of two selected M3DIS model atmospheres, to highlight how highly-corrugated these surfaces are and why 3D RT calculations in such physical structures require special carefully-tested methods that can deal with strong curvature and non-linearities.


Synthetic stellar spectra computed for different snapshots are then averaged in time \citep[e.g.][]{Bergemann2019,Gallagher2020}, applying the Ergodic hypothesis (credited to Ludwig Boltzmann, \citealt{Boltzmann1964}), which postulates --- for long time series and large volumes --- the equivalence between averaging in time and averaging in space. 

The horizontal and vertical resolution of the 3D cubes in the $(x,y,z)$ space is relevant in the context of  asymmetries of spectral lines. The detailed line properties are sensitive to the statistics of convective up- and down-flows, and hence usually careful tests by down-sampling of the 3D RHD models are performed \citep{Magic2013, Steffen2015, Nordlander2017}. For example, \citet{Rodriguez2024} showed that in LTE, the error introduced by down-sampling the 3D cubes depends on stellar parameters, but for weaker lines (reduced log$_{10}$(EW)$< -5$) is typically within 0.02 dex to 0.03 dex at a vertical resolution of 120 grid points for both turnoff stars and red giants. On the other hand, these asymmetries have a minimal effect on the integrated radiative rates in the SE calculations. Therefore, it is possible to compute the departure coefficients using  downsampled 3D snapshots, e.g., at horizontal resolution of $(x,y)=(30,30)$ \citep{Bergemann2019}. These can then be interpolated and applied to the full-sized 3D snapshot to perform the spectrum synthesis. While this approach applies to the horizontal resolution, it does not necessarily apply to the vertical resolution of the boxes. The steep drop in opacity in the photosphere (caused by the continuum $H^{-}$ cooling in FGK-type stars) has to be resolved as well as possible, therefore it is best not to reduce the vertical resolution of the snapshot in the vertical. Potentially one can also utilise static mesh refinement, which keeps the layers of the photosphere at maximum detail, but interpolates the layers at large optical depths onto a coarse grid as the temperature structure becomes more and more homogeneous deeper in the atmosphere.

Finally, we would like to point out that one very convenient approach is to collapse all the information from the 3D simulation cubes into a 1D depth-dependent structure. These models are known as average 3D, $<$3D$>$ models \citep{Bergemann2012a, Magic2013b} and they drastically simplify the RT calculations. This is done with the goal to retain the bulk average properties of the system, in particular, the average behavior of $T$, $\rho$, velocity vector, but not retaining their horizontal variability (in the x-z plane). Various methods of averaging were studied in the literature, including averaging over iso-surfaces of geometrical depth, optical depth, column-mass density, as well as logarithmic, weighted by the 4th power of temperature \citep{Magic2013b}. Sometimes a depth-dependent velocity profile with a value of one standard deviation of the 3D velocity components \citep{Uitenbroek2011} is adopted. \citet{Magic2013b}, find that averages on constant column-mass density work more reliably for LTE RT calculations, however, for NLTE calculations it has become more common to use averages on surface of optical depth at $500$ nm or on Rosseland optical depth. Applications of such models are still rare but they have already been successfully in the context of solar and stellar spectroscopy \citep{Bergemann2012a, Nordlander2017, Magg2022, Gerber2023}, predictions of adiabatic oscillation frequencies from solar models \citep{Magic2016}, and stellar structure and evolution calculations \citep{Jorgensen2019, Zhou2025}.
\subsection{The equation of gas state}
The Equation of State (EoS) defines the relationship between various state variables, which describe the properties of a plasma, gas or more generally, any kind of medium. These state variables include temperature $T$, density $\rho$, pressure $P$, internal energy and more, therefore the EoS is not strictly a single equation.

While the formal solution of the RT equation is similar in all types of astrophysical media, the choice of the EoS defines the applicability of a numerical code to a certain $T$-$\rho$ regime, since it plays a central role when determining the radiative opacities and hence stability with respect to convective energy transport \citep{Cox1965, Mihalas1978, Hummer1988}. The most common and simplest EoS is the ideal gas law:
\begin{equation} \label{eq:Ideal}
    P_{\rm gas} = n{k_B}T,
\end{equation}
where $P_{\rm gas}$ is the gas pressure, $n\equiv N/V$ is the number density of particles in the gas (per unit volume), $k_B$ is the Boltzmann-constant, and $T$ is the gas temperature. The latter is sometimes referred to as kinetic temperature as it defines the kinetic energy of particles, assumed to be Maxwellian in stellar atmospheric conditions. But the latter is not the case for kilonovae, Sect. \ref{sec:kilonovae}, where electrons are produced in radioactivity and thus do not follow the Maxwell-Boltzmann distribution. 

The relation \ref{eq:Ideal} appears simple, but it hides the fact, that $n$ has a number of dependencies, including a strong exponential $T$ dependence. For the gas composed of free electrons $e$, ions (atoms in different ionization stages), and molecules, the total number density is:
\begin{equation}
    n = n_e + n_{\rm ions} + n_{\rm molecules},
\end{equation}
%

The number densities of all particles in the eq. \ref{eq:Ideal} depend on $T$, chemical composition of the gas, and other physical quantities. Different particles can combine in one way or. If the gas is in thermodynamic equilibrium, the chemical equilibrium constant $K_{A+B}$ may be used. These constants yield the ratios of particles from a given process of dissociation or attachment, $A + B \rightleftharpoons AB$.

\begin{equation}
    K_{A+B} = \left[\frac{n_A n_B}{n_{AB}}\right]_{\rm LTE}
\end{equation}

In TE, the relative number of particles at a given temperature $T$ are given by:
\begin{equation}
    \left[\frac{n_A n_B}{n_{AB}}\right]_{\rm TE} =
    \frac{U_A U_B}{U_{AB}}
    \left(\frac{2\pi  k{_B} T}{ h^2}\right)^{3/2}
    \left(\frac{m_A m_B}{m_{AB}}\right)^{3/2}
    e^{-E_D/k{_B}T},
\end{equation}
where $U$ are the temperature dependent partition functions of the respective species, $m$ are their masses, $E_D$ is the dissociation energy of the compound specie (molecule), and the rest of the constants have their usual meanings. This form is used in the majority of atmosphere codes for stars and exoplanets (see Sect. \ref{sec:codes} for more details).

The fundamental Saha equation describes the relationship between atoms in different ionization stages $I$ in equilibrium at a given $T$ value:
\begin{equation} \label{eq:Saha}
    \left[\frac{n_e n_{I+1}}{n_I}\right]_{\rm LTE} = \frac{2 U_{I+1}}{U_I} \left(\frac{2\pi m_e k_B T}{h^2}\right)^{3/2} e^{-E_I/k{_B}T},
\end{equation}
where $E_I$ is the ionisation threshold energy. 

Given estimates of $K_{A+B}$ for the most frequent interactions, one can assemble a system of non-linear equations and solve it iteratively to obtain the number densities of the different particles. This system of equations is closed through particle conservation. The number of free electrons changes due to ionisation of atoms, and thus $n_e$ is obtained by summing the number of ionised particles multiplied by their corresponding ionization stage. To determine an electron number density, which is consistent with $T$ and $\rho$, an iterative process is used, starting with an initial guess for $n_e$, since it is also required in the Saha equation.

The other fundamental equation, the Boltzmann equation, relates the ratio of bound energy states (states within the same ionization stage) is:
\begin{align}
    \left[\frac{n_j}{n_i}\right] = \frac{g_j}{g_i} e^{-(E_j - E_i)/k_BT}, 
    \label{eq:Boltzmann}
\end{align}
where $E_i$ is the excitation energy of the level $i$ and $E_j$ is the excitation energy of the level $j$, and $g_i$ are the statistical weights of both levels, respectively. This formula also applies to energy states above the continuum, but it needs to be modified slightly for states very close to the ionization limit at low temperatures.

The choice of partition functions $U_I$ of the ions has an impact on the EoS \citep{Graboske1969}. They influence the state variables through Eq.~\eqref{eq:Saha} and they enter the opacities when computing the number densities of individual atomic energy states in LTE.  The partition function of the ionization stage $I$ of an ion is give by the sum:
\begin{align}
    U_I = \sum_i g_i e^{-(E_i - \Delta E)/k_BT}, \label{eq:Partition}
\end{align}
here, the $\Delta E$ term is introduced since ionisation and excitation energies are effectively altered by the presence of charged particles in the surrounding medium. Increasing electron number density lowers the ionisation threshold and some highly excited energy states might become unbound. This phenomenon is known as Debye shielding, Debye screening or pressure ionisation. Without this depression in the ionisation potential one would face infinite partition functions as Eq.~\eqref{eq:Partition} diverges when leaving out $\Delta E$ and including more and more excited energy states. The radius of the atoms would become larger than the mean distance to the neighboring atoms, which is obviously not the case in reality. Debye shielding can be treated to various degrees of physical realism and is mostly important in the interiors of stars. 

The atomic partition function calculations by \cite{Irwin1981} for example use a static value $\Delta E$ = \SI{0.1}{\eV}. This means they truncate the sum in Eq.~\eqref{eq:Partition} to only include energy levels with $E_i < E_I - \Delta E$. The high-energy states were given lower weights when fitting polynomials to their calculations though. Therefore, the effective value of $\Delta E$ in the published polynomial expressions of the partition functions might vary.  \cite{Irwin1981} also states that many of the atomic partition functions used by \cite{Kurucz1970} are based on tables by \cite{Drawin1965}, which also employ $\Delta E$ = \SI{0.1}{\eV}. Some of the other partition functions by \cite{Kurucz1970} were calculated using values for $\Delta E$  based on the Debye--H\"uckel theory
\begin{align}
    \Delta E = Z\frac{q_e^2}{r_D}, && r_D = \sqrt{\frac{k_B T}{4\pi q_e^2}} \left[n_e + \sum_i Z_i^2 n_i\right]^{-1/2},
\end{align}
where $Z$ is the ionisation stage, so $Z=0,1,2$ for neutral, singly ionised and doubly ionised species respectively, $r_D$ is the Debye length and $q_e^2 = e^2/(4\pi\epsilon_0)$, with the electron charge $e$.  If there are only singly ionised species or hydrogen is by far the most dominant ion, then  $n_e + \sum Z_i^2 n_i \approx 2n_e$, therefore
\begin{align}
     r_D \approx \sqrt{\frac{k_B T}{8\pi q_e^2 n_e}} && \mathrm{and} && \Delta E \approx Z\sqrt{\frac{8\pi n_e}{k_B T}}q_e^3,
\end{align}
The latter approximation is employed by \cite{Griem1964},  whose formulation has been widely adopted in the stellar atmospheres community, especially those codes descendant from the Uppsala opacity package \citep{Gustafsson1975}, such as \texttt{MARCS} and \texttt{STAGGER}.

Using this depression in the ionisation potential results in a thermodynamically inconsistent EOS however, as the populations are corrected, but the same does not necessarily apply for partition functions, internal energies and gas pressures. The abrupt switching between bound and free states also introduces undesirable discontinuities in the thermodynamic as well as optical properties of the medium \citep{Hummer1988}. 

A more accurate treatment of the shielding effect is handled within the occupation probability formalism \citep{Hummer1988, Mihalas1988, Däppen1988}, which is also known as the MHD (Mihalas--Hummer--D\"{a}ppen) equation of state (EoS). In this formalism the energy shift $\Delta E$ is interpreted as the occupation probability $w_i$ of a given state and the partition function is reformulated as
\begin{align} \label{eq:Partition2}
    U_I = \sum_i w_i g_i e^{-E_i/k_bT}.
\end{align}
$w_i$ describes the complement probability of the event that an energy level is disrupted by interactions with other particles in the medium and an electron becomes unbound. The occupation probability drops rapidly close to the ionisation threshold and allows for a smooth transition between bound and free states. Equations \eqref{eq:Partition} and \eqref{eq:Partition2} are mathematically equivalent, so reformulating the partition function is not the reason, why the MHD formalism is more accurate. Unfortunately, $w_i$ depends on the number densities of particles, which act as potential collision partners. This means $U_i$ depends on the chemical composition of the medium and its computation becomes significantly more involved.
In all cases the quality of the partition functions depends on the available atomic data, such as values of $g_i$ and $E_i$. One consequence of using the probability formalism of MHD is that it leads to a smooth transition between lines and continuum around, for example, the Balmer Jump. Such a treatment is particularly important for modeling white dwarfs with their large surface gravities. The MHD approach apparently has some limitations in certain regimes, and other approaches are being developed.

Considerable effort was devoted to improving the partition functions (see \citealt{Alimohamadi2022} for a review). Partition functions and dissociation constants for vast number of molecules and all stable atoms are provided by \citet{Barklem2016}. With the intention to provide chemical composition independent partition functions they neglect the effects of pressure ionization. According to \citet[][their Fig. 8]{Barklem2016}, the uncertainties in the partition function of neutral atoms due to pressure ionisation are $>1\%$ above 6000 K and $>10\%$ above 8000 K.

In summary, astrophysical equations of state can be divided into two groups. Firstly, composition dependent EoSs used in spectrum synthesis codes, which typically neglect or approximate high-density effects like pressure ionisation. Secondly, EoSs used in atmosphere modelling, which are physically more complete but usually precomputed for specific chemical compositions for reasons of efficiency. In this context, we note that the approaches by \citet[][Stagger 3D RHD code]{Zhou2023} and \citet[][M3DIS 3D RHD code]{Eitner2025} are more flexible and they allow for custom chemical abundances in the EoS calculations. The former is based on the FreeEOS package, whereas the latter relies on the EoS framework from the MARCS/Turbospectrum codes. The differences between the two approaches need to be kept in mind, in particular, in studies of highly chemically peculiar environments.

\subsection{Models of atoms and molecules}\label{sec:models}

LTE and NLTE calculations rely on atomic and molecular structure. Here we provide a brief overview of sources that are typically used in calculations of radiation transfer in conditions of stellar and planetary atmospheres, as well in conditions of expanding shells, such as, e.g. kilonovae.

In LTE, the required inputs are vastly simpler compared to NLTE, as only a few parameters are required to describe the radiation field in a given transition. These are: wavelengths of transitions, energies and statistical weights of their lower and upper energy states, and the corresponding transition probabilities. For LTE radiation transfer, all required information is usually provided in public databases, such as:

\begin{itemize}

\item VAMDC (Virtual Atomic and Molecular Data Center) is a catalogue of links and references to databases of atomic and molecular data worldwide; whereas it does not include the data, it provides most up-to-date URL links to other databases \url{https://species.vamdc.org}

\item CHIANTI database (An Atomic Database for Spectroscopic Diagnostics of Astrophysical Plasmas) presents tables with energy states, radiative bound-bound transition data, electron collision data for many ions up Zn (from neutral to 28-th ionization stage), but the coverage is incomplete for elements above Ca. \url{https://db.chiantidatabase.org/}

\item NIST (National Institute of Standard and Technology, at the time of submission of this review, Standard Reference Database 78) includes ionization energies, levels, and lines for all elements from H to U \url{https://physics.nist.gov/PhysRefData/ASD/lines_form.html}
     
\item VALD (The Vienna Atomic Line Database) includes linelists for atoms and di-, tri-atomic molecules, also hyperfine structure and isotopic shifts \url{http://vald.astro.uu.se/}
     
\item Exomol (High temperature molecular line lists for modelling exoplanet atmospheres) includes linelists for  \url{https://www.exomol.com/} 

\item Kurucz database includes energy states, linelists, hyperfine and isotopic structure for neutral, singly-ionized atoms of most chemical elements, as well as some data for higher-ionized atomic species, and selected di- and tri-atomic molecules \url{http://kurucz.harvard.edu/}

\item DREAM (Database on Rare Earths At Mons University) includes atomic data for lanthanides (Z$=57$ to $71$ in up to three ionization stages) \url{https://agif.umons.ac.be/databases/dream.html} \citep{Quinet2020}

\item MOOG \texttt(linemake) (Atomic and Molecular Line List Generator for the MOOG spectrum synthesis code) includes critically-reviewed atomic and molecular linelists for neutral and singly-ionised atoms of most chemical elements \url{https://github.com/vmplacco/linemake} \citep{Placco2021}

\item HITRAN (high-resolution transmission molecular absorption database) \citep{GordonRothman2022}. \url{https://hitran.org/home/}.

\end{itemize}
The following compilations may also serve as a useful and critical resource of atomic and molecular data: MOOG data \url{https://www.as.utexas.edu/~chris/lab.html}, the Carsus package \url{https://github.com/tardis-sn/carsus}, and the CMFGEN package \url{www.pitt.edu/~hillier}.

\begin{figure}[ht]
    \centering
    \includegraphics[width=\textwidth]{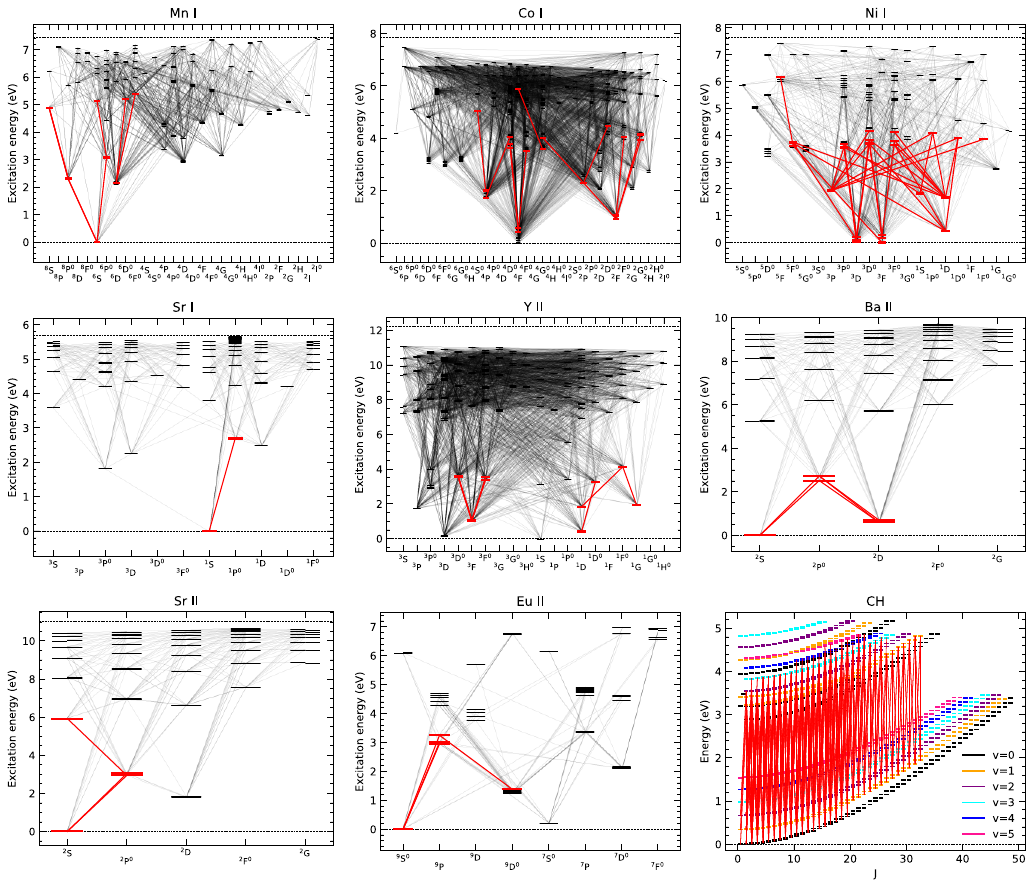}
    \caption{NLTE model atoms for selected atoms and molecules (see figure insets). Figure reprinted with permission from \cite{Storm2025}, copyright by the author(s).} 
    \label{fig:nlteatoms}
\end{figure}

In NLTE RT, in contrast, much more comprehensive atomic or molecular datasets are required as input. These, in addition to the parameters described above, include detailed wavelength-dependent photo-ionization cross-sections for each energy state, collisional rate coefficients for excitation, charge-transfer, and ionising reactions, photo-dissociation cross-sections, and possibly other types of processes, such as di-electronic and three-body recombination. Some examples of NLTE atomic models are shown in Fig.~\ref{fig:nlteatoms}.

Recently, quantum mechanical (QM) calculations of collision processes became available for many atoms. The status of such data for processes in H I and proton induced inelastic collisions was reviewed in \citet{Barklem2016b}. These include transitions in neutral and singly-ionized atoms of Li I, Na I, O I, Fe I, Ni I, Y I \citep[e.g.][]{Belyaev2003, Barklem1998, Barklem2007, Belyaev2010, Barklem2018, Voronov2022, Wang2024}. QM data were also presented for transitions involving different isotopes, e.g. \citet{Belyaev2021} provided rate coefficients for transitions between $^{6}$Li and $^{7}$Li induced by collisions with Hydrogen (H I), Deuterium ($^2$H), and Tritium ($^3$H). These rely on methods of nuclear dynamics, such as probability currents and multichannel linear combination of atomic orbitals (LCAO), which involves different levels of treatment of non-adiabatic regions (in the simplest case, ion-covalent interactions only). But also more sophisticated approaches are available that consider large and small inter-nuclear distances (Landau--Zener approach, \citealt{Belyaev2014}). Also potential energy surfaces for interim states (quasi-molecule) are needed, which are typically obtained via quantum chemistry calculations. In addition, the recipe by \citet{Kaulakys1991a, Kaulakys1991b} can be used for very high-excitation (above $\sim 5$ eV) Rydberg states. In certain instances, e.g. Li and Na, validation on experimental data has been carried out \citep{Barklem2003, Barklem2021}.

NLTE atomic models have been publicly released as part of several NLTE RT codes. These include the \Tlusty database: \url{} \citep{Lanz2003} and Turbospectrum v20. database  \url{https://keeper.mpdl.mpg.de/d/6eaecbf95b88448f98a4/} \citep{Gerber2023}.

\section{Numerical 3D radiative transfer} \label{sec:num3DRT}
\subsection{Motivation and equations} \label{subsec:motivationeq}
The development of astrophysical 3D RT solvers and RT codes has been tightly coupled to the development of 3D RHD models of stellar sub-surface convection and of model stellar spectra. The latter can be used to determine fundamental stellar parameters and chemical composition by comparing synthetic observables with the observed data (spectra, broad-band photometry, light-curves, etc).

In the following section, we limit ourselves to 3D RT codes and methods, which found application in stellar diagnostic spectroscopy and 3D NLTE radiative transfer. We thus, do not cover comprehensively the RT methods that have been developed as components of RHD solvers. The latter usually employ a simplified treatment of RT equations, because these dominate the computational cost when progressing the RHD solvers by one timestep \citep[e.g.][]{Nordlund2019}. The scaling times of calculation schemes are described in Sec. \ref{sec:timing} below.

We note that commonly radiative transfer is assumed to be \textit{quasi-static}, that is, the time derivative is neglected. This is a good approximation when the photon propagation time is much shorter shorter than the fluid motion time \citep{Mihalas1988, Hubeny2014}.
Light is assumed to propagate instantly through the system. 

\newgeometry{left=0.5in, right=0.5in, top=0.75in, bottom=0.75in}
\begin{landscape}
\begin{table}[htbp]
\caption{Comparison of available \textbf{1D and 3D RT codes for different types of objects}. Recent interest in inversion codes has revived forward modelling codes. \texttt{RH} (STIC), \texttt{NICOLE}, \texttt{SNAPI} and \texttt{PORTA} are used in inversion codes. \texttt{Balder} is a branch of the original code \texttt{MULTI3D} and is listed in the same row. Cool stars are defined as those with Teff below 8000 K (following \citealt{Gustafsson2008}). \textbf{MC stands for Monte-Carlo radiation transfer}.}
\label{tab:codes}
{\small
\begin{tabular}{lcclp{6cm}}
\toprule
Code  & open access & dim. & physics  & reference \\
\midrule       
\textbf{SNe, Kilonovae}  & & & & \\
\texttt{SUMO} (Supernova Modeling, MC)  & no & 1D / Sobolev & NLTE  & \citet{Jerkstrand2011, Pognan2023} \\
\texttt{EXTRASS}    & no  & 3D / Sobolev & NLTE & \citet{vanBaal2023} \\ 
\texttt{TARDIS}     & yes & 1D / Sobolev & LTE/NLTE(approx.) & \citet{Kerzendorf2014} \\
\texttt{ARTIS}      & yes & 3D / Sobolev & NLTE & \citet{Kromer2009, Shingles2023} \\
   & & & & \\
\textbf{Exoplanet atmospheres}  & & & & \\
\Cloudy  & yes & 1D & NLTE &  \citet{Ferland2017, Ferland2013} \\
   & & & & \\
\textbf{Hot stars}  & & & & \\
\PhoenixD            & no  & 3D & NLTE & \citet{Hauschildt2006, HauschildtBaron2014} \\
\texttt{Fastwind}    & no  & 1D & NLTE & \citet{Puls2005} \\
\Tlusty              & yes & 1D & NLTE & \citet{Hubeny2017} \\  
\texttt{CMFGEN} (also for SNe)     & yes & 1D & NLTE & \citet{Hillier1998, HillierDessart2012} \\
\texttt{PoWR}        & yes & 1D & NLTE & \citet{SanderShenar2015} \\
\texttt{METUJE}          & yes & 1D & NLTE & \citet{KrtickaKubat2010} \\
   & & & & \\
\textbf{T Tauri stars, YSO} & & & & \\
\texttt{MCFOST-art}       & no & 1D/3D & NLTE & \citet{Pinte2009,Tessore2021} \\
   & & & & \\
\textbf{FGKM stars}  & & & & \\
\texttt{Magritte}        & yes & 3D & NLTE, winds & \citet{deCeuster2022} \\
\texttt{RH}              & yes & 1.5D & NLTE/polarized  &  \citet{Uitenbroek2006, PereiraUitenbroek2015} \\
\texttt{Nicole}          & yes & 1.5D & NLTE/polarized  & \citet{SocasNavarro2015}\\
\texttt{Optim3D}         & no  & 3D & LTE  & \citet{Chiavassa2009} \\
\texttt{Iris}            & no  & 3D & LTE  & \citet{Ibgui2013} \\
\texttt{Porta}           & yes & 3D & NLTE/polarized & \citet{StepanTrujillo2013} \\
\texttt{Scate}           & no  & 3D & LTE/scattering & \citet{Hayek2010} \\
\texttt{Linfor3D}, \texttt{NLTE3D}         & no  & 3D & LTE / NLTE      & \citet{Cayrel2007, Steffen2015} \\
\texttt{Multi3D}/Dispatch, \texttt{Balder}         & no  & 3D & NLTE & \citet{Leenaarts2009b, Amarsi2018, Eitner2024} \\
%
%
%
\bottomrule 
\end{tabular}
}
\end{table}
\end{landscape}%
\restoregeometry

\subsection{Time-independent formal solution}
Standard textbooks and papers, including \citet{Mihalas1978}, \citet{Carlsson2008}, \citet{Crivellari2019}, give an excellent overview of numerical approaches in one- and multi-dimensional radiative transfer. A compact overview of the history of radiation transfer and its basis can be found in \citet{Shore2002}. 

The main work in 3D radiative transfer consists of solving the time-independent monochromatic equation of radiative transfer along the path of the photons, which is essentially the same problem as in the 1D case. 
\begin{align} \label{eq:RTE}
    \frac{dI_{\nu}}{d\tau_{\nu}} &= I_{\nu} - S_{\nu},
\end{align}
where $I_{\nu}$ is the specific intensity [usually in units of erg/cm$^2$/s/\AA/ster] and $S_{\nu}$ is the source function defined as $S_{\nu} = \eta_{\nu}/\chi_{\nu}$, the ratio of the emissivity $\eta_{\nu}$ [erg/cm$^3$/s/\AA/ster] and the extinction coefficient $\chi_{\nu}$ [\SI{}{\per\cm}]. 

The formal solution is obtained by integrating the optical depth $\tau_{\nu}$ and source function along the direction of propagation.
\begin{align}
    \label{eq:RT1}
    \tau_{\nu}(s) &= \tau_{\nu,0} - \int_0^s \chi_{\nu} ds \\
     \label{eq:RT2}
    I_{\nu}(\tau) &= I_{\nu,0} e^{-\tau_{\nu}} + \int_{0}^{\tau_{\nu}} S(\tau_{\nu}) e^{-(\tau_{\nu} - \tau'_{\nu})} d\tau'_{\nu} .
\end{align}

Since there are usually no analytical expressions of $\chi$ and $S$ along the ray, we have to discretise the integrals above to match the spacing of points on the simulation grid. Then we can use simple functions to estimate how $\chi$ and $S$ vary between the grid points, i.e., interpolation. These simple functions can be integrated analytically and give us an approximation of the original integral. \cite{Janett2017a} point out that the accuracy of the conversion to optical depth in Eq.~\eqref{eq:RT1} is as important as the interpolation of the source function in Eq.~\eqref{eq:RT2}. 

In order to minimise the computational cost, one has to strike a balance between fine grid spacing with simple interpolation techniques or a coarser grid with higher-order interpolation methods. The latter usually inhibits efficient vectorisation of the radiative transfer solver and increases the amount of communication in parallelised codes, which divide the simulation grid between multiple MPI ranks. Increasing the number of grid points, on the other hand, increases memory requirements and overall execution time (including opacity and possibly NLTE calculations). We discuss these considerations further in Section \ref{sec:Formal}, reviewing common choices made within the community.

\subsection{1.5D RT approach}
We begin with the so-called "1.5D" RT approach \citep{KiselmanNordlund1995}, which is a very convenient method that circumvents the problems of full angle-coupled 3D RT in a 3D data cube. In addition to much increased computational cost - which is arguably the main bottleneck, the additional strong non-monotonicity due to horizontal inhomogeneities (with opacity varying by a factor of 3-10) even at the same geometric height (Fig. \ref{fig:optdepth}), may imply significantly more numerical problems with convergence of the intensities (via the $\Lambda$ iteration) and of population numbers in the statistical equilibrium solution). This effectively leads to divergence in certain cases with strong variations in physical structure due to large convection cells, as is the case of red giants and red supergiants. Such problems are - to the best of our knowledge - not explicitly documented in academic studies, but - according to our experience with 3D RT with the MULTI3D code - they often lead to preference of down-sampling and preferences for linear-type interpolation in 3D RT calculations \citep{Bergemann2019, Gallagher2020}.

In the 1.5D approximation, every vertical column in the 3D cube from the 3D RHD simulation is treated as a separate 1D  atmosphere, however, with realistic physical T-P$_{\rm gas}$-velocity structure from the 3D RHD simulations. This way there is no interpolation of local matter quantities between the columns and no cross-talk of the radiation field either. The emerging radiation fluxes computed for each column are then averaged at the top plane of the atmosphere to represent the flux of the full 3D RT solution. Such a 1.5D approach is used, for example, in NLTE codes \RHD \citep{PereiraUitenbroek2015} and \Nicole \citep{SocasNavarro2015}.

The main advantage of the 1.5D approach is that it may yield a major speed up in the 3D RT computation, if the interpolation of off-grid quantities is the main bottleneck. This is indeed useful for most NLTE models less or similar to the size of Mn \citep{Bergemann2019}. Only for very large NLTE statistical equilibrium problems, such as e.g. molecular calculations \citep{Popa2023}, this approach may not yield a significant gain, as the interpolation is only one of many computationally expensive operations.

The consequences of the 1.5D approach in RT are not easily predicted and are typically tested against selected 3D RT corner cases  for a concrete science problem under investigation. \citet{SocasNavarro2015} suggest that it works well in LTE and also for strong lines in NLTE, where the mean free path is small. \cite{KiselmanNordlund1995} find that it has little influence on the NLTE source function of the \SI{777}{\nm} oxygen triplet in the Sun, but recommend using full 3D radiative transfer when computing the line profiles. \cite{Amarsi2016} repeat this test with multiple FGK-type stars and show that the approximation yields an error of $0.01$ dex in the oxygen abundance where the \SI{777}{\nm} triplet forms higher in the atmosphere, typically solar-metallicity turn-off stars. \citet[][their Fig. 13, top panel]{Bergemann2019} compare 3D and 1.5D for Mn and find that both yield very similar (the difference of the order of 1 $\%$ or less in the line cores of Mn I lines) line profiles suggesting that 1.5D NLTE RT is fully sufficient for spectroscopic diagnostics of late-type stars. We note, however, that here are examples where full 3D yields significantly different results compared to 1.5D. This is the case for hyper- or mega-metal-poor stars \citep{Beers2005}, such as SMSS0313-6708 with a metallicity [Fe/H]$\lesssim -6$ analyzed in \citet{Nordlander2017b}. The latter study reported differences of up to 0.22 dex between 1.5D NLTE and 3D NLTE for Fe lines in this star.


Since continuum emission in FGKM-type stellar atmospheres is fairly isotropic, the 1.5D RT method produces reasonable estimates of mean (angle-averaged) intensities $J_{\nu}$ even when integrated over only a few number of angles. This implies that 1.5D approach can be safely used in NLTE SE calculations, where only $J_{\nu}$ (usually integrated over large frequency ranges) are required for the integrals of radiative rates. 

\subsection{Interpolation of extinction, source function and intensity to the photon path} \label{sec:interp2D}
The main complexity of multi-D radiative transfer in  hydrodynamic atmospheres, in contrast to 1D RT in hydrostatic models with monotonic behavior of all quantities, is that all quantities appearing in the radiative transfer equation must be mapped to the path of the photons. This mapping is necessary before the formal solution of the radiative transfer equation can be carried out and it requires a decision on the choice of rays, which are to be followed through the simulation domain. 

\begin{figure}[ht]
	\includegraphics[width=0.5\columnwidth]{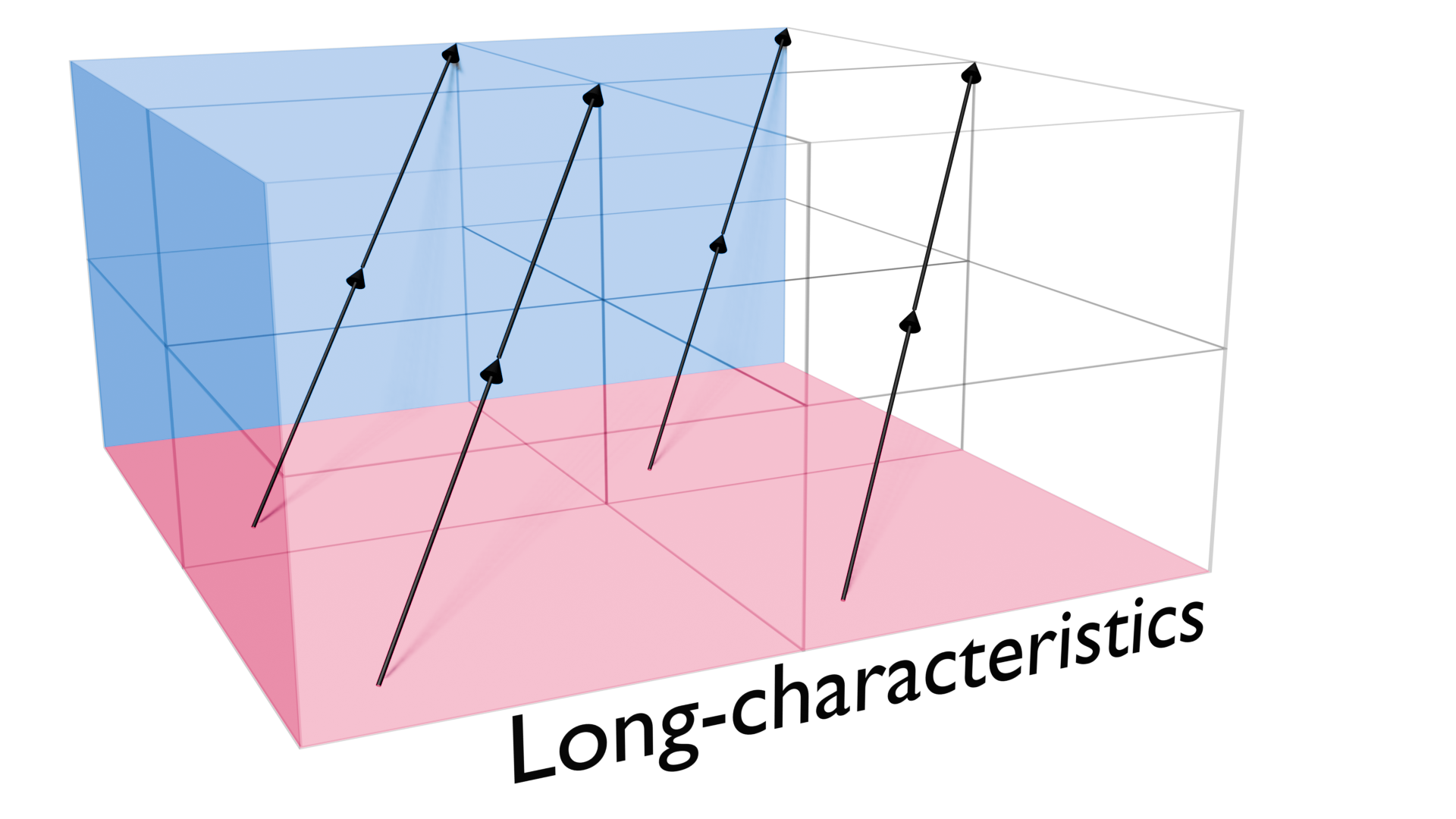}
	\includegraphics[width=0.5\columnwidth]{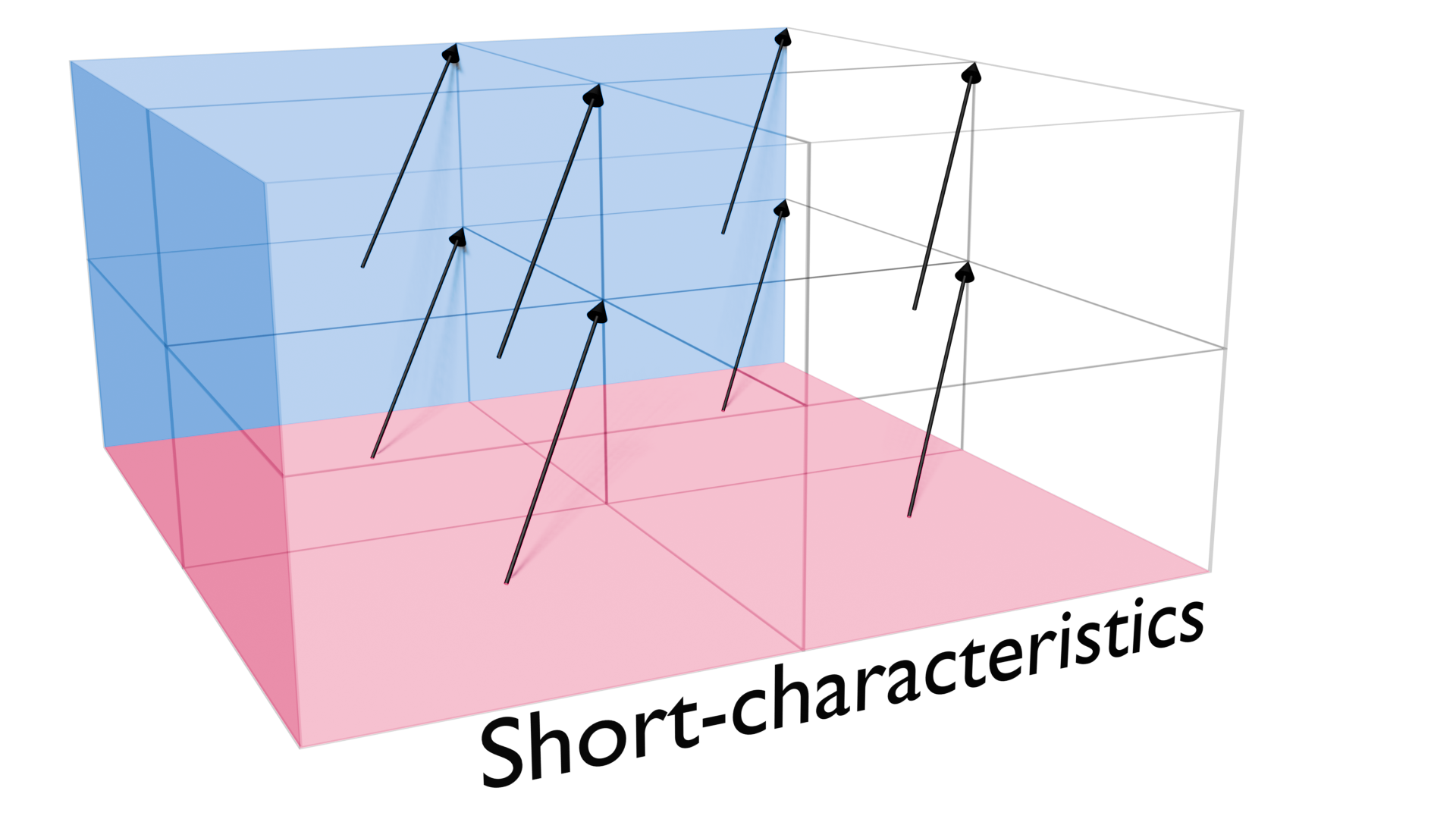}
    \caption{Methods of long and short-characteristics. In the case of short-characteristics (right-hand side panel) the upstream intensity $I_{+}$ is interpolated at the base of each ray, whereas in the case of long-characteristics (left-hand side panel) the resulting intensity has to be interpolated on the grid points.} \label{fig:Characteristics}
\end{figure}

The methods of \textit{long-characteristics} and \textit{short-characteristics} have proven dominant for this purpose in the context of stellar atmospheres \citep[e.g.][]{Carlsson2008}. In the context of both schemes, it is useful to introduce the concept of \textit{upstream, or "U"} and \textit{downstream, or "D"} directions. In this analogy, the flow of photons resembles a stream, with quantities positioned either upstream (the direction from which the photons come) or downstream (the direction they flow in). Both of these methods are illustrated in Fig.~\ref{fig:Characteristics} and will be described below. 

The method of long-characteristics (LC), also known as full characteristics, involves following a ray of light from a fixed boundary condition through the entire simulation box until it exits the atmosphere at the top \citep{JonesSkumanich1973, Jones1973}. The extinction $\chi$ and the source function $S$ have to be interpolated between the grid-points, but the intensity $I$ can be re-used from the upstream ray, because the rays connect. However, if the value of $I$ is needed on the grid points, one would either have to trace as many rays through the entire box as there are grid points, which is very costly, or one would have interpolate $I$ after all and map it back onto the grid-points \citep[e.g.][]{Skartlien2000}.
LC is the method of choice, when high-precision spectral line profile are desired as it is the least diffusive way of transferring radiation through the atmosphere.

A widely-used alternatively is the method of short-characteristics (SC) \citep{Mihalas1978, Kunasz1988}, also shown in Fig.~\ref{fig:Characteristics}. In this approach, one follows disconnected rays, which only extend from one layer to the next. The basic quantities $\chi$, $S$ and $I$, are interpolated between the grid points of the upstream plane. This leads to $I$ being blurred between each layer and strong sources of intensity can be diffused outwards horizontally, as shown in idealized searchlight beam tests \citep{Hayek2010, Ibgui2013}. For a more detailed description of the SC method in three dimensions see \cite{Väth1994}. The SC method is empirically favoured by the community, if the mean intensity on the grid points is required. This is for example the case when the scattering problem or the NLTE statistical equilibrium are tackled. The SC method is more easily vectorised than the LC method, especially on regular grids. 

Both SC and LC are typically used to interpolate the frequency-dependent quantities $\chi$ and $S$ to the photon path. On the other hand, one can also map the frequency independent quantities, such as $T$, $\rho$ and velocities onto the photon path before computing the corresponding values for $\chi$ and $S$. In slab geometries this represents a pre-tilting of the simulation box before the post-processing. On ray aligned grids, SC and LC are equivalent and no decision between the two methods is required. This tilting approach minimises horizontal diffusion in the emerging intensity as is shown in Fig. 3 in \citet{Peck2017}, where they compare emergent intensities based on the classical short-characteristics and the pre-tilted approach. Furthermore, it can reduce the number of interpolations as no frequency dependent data has to be interpolated.

All FGKM-type codes listed in Table \ref{tab:codes}, except for \Magritte \citep{deCeuster2022} and \NLTED \citep{Steffen2015}, use the SC method to compute the the mean intensity $J$ at each grid point. This requires repeated usage of the SC method for multiple directions. The resulting intensities at the grid points can be integrated via a weighted sum to yield $J$, see section \ref{sec:AngleQuadrature}. The value of $J$ has an impact on $S$ of each grid point and radiative transfer can be repeated with updated $S$ values. Coherent and isotropic scattering is additionally sometimes included in the continuum \citep[e.g.][]{Cayrel2004}, therefore the diffusion of spectral line shapes is often no great concern. 

SC and LC are equivalent for 1D homogeneous atmospheres or if the rays cross a grid point in each horizontal layer, i.e. when there is no need for horizontal interpolation. 
\citet{Hauschildt2006} have shown that for strongly scattering atmospheres, the method of LC outperforms SC in terms of accuracy in $J$, however, this test was carried out on a very high angular resolution ($64^2$ angles). \citet{Baron2007} point out that SC might make up for the lack of accuracy with an increased grid resolution, since their SC implementation has lower memory requirements.

\subsection{The problem of negative opacities}
\begin{figure}[ht]
    \centering
    \includegraphics[width=\columnwidth]{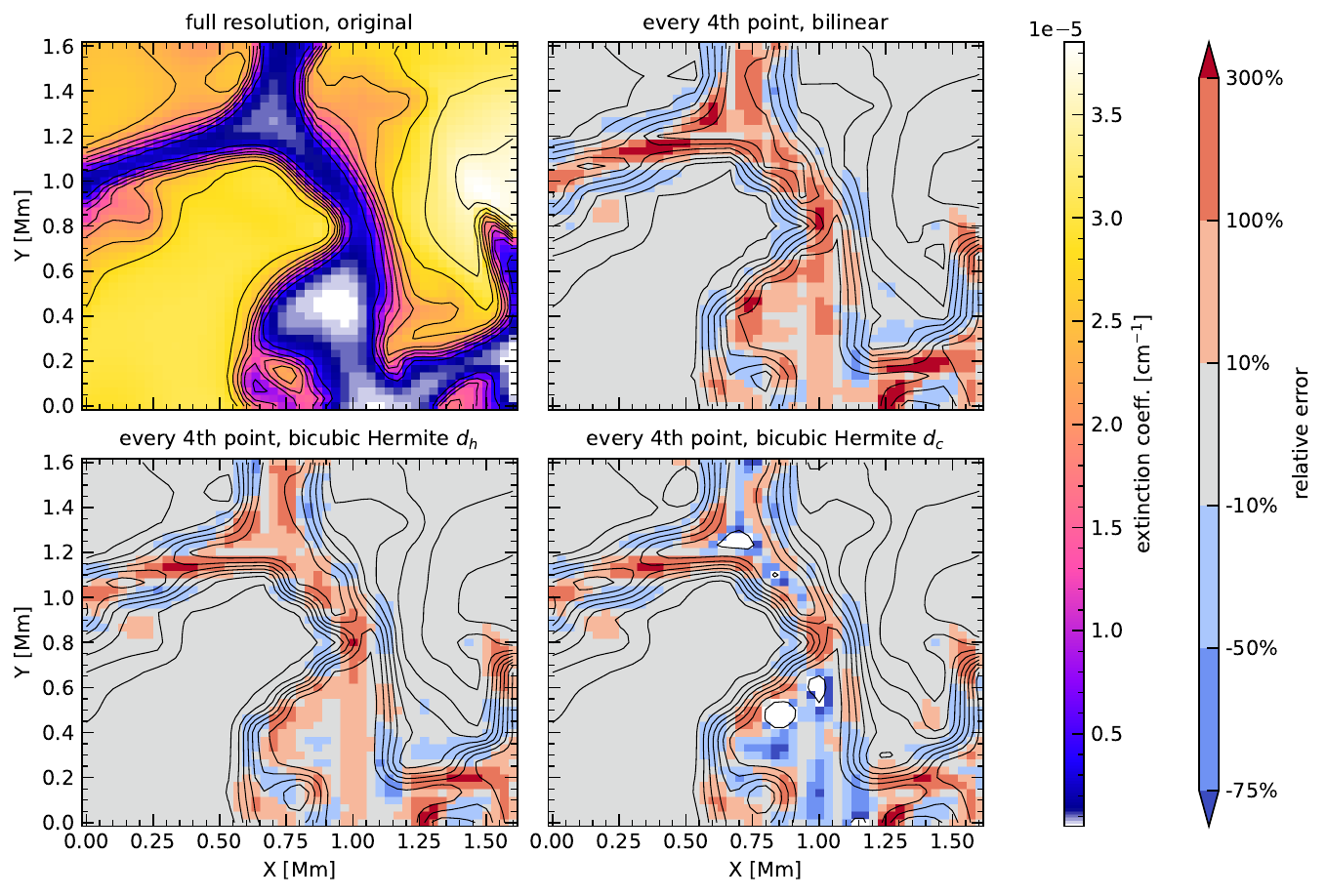}
    \caption{We compare several 2D interpolation schemes applied to the extinction coefficient $\chi$ in a solar atmosphere model. The top-left panel displays a horizontal slice at the original radiation-hydrodynamics resolution using both colours and contours. In the other three panels, we downsampled the data by selecting every fourth point in each spatial dimension and then interpolated back to the original resolution. The colours in these panels represent the relative error of the interpolated $\chi$ compared to the full-resolution values. White regions in the bottom-right panel denote critical interpolation undershoots where the results are negative.}
    \label{fig:offgridInterp}
\end{figure}

For both SC and LC schemes, an interpolation scheme is required in order to map $\chi$, $S$, and potentially $I$ to the photon path. In order to minimise the number of interpolations, most codes interpolate these quantities at the positions where the rays intersect the X-Y, X-Z or Y-Z planes. This reduces the problem to a 2D interpolation. An exception is the "sub-griding" technique, where one increases the grid resolution through interpolation before the formal solution, which represents a 3D interpolation. While linear interpolation is a simple and robust choice, higher-order methods, like parabolic and cubic methods, are also used in the stellar atmospheres community, as their accuracy improves faster with increased resolution compared to linear interpolation \citep{DeLaCruzRodriguezPiskunov2013}. However, the main  drawback of such more complex interpolation schemes, is that they can introduce spurious extrema in the interpolated quantity between the grid points. This is dangerous in physical modeling, because these extrema may result in negative values in  interpolated quantities, thereby leading to numerical problems and code crashes, which may be extremely difficult to trace owing to a highly multi-dimensional nature of the simulated system. For instances, this sometimes leads to \textit{negative radiative opacities} even when the underlying physical quantities (temperature, pressure, velocities, etc) are monotonically varying (in one of the dimensions) and strictly positive (see Fig. \ref{fig:offgridInterp}). \citet{Auer1994a} show that parabolic interpolation can also lead to negative intensities near sharp beam edges, which is non-physical.

A number of different interpolation methods exist which avoid any additional extrema in the interpolant. \citet{Steffen1990} introduced a strictly monotonic interpolation scheme for the use in numerical hydrodynamics. Similar methods, such as 
 piecewise monotonic interpolation were brought to the attention of the wider astrophysical community by \citet{Auer2003}. 

While $n$ data points can be used to construct a unique Lagrange polynomial of the order $n-1$, one can also construct higher-order polynomials by fixing the derivatives at the data points. Constructing an interpolant polynomial by making use of fixed derivatives is known as Hermite interpolation. One can use parabolic or cubic interpolation between two points $i$ and $i+1$ by fixing the derivatives at either a single or both points respectively. By supplying suitable estimates of these derivatives one can easily avoid spurious extrema in the interpolant. 

B\'ezier interpolation between two points, on the other hand, is essentially recursive linear interpolation. By introducing a number of control points one can construct a curve between the two anchor points $i$ and $i+1$. B\'ezier curves cannot necessarily be expressed as polynomials, unless one places the x-positions of the control points uniformly between the two anchor points. In this case the gradient at the anchor points is given by the gradient of the line connecting the anchor and the adjacent control point. Therefore the polynomial type of B\'ezier curve can also be rewritten as a Hermite polynomial. B\'ezier curves have the nice property, that they do not leave the bounding box spanned by the control and anchor points, which allows for suppression of under- or over-shoots in the interpolant function. If there is no over- or under-shoot, however, the position of the control point is determined from the gradients at the data points. Differences between the properties of Hermite and B\'ezier interpolants are discussed in further detail in \citet{Ibgui2013}. 

Additionally, other higher-order interpolation schemes have been proposed for radiative transfer, but to our knowledge they have not found any use in the 3D community yet.  \citet{Janett2021} point out that strictly monotonic schemes sacrifice accuracy when enforcing monotonicity (which may not necessarily be a drawback), but such an approach may yield a  linear interpolation near smooth extrema of the interpolated function. Therefore they argue for the use of Weighted, Essentially Non-Oscillatory (WENO) techniques, which do not suffer from this loss in accuracy while still suppressing oscillations near discontinuities. The fourth-order WENO interpolation proposed by \citet{Janett2021} has been implemented in the 1D NLTE code LIGHTWEAVER \citep{Osborne2021}.

The B\'ezier/Hermite interpolation techniques discussed in the literature vary mostly in how the derivatives at the data points are estimated. Here we discuss a few common choices. Considering the three consecutive points $i-1$, $i$ and $i+1$, we can derive the two interval spacings $s_-$ and $s_+$

\begin{align}
    s_{-} = x_{i} - x_{i-1}, && s_{+} = x_{i+1} - x_{i},
\end{align}

as well as the \textit{backward difference} $d_-$ and \textit{forward difference} $d_+$
\begin{align}
    d_{-} = \frac{y_{i} - y_{i-1}}{s_{-}}, && d_{+} = \frac{y_{i+1} - y_{i}}{s_{+}}.
\end{align}

For the following discussion we will introduce a few approximate derivatives for the central point $i$. A simple approximation is the \textit{central difference}, which is simply the gradient between the two surrounding points to $i$.
\begin{align} \label{eq:centralDiff}
    d_{c} = \frac{y_{i+1} - y_{i-1}}{x_{i+1} - x_{i-1}} = \frac{s_- d_- + s_+ d_+}{s_- + s_+}
\end{align}

The gradient at point $i$, when fitting the Lagrangian parabola through the 3 points is given by

\begin{align} \label{eq:parabolicDiff}
    d_{p} = \frac{s_- d_+ + s_+ d_-}{s_- + s_+}.
\end{align}

We note that \eqref{eq:centralDiff} and \eqref{eq:parabolicDiff} are identical in the equidistant case, where $s_-=s_+$. The central difference $d_c$ and the parabola gradient $d_p$ are essentially weighted means of $d_-$ and $d_+$
\begin{equation} \label{eq:weightedDiff}
    d_w = \alpha d_- + (1-\alpha) d_+,
\end{equation}
which only differ in their choice of the weight $\alpha$. Alternatively one can use the weighted harmonic mean of $d_-$ and $d_+$ 
\begin{align} \label{eq:harmonicDiff}
    d_{h} = 
    \begin{cases}
    \left(\frac{\displaystyle \alpha}{\displaystyle d_-} + \frac{\displaystyle 1 - \alpha}{\displaystyle d_+}\right)^{-1}
    , & \text{if } d_+ d_- > 0, \\
    0
    , & \text{if } d_+ d_- \leq 0,
    \end{cases}
\end{align}
\citet{Brodlie1980} and \citet{FritschButland1984} suggest the following weight $\alpha$ to minimize artifacts, such as unphysical "wiggles" in the interpolant:
\begin{align} \label{eq:Fritsch}
    \alpha = \frac{1}{3} \left( 1 + \frac{s_+}{s_- + s_+}\right).
\end{align}

The weighted harmonic mean has the interesting property that $d_h$ approaches zero if either $d_-$ or $d_+$ approach zero. Therefore there is a smooth transition between the 2 cases of Eq.~\eqref{eq:harmonicDiff}. 
Finally, \citet{Steffen1990} propose yet another approximation of the gradient:
\begin{align} \label{eq:Steffen}
    d_{p^*} = ({\rm sign}(d_-) + {\rm sign}(d_+))
    \; {\rm min}(\lvert d_-\rvert, \lvert d_+\rvert, \lvert d_p / 2\rvert),
\end{align}
where sign($d_-$) is $-1$ for negative $d_-$ and $+1$ if it is positive. This statement has intriguing consequences. If $d_-$ and $d_+$ have opposite signs, the result is zero as for $d_h$. If the absolute values of $d_-$ or $d_+$ are smaller than $\lvert d_p / 2\rvert$, this means an extreme will be present in the backward or forward interval if one fitted a parabola through the three points. If this is not the case $d_{p^*}$ is simply equal to $d_p$. As for Eq.~\eqref{eq:harmonicDiff} the resulting cases of $d_{p*}$ transition smoothly.

The previously mentioned techniques can be categorised into those where interpolation weights depend only on the grid spacing and those where weights are determined by the values to be interpolated. One disadvantage of the latter category is that weights cannot be precomputed, which may limit their computational efficiency or suitability depending on the specific application. Linear interpolation as well as Hermite interpolation using $d_c$ and $d_p$ is purely based on the distance between points. The methods employing $d_h$, $d_{p^*}$ or Bézier control points, which were designed to prevent wiggles and ensure monotonicity, generally fall into the data dependent category.



It is not straightforward to review the interpolation techniques used in 3D radiative transfer codes, because many codes are not public. For the proprietary codes, we had to rely mostly on the papers where they were used for the first time, which are not always concerned with the technical details of the codes. Yet, we have compiled a summary of interpolation schemes used in the cool stars community in table \ref{tab:interpolation_summary}.


\newgeometry{left=0.5in, right=0.5in, top=0.75in, bottom=0.75in}
\begin{landscape}
\tiny
\begin{table}
\centering
\caption{Summary of interpolation schemes employed by various radiative transfer codes for (i) interpolation of opacity and source function \emph{to} the photon path and (ii) interpolation of opacity and source function \emph{along} the photon path. SC~$=$~short-characteristics solver; LC~$=$~long-characteristics solver. All presented B\'{e}zier schemes are truly Hermite schemes, since the control points are positioned half way between the grid nodes. We are aware that likely some of the here accumulated information does not represent the current status of the codes.}
\label{tab:interpolation_summary}
\setlength{\tabcolsep}{5pt}
\renewcommand{\arraystretch}{1.35}
\begin{tabular}{%
  >{\raggedright\arraybackslash}p{2.0cm}  
  >{\raggedright\arraybackslash}p{6.5cm}  
  >{\raggedright\arraybackslash}p{6.5cm}  
  >{\raggedright\arraybackslash}p{6.5cm}  
}
\toprule
\textbf{Code} & \textbf{Interpolation to the path} & \textbf{Interpolation along the path} & \textbf{Comments} \\
\midrule

\texttt{MAGRITTE} \citep{deCeuster2022}
  & LC: No interpolation; $\chi$ and $S$ assumed constant within each Voronoi grid cell.
  & Linear in $\chi$ and $S$ with sub-griding for steep velocity gradients.
  & Accuracy is very sensitive to spatial resolution and boundary conditions. \\

\texttt{OPTIM3D} \citep{Chiavassa2009}
  & SC: Bilinear in $\ln\chi$ and $S$.
  & Linear in $\ln\chi$ to convert to optical depth; linear in $S$ on the $\ln\tau$ scale.
  & \\

\texttt{IRIS} \citep{Ibgui2013}
  & SC: Monotonic bicubic Hermite using $d_h$.
  & Monotonic cubic Hermite; gradient at $i{-}1$ fixed to $d_-$, gradient at $i$ set to $d_h$ with $\alpha$ from eq. \ref{eq:Fritsch}.
  & \citet{StepanTrujillo2013} argue that fixing the gradient at $i{-}1$ to $d_-$ may reduce accuracy of the formal solver. \\

\texttt{ASS$\epsilon$T} \citep{Koesterke2008}
  & SC and LC: Monotonic cubic B\'{e}zier (Hermite), details unclear.
  & Monotonic cubic B\'{e}zier (Hermite), details unclear.
  & Doppler shifts are applied through interpolation in frequency using the same B\'{e}zier scheme.\\

\texttt{PORTA} \citep{StepanTrujillo2013}
  & SC: Choice between bilinear and biquadratic.
  & Monotonic quadratic B\'{e}zier (Hermite) with $d_{p^*}$, equivalent to \texttt{Linfor3D}.
  & Bilinear is stated to be sufficient when using original resolution of model atmosphere.\\

\citet{Hennicker2020}
  & SC: Monotonic quadratic B\'{e}zier (Hermite); asymmetric point selection (left/below interpolation interval). Monotonicity is only enforced in the interpolated interval.
  & Monotonic quadratic B\'{e}zier (Hermite) using $d_{p^*}$ smoothly transitioning to linear in $S$ when there are issues with convergence.
  & Their discussion implies that enforcing monotonicity can introduce convergence-preventing oscillations. \\

\texttt{MULTI3D} \citep{Leenaarts2009b}
  & SC: Lagrangian bicubic Hermite, via a precomputed $4{\times}4$ convolution kernel. LC: bilinear.
  & Schemes used in \texttt{SCATE} or \texttt{IRIS} codes, also linear scheme can be used.
  & Undershoots in the bicubic scheme are suppressed by limiting interpolated values to the minimum of the 4 corner points. \\

\texttt{BALDER} \citep{Amarsi2018}
  & SC: same as \texttt{MULTI3D}. LC: monotonic bicubic Hermite, details unclear.
  & Choice between linear and the schemes used in \texttt{SCATE} or \texttt{IRIS}.
  & \\

\texttt{SCATE} \citep{Hayek2010,Hayek2011}
  & SC: bicubic Hermite with $d_h$. LC: bicubic logarithmic in ($T$, $\rho$, velocities) onto a ray-aligned pre-tilted grid.
  & Monotonic quadratic B\'{e}zier (Hermite) \citep{Auer2003} with $d_{p^*}$.
  & They neglect $d_+$ in the min() function, allowing over/undershoots in $[i,\,i{+}1]$ \\

\texttt{Linfor3D} (priv.\ comm.\ M.\ Steffen)
  & LC: Lagrangian bicubic in ($T$, $\rho$, velocities) onto a ray-aligned pre-tilted grid.
  & Monotonic quadratic Hermite scheme with $d_{p^*}$, equivalent to \texttt{PORTA}.
  & \\

\texttt{PHOENIX/3D} \citep{Hauschildt2006}
  & LC: details unclear.
  & Linear in $\chi$; linear in $S$, switching to Lagrangian quadratic in the optically thick regime.
  & \\

\bottomrule
\end{tabular}
\end{table}
\end{landscape}
\restoregeometry

Fig.~\ref{fig:offgridInterp} compares bilinear, bicubic (using $d_c$) and monotonic bicubic (using $d_h$) interpolation methods in a downsampled solar atmosphere model. \citet{Peck2017} emphasize that enforced monotonicity in interpolation also results in the non-conservation of the radiative energy, however, both \cite{Hayek2010} and \cite{Ibgui2013} suggest that their monotonic cubic Hermite schemes conserves the photon count in their searchlight beam tests.

In summary, each 3D RT code has a unique way of determining opacity, source function and intensity values along the photon path. Another question left unanswered is what quantities are best interpolated. Should one interpolate for example in $\chi$ or $\ln\chi$ or only in frequency independent variables? The range of techniques in use likely reflects the pros and contras of each in application to a give science question at hand.  
The proprietary nature of the majority of these codes prevents meaningful comparisons of the accuracy and efficiency of the different implementations. Comprehensive tests of this kind are welcome, along with performance benchmarks, in order to identify reliable as well as efficient interpolation methods. 
\subsection{Interpolation of opacity and source function along the photon path} \label{sec:Formal}
The Feautrier methods \citep{Feautrier1964}, which are popular amongst 1D codes \citep{Carlsson1986, Uitenbroek2001, HubenyLanz2017b}, requires big matrix inversions and storing all values of $\chi$ and $S$ along the entire photon path \citep{Fabiani2003}. This is incompatible with many 3D codes as they are often domain decomposed and only pass information between a limited number of neighbouring grid points. An exception to this is the 3D NLTE code \Magritte \citep{deCeuster2020}, which assumes shared-memory across the entire domain and uses a second-order Feautrier scheme. Also the 3D NLTE departure coefficient determination code \NLTED relies on a Feautrier scheme, but little detail on this is given in \cite{Steffen2015} and it is not clear whether this code is parallelised at all. Unfortunately, we were not able to find any information on the numerical details of the radiative transfer solver in the widely used \Cloudy code \citep{Ferland2017}. This might be due to the fact, that the references to older papers on the code have been linked to the online documentation of the code. This documentation focuses on guiding the user rather than giving insights on the technical details however.

All codes presented in table \ref{tab:interpolation_summary} are time-independent radiative transfer codes, used for post-processing 3D hydrodynamical simulations. They all rely on a 3-point scheme, which considers for each point $i$ exactly one upstream point $i-1$ and one downstream point $i+1$ when constructing the polynomial that approximates the interval between $i-1$ and $i$. This reflects the importance of minimising the amount of communication between ranks in MPI parallelised programs. \cite{Auer1994a} state that parabolic or higher order interpolation between $S_{i-1}$ and $S_i$ is important to recover the diffusion approximation in the optically thick regime. For this reason switches the \PhoenixD code \citep{Hauschildt2006} from linear to parabolic interpolation in $S$ in the optically thick regime. For these kind of adaptive schemes, \cite{HolzreuterSolanki2012} point out that changes in the source function values can result in switching interpolation techniques between iterations when one solves for the source function in NLTE. These ``flip-flop'' situations are non-linear and can lead to oscillations which prevent convergence.

Monotonic schemes are clearly very popular for the interpolation along the photon path amongst the presented codes. However, it appears from the discussion in \citet{Hennicker2020}, concerning 3D scattering problems in the winds of hot stars, that enforcing monotonicity can introduce convergence preventing oscillations in an iterative scheme. This is why they smoothly transition to a linear scheme if convergence is otherwise not reached.

Though not a 3D code, we note that \cite{Auer2003} approximate the optical depth integral by linearly integrating the logarithmic density $\ln\rho$ and multiplying it with $(\kappa_{i}+\kappa_{i-1})/2$, where $\kappa=\chi/\rho$.
\citet{Auer1984} also suggests the Euler-Maclaurin summation formula as a quick and convenient method to improve the accuracy of the optical depth scale. It uses the first derivatives to provide a correction to the trapezoidal rule.

\begin{figure}[ht]
	\includegraphics[width=\columnwidth]{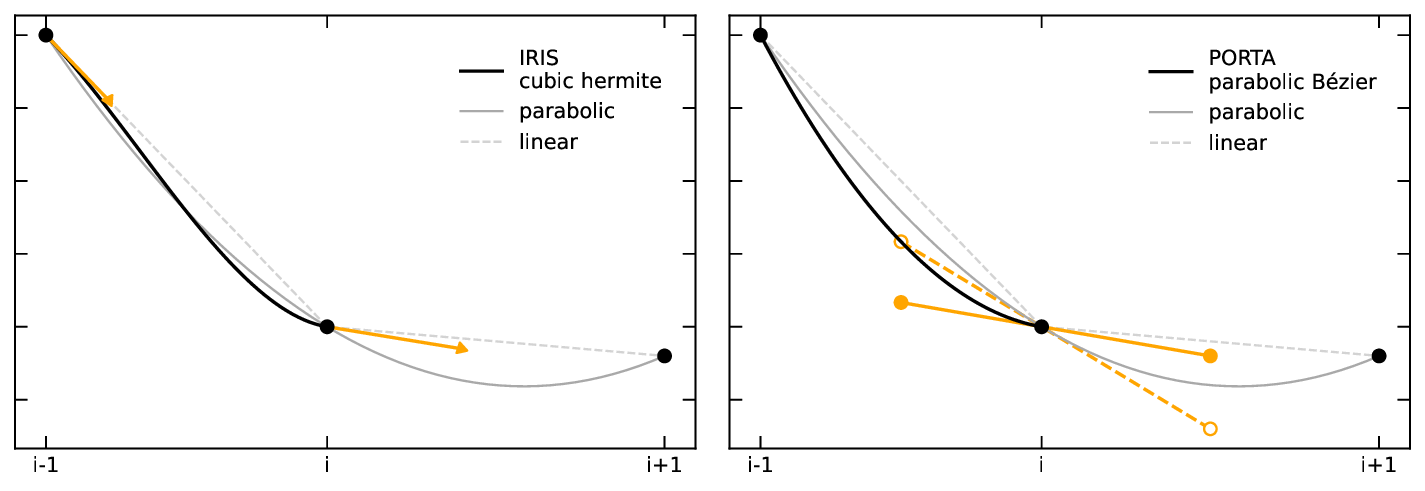}
    \caption{Piecewise monotonic interpolation methods used in the formal radiative transfer solvers in the IRIS and PORTA codes. Linear and parabolic interpolation is overplotted for comparison. The derivatives at the grid points are shown as orange arrows for the Hermite interpolation in the left panel. The linear gradient is used at the $i-1$ position, while the weighted harmonic mean is used at $i$. In the right panel the control points of the B\'ezier curves are plotted as orange points. The unfilled points are the default control points, which yield the Lagrangian parabolic curve. Even though the parabolic curve does not undershoot in the interval $[i-1, i]$, the control points are corrected. The filled points are the control points after overshoot correction, since the control point at $i+1/2$ is outside the bounding box spanned by the three points.}
    \label{fig:monotonic}
\end{figure}

The \Iris and \Porta/\LinforD implementations are shown in Fig.~\ref{fig:monotonic}. 
A concise list of proposed solvers for polarized radiative transfer is found in Table~1 of \cite{Janett2017a}.

Summary: The performance of a given interpolation scheme — just as any other ingredient in 3D NLTE RT — can and shall be evaluated numerically with respect to a given astrophysical problem. Codes that are used for atmospheres of FGKM-type stars, such as \PhoenixD, \OptimD, \Scate, \MultiD, \Balder and \LinforD, are tailored to physical conditions of FGKM-type stellar atmospheres and they have been tested for corresponding astrophysical applications. Other codes, such as \Cloudy and \Magritte use much more basic methods, but these two codes are used in very diverse temperature-density regimes spanning AGN, exoplanet atmospheres, ISM, etc (Fig. \ref{fig:maindiagram}).

We trust the authors that their interpolation schemes are sufficient in their particular regimes, even though it would be better to see this proven and published. Unfortunately, it seems that the community has a preference to develop a new code every time a new PhD thesis begins, which likely reflects the progress in computer sciences and availability (or progressive phase-out) of different compilers. Therefore we cannot recommend an all-purpose interpolation scheme, which is efficient and sufficiently accurate for all applications. Linear interpolation is still widely used as it remains the most efficient and robust method. Overall, it is desirable that multiple interpolation schemes are implemented, so that specific choices can be tested for real astrophysical applications.

%
%
%
\subsection{Angle quadrature} \label{sec:AngleQuadrature}
\begin{figure}[htbp]
    \centering
    \includegraphics[width=\columnwidth]{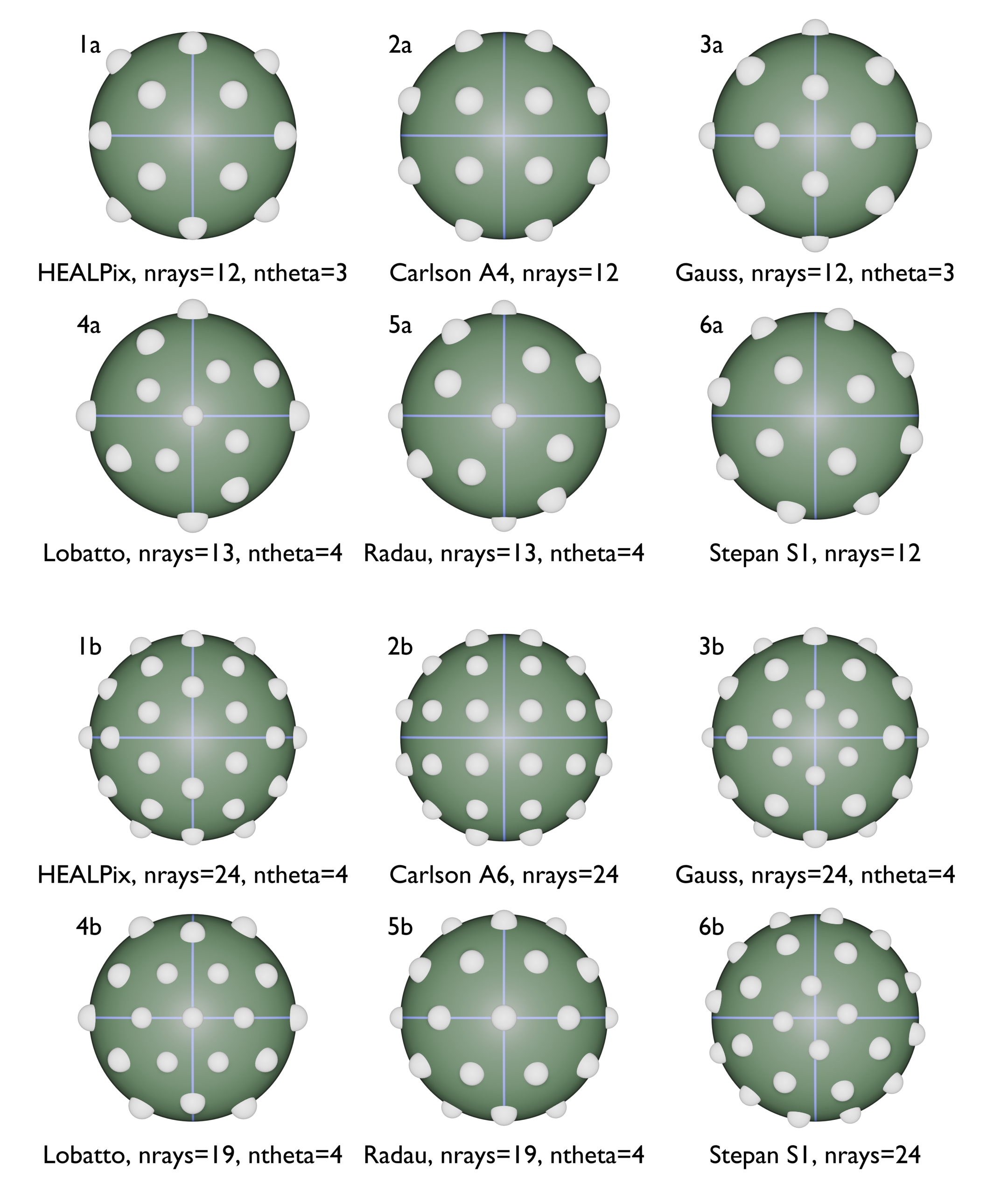}
    \caption{Visualisations of popular angle quadratures used for 3D radiative transfer. Here we only show the top hemisphere, i.e. only ``up'' pointing rays. ``Down'' pointing rays are simply mirrored version of the ``up'' rays. The number of upward rays $n_{\rm rays}$ and theta angles $n_{\theta}$ have been chosen in order to highlight the differences between the quadrature schemes. They do not necessarily represent the most commonly used choices of $n_{\rm rays}$ and $n_{\theta}$. The white spheres represent the discretised set of angles, i.e. rays, for which the radiative transfer equation is solved. The volume of each white sphere is proportional to the weight given to that particular ray. The blue lines mark the position of the x and y axes. Neighbouring rings of constant $\theta$ have been offset by $90^{\circ}/(n_\theta -1)$ with respect to one another in the case of the Gauss, Lobatto and Radau quadratures.}
    \label{fig:AngleQuad}
\end{figure}

Radiative transfer codes based on the methods of long or short-characteristics solve the radiative transfer equation for a discrete set of geometric rays, see Fig.~\ref{fig:AngleQuad}. The \textit{angle quadrature} defines which rays have been chosen and how they are weighted when integrating intensities.
The following section is tailored to Cartesian grids, where the same angle quadrature can be used throughout the entire spatial domain. The situation is significantly more complex in spherical geometries, where a single straight ray intersects horizontal layers at different angles and some rays never reach certain depths. In this case, the angle quadrature  becomes spatially dependent \citep[e.g.][]{Nordlund1984}. Locally, however, the employed rays still have to be judged by the criteria outlined below. To the best of our knowledge, the vast majority of RT codes used in stellar and plan community rely on plane-parallel RT solver, with some exceptions, where spherical symmetry may become important (such as, AGB and red supergiant stars, see Sec. \ref{sec:atmospheres}). A detailed examination of the relevance of spherically-symmetric RT in the model atmospheres structure of stars can be found in \citep{Heiter2006}, but studies on this subject are sparse and therefore here we limit the review to cartesian grids.

For NLTE, for example, the mean intensity $J_{\nu}$ is the quantity that is needed in order to solve the rate equations. The value of $J_{\nu}$ is the integral of the directed intensities over the unit sphere

\begin{align}
    J_{\nu} = \frac{1}{4\pi}\int I_{\nu} \,d\Omega
    &= \frac{1}{4\pi}\int_0^{2\pi}\!\!\!\int_0^{\pi} I_{\nu} \sin{\theta} \, d\theta \, d\phi \\
    &= \frac{1}{4\pi}\int_0^{2\pi}\!\!\!\int_{-1}^{1} I_{\nu} \, d\mu \, d\phi  \label{eq:Jnu}
\end{align}
where we introduced $\mu$=$\cos\theta$ for convenience. The resulting monochromatic net flux from an astronomical object $F_{\nu}$ is defined as:
\begin{equation} \label{eq:Flux}
    F_{\nu} = \int I_{\nu} \mu \,d\Omega = \int_0^{2\pi}\!\!\!\int_{-1}^{1} I_{\nu} \mu \, d\mu \, d\phi
\end{equation}

and the emerging flux is given by integration over the interval where $\mu$ is positive, therefore:

\begin{equation} \label{eq:Flux+}
    F_{\nu}^+ = \int_0^{2\pi}\!\!\!\int_{0}^{1} I_{\nu} \mu \, d\mu \, d\phi
\end{equation}
For efficient evaluation of the integrals we discretise the intervals and replace the integrals by a weighted sum
\begin{equation}
    J_{\nu} \approx \sum_{i}^{n} w_{i} I_{\nu},
\end{equation}
where $n$ is the number of discrete rays and $w_i$ are their respective weights. As the computational cost of any radiative transfer application scales nearly linearly with $n$, it is essential to strike an optimal balance between the accuracy of the integral and the number of rays. However, even for a given $n$, it is not trivial to find the set of rays and weights, which yields optimal accuracy. For a 1D integration, Gaussian quadrature and its variants yield a good accuracy and the exact choice is not important. Therefore 1D radiative transfer codes are using Gaussian quadratures for approximating $J_{\nu}$ \citep[e.g.][]{HubenyLanz2017c}. For plane-parallel atmospheres, we can write
\begin{align}
    \label{eq:Jnu1D}
    J = \frac{1}{2}\int_{-1}^{1} I(\mu) \, d\mu &\approx \frac{1}{2}\sum_{i}^{n} w_{i} I(\mu_i) \\
    &\approx \frac{1}{2} (w_1 I(\mu_1) + w_2 I(\mu_2) +... + w_n I(\mu_n)), \label{eq:GaussSum}
\end{align}
where we have dropped the frequency index for clarity. Gaussian quadratures yield the values for $w_i$ and $\mu_i$, which approximate $I(\mu)$ exactly if it was a polynomial of order $2n-1$ or lower. This is because Eq.~\eqref{eq:GaussSum} has $2n$ unknowns, namely $w_1, w_2, ..., w_n$ and $\mu_1, \mu_2, ..., \mu_n$, assuming that we are free to evaluate $I(\mu)$ at any $\mu$ we like. For the case of n=2 we have 4 unknowns and we can set up the following system of 4 equations:
\begin{equation}
\begin{split}
    \smallint \!_{-1}^{1} \, \mu^0 \, d\mu &= w_1 + w_2 \\
    \smallint \!_{-1}^{1} \, \mu^1 \, d\mu &= w_1 \mu_1 + w_2 \mu_2 \\
    \smallint \!_{-1}^{1} \, \mu^2 \, d\mu &= w_1 \mu_1^2 + w_2 \mu_2^2 \\
    \smallint \!_{-1}^{1} \, \mu^3 \, d\mu &= w_1 \mu_1^3 + w_2 \mu_2^3
\end{split}
\end{equation}
where we have approximated $I(\mu)$ by simple polynomials of up to third order in $\mu$. Solving this system yields the Gauss--Legendre quadrature weights and abscissas for n=2 points. One common variant is the Gauss--Lobatto quadrature which fixes the endpoints to $\mu_1$=-1 and $\mu_n$=1, i.e. the vertical rays.
Considering an isotropic radiation field we expect $J_\nu = I_\nu$ , $F_\nu=0$ and $F_{\nu}^+ = I_\nu/2$. The construction of any quadrature scheme for radiative transfer is motivated by meeting the following criteria:

\begin{align} \label{eq:moments}
     \sum_{i}^{n} w_i = 2, &&
     \sum_{i}^{n} w_i \mu_i = 0, &&
     \sum_{i}^{n} w_i \mu_i^2 = \frac{2}{3}
\end{align}

Equation \eqref{eq:Jnu1D} is only fulfilled when the first criterion holds. The second criterion demands that $F_\nu^+$ and $F_\nu^-$ cancel each other, i.e. $F_\nu=0$. The third criterion is derived from conservation of the third moment of the intensity $K_\nu$, which is related to radiative pressure, in isotropic radiation. The case $F_{\nu}^+ = I_\nu/2$ is  not reproduced when using Gaussian quadratures in the $\mu$ interval $[-1,1]$, which is a strong limitation of these schemes. A simple solution to this issue has been found by \cite{Sykes1951}, who propose to use double-Gaussian quadratures. They map the $\mu$-values and their weights to the $\mu$ interval $[0,1]$ and simply ``mirror'' them for the interval $[-1,0]$. For the Gauss--Lobatto quadrature this has the inconvenient result of  including $\mu=0$, which is problematic in the 3D case if one has periodic boundary conditions in the horizontal. An alternative can be found in the double-Gauss--Radau quadrature, which only fixes one end-point, i.e. $\mu=1$, and results in two fixed endpoints $\mu_1=-1$ and $\mu_n=1$ after the mirroring. Unfortunately, many authors do not specify, whether they are using Gaussian quadratures in the interval $[-1,1]$ or the mentioned double-Gaussian quadratures.

Thus, in 1D we can integrate the intensity efficiently with high order accuracy using only a few rays. This is superior to integration using the trapzoidal or Simpson's rule, which require significantly more points for the same accuracy. Unfortunately, Gaussian quadratures do not generalise to higher dimensions. Therefore the choice of rays in 3D radiative transfer is significantly less straight forward and quite diverse angle quadratures are used in the literature.

One popular strategy to obtain a discretised set of rays was proposed by \cite{Carlson1963}, also see \cite{Bruls1999} for a more recent derivation. The so-called \textit{Carlson A} quadratures are defined in a single octant and are invariant under rotations of $90^{\circ}$ along the x,y and z axis. Furthermore, the Carlson A rays are equidistant in $\mu^2$, although only within one octant and not across octant boundaries. $\mu_1$ is therefore a free parameter and related to the spacing $\Delta\mu$. The $\phi$ angles follow from the invariance under $90^{\circ}$ rotation. This gives rise to the triangular patterns in the octants of panels 2a and 2b in Fig.~\ref{fig:AngleQuad}. This scheme has been implemented in many radiative transfer codes including \MultiD, \Scate, \Balder and \Iris.

Another approach is to split the 2D integral in Eq.~\eqref{eq:Jnu} into two 1D integrals. The codes \Porta and \Asset use a Gauss--Legendre quadrature for the integration in $\mu$ and trapezoidal integration in $\phi$. We suspect that they  employ the double-Gauss--Legendre quadrature mentioned above. The integration weights for each ray are therefore the tensor-product of the 1D quadrature weights. Double-Gauss--Legendre quadratures are presented in panels 3a and 3b of Fig.~\ref{fig:AngleQuad}. For better comparison with the other schemes we have taken the liberty to rotate rings of constant $\mu$ by $90^{\circ}/(n_\mu -1)$ with respect to their neighbours. \Balder implemented a similar quadrature, only that they choose a Gauss--Lobatto quadrature in $\mu$, which includes the vertical rays \citep{Amarsi2017}. Their quadrature choice is shown in panel 4a of Fig.~\ref{fig:AngleQuad}, although their particular quadrature is rotated by another $\pi/12$, probably in order to avoid the alignment of the rays with the x and y axes. Unfortunately, they do not quantify the gain from using this quadrature instead of the Carlson quadratures used in their previous studies. \cite{Stepan2020} propose another type of quadrature which is based on spherical harmonics. Employing their quadrature schemes, two of which are presented in panels 6a and 6b, they find similarly accurate results using up to 30\% less rays compared to the tensor product approach of Gauss--Legendre and trapezoidal integration. It should be noted, that their scheme, however, does not satisfy the $F_{\nu}^+ = I_\nu/2$ condition in an isotropic radiation field.

The previously mentioned double-Gauss--Radau quadrature is used in the MHD codes \Stagger, \Bifrost and \Dispatch \citep{Trampedach2013} and is also available in \LinforD. It is shown in panels 5a and 5b of Fig.~\ref{fig:AngleQuad}.
 
The \Magritte code uses the HEALPix discretisation of the sphere \citep{Gorski2005}, shown in panels 1a and 1b of Fig.~\ref{fig:AngleQuad}. This quadrature is equally spaced in $\mu$ and $\phi$, however every second ring of constant $\mu$ is rotated by $\pi/n_\phi$. This has the result that all rays are equally weighted.

The current state of the literature does not allow to utter a definite recommendation for which angle quadrature is the most accurate. There appears to be some agreement, that the Gauss--Legendre tensor product quadratures are sub-optimal \citep{Ibgui2013, Stepan2020, JaumeBestard2021} and that the double-Gauss--Radau quadrature is an improvement over the Carlson quadratures for the case of stellar atmospheres \citep{Amarsi2017}. Naturally, it would be good to see a comparison of the double-Gauss--Radau quadrature and the ``near optimal'' quadratures proposed by \cite{Stepan2020} and \cite{JaumeBestard2021}. Furthermore it would be interesting to see how Sobolev--Lebedev quadratures as those proposed by \cite{Ahrens2009} , which are invariant under the icosahedral rotation group, perform in comparison.

Generally, it seems like one needs relatively few rays to approximate $J_\nu$ to sufficient accuracy, when calculating the impact of the radiation field on the source function. Significantly more rays are recommended for the calculation of spectral lines in $F_\nu^+$ in order to reproduce the centre-to-limb variation and therefore fluxes from stellar atmospheres. \cite{Bergemann2019} use the Carlson A4 quadrature, so 12 rays per hemisphere, for the statistical equilibrium calculation of Mn.  Using \Balder \cite{Wang2021} state the use of 26 rays for the statistical equilibrium, i.e. $J_\nu$, calculation of atomic Li. This corresponds to the 13 rays on the hemisphere in panel 4a. However, for the emerging fluxes they use a total of 57 rays distributed over a hemisphere. For the considerably more complex atoms Fe and K, they revert to 5 rays on the hemisphere for $J_\nu$ and 29 rays for $F_\nu^+$ in \cite{Wang2022}. Many angles for the $J_\nu$ calculation are used in \cite{Koesterke2008}, which use 24 rays on the hemisphere ($n_\mu$=3, $n_\phi$=8). For the flux calculation they use essentially the same quadrature but reduce $n_\phi$ to 4 for the angles with the lowest $\mu$ value. This saves them 4 rays and yields 20 rays in total. \cite{StepanTrujillo2013} do a benchmark $J_\nu$ calculation with 80 rays on the hemisphere, but do not explain why they deem this high number of angles necessary.

\subsection{Escape probability} \label{sec:Escape}
For completeness, we should state that there is an approximate way of obtaining angle-dependent intensities, which avoids solution of the RTE. The emergent intensity along a given ray is
\begin{align}\label{eq:eq_mb1}
    I_\nu(\tau_\nu=0) &= \int_0^\infty S_\nu \, e^{-\tau_\nu}\,d\tau_\nu \\
    &= \int_0^\infty S_\nu \, p_\nu\,d\tau_\nu,
\end{align}
where the probability of a photon escaping the medium along this ray is defined as:
\begin{equation}
    p_\nu = e^{-\tau_\nu}.
\end{equation}
This quantity is very useful and it forms the basis of the \textit{Sobolev} approximation, which is a classic example of escape probabilities for lines \citep{Sobolev1957, Karp1977}. This approximation is often used in the calculations of RT for supernovae and kilonovae (Sec. \ref{sec:kilonovae}).

It follows from integrating Eq.~\eqref{eq:RTE}:
\begin{equation}\label{eq:eq_mb2}
    I_\nu(\tau_\nu=0) = \int_0^\infty (S_\nu-I_\nu)\,d\tau_\nu.
\end{equation}

Equating both expressions for the emergent intensity, eq. \ref{eq:eq_mb1} and eq. \ref{eq:eq_mb2}, we obtain an expression for the intensity $I_\nu$ at the optical depth $\tau_{\nu}$:
\begin{align} \label{eq:pesc1}
    \int_0^\infty S_\nu\,p_\nu\,d\tau_\nu &= \int_0^\infty (S_\nu-I_\nu)\,d\tau_\nu, \\
    I_\nu(\tau_\nu)  &\approx S_\nu(1 - p_\nu) \label{eq:pesc2}
\end{align}
While \eqref{eq:pesc1} is exact, the latter is only an approximation as it does not have to hold for each depth point in the atmosphere, unless $S_\nu$ is constant throughout the atmosphere. 

If there is a significant external source of intensity $I^{\rm ext}_\nu$, the expression \label{eq:pesc1} can be re-written as:
\begin{align}
    I_\nu(\tau_\nu) &\approx S_\nu(1 - p_\nu) + p_\nu \, I^{\rm ext}_\nu.
\end{align}
If $I^{\rm ext}_{\nu} \gg S_{\nu}$, as it is typically assumed in irradiated planetary atmospheres, this is a reasonable approximation. In this form, only the optical depth $\tau_\nu$ shall be integrated numerically, but there is no need to integrate $S_\nu$ along the ray. The \Cloudy code uses Eq.~\eqref{eq:pesc2} to calculate the mean intensity for each spectral line at each point in the model.
\subsection{Boundary conditions}

The boundary conditions of the formal solution generally depend on the astrophysical application. 

Box-in-a-star simulations of late-type (FGKM) stellar atmospheres and sub-surface convection employ periodic boundary conditions in the horizontal direction (that is perpendicular to the radial direction). Therefore these boundary conditions are adopted for the radiative transfer calculations in 3D boxes of stellar atmospheres as well. The origin of each ray is traced to either the top or bottom boundary. The boundary condition at the start of each ray does not differ from the boundaries applied in 1D radiative transfer codes. If the ray starts in the optically thick regime (bottom boundary), the starting intensity is assumed to be the Planck function at the bottom boundary temperature  \citep[e.g][]{Mihalas1973, Rutten2003, Ibgui2013}, i.e. $I_\nu=S_\nu=B_\nu$. For the top boundary, when the ray starts in empty space, one can for example assume that the stellar atmosphere is non-irradiated, i.e. the incoming intensity at the top is zero. 

Alternatively, incoming radiation from a binary companion, from the host star of an exoplanet, or even the chromosphere and corona can be modeled by adjusting the incoming (downward) intensity at the upper boundary. In the standard approach, this is done using the following approach:
\begin{equation}
    I_{\nu\mu}^{-} = I_{\rm \nu,\,irr} \cdot e^{-\tau_{\rm \nu,\,top}/\mu} + S_{\rm \nu,\, top} (1-e^{-\tau_{\rm \nu,\,top}/\mu})\, ,
\end{equation}
where $I_{\rm \nu,\,irr}$ is computed from the flux of the star times the stellar radius and diluted by the square of the orbital distance of the planet.
Extrapolations of $I_{\nu}$ and opacity are also possible and sometimes necessary when smooth gradients in $J_{\nu}$ are needed. This is for example the case, when calculating heating or cooling rates for 3D RHD stellar atmospheres.

For RT calculations at atmospheric conditions of massive OBA-type or WR stars, it is common to adopt the diffusion approximation at the inner boundary. In modeling stellar winds, the inward intensity at the outer boundary may also be specified. However, this may lead to a discontinuous change in the radiation field at the outer boundary and therefore to problems with the statistical equilibrium solver, which may require an approximate treatment of $I^{-}$ or extrapolations of the atmospheric structure.

\subsection{Opacities and emissivities}
As is clear from Eq.~\eqref{eq:RTE}, radiation transfer depends highly on opacities and emissivities, which are the two critical parameters that describe the local state of matter at each point in the atmospheres. For convenience, different definitions of opacities are used in the literature, which all depend simply on the expression of \textit{monochromatic extinction coefficient} representing the cross-section $\sigma_{\nu}$ in cm$^2$ per particle, but it can be given (or tabulated) either per unit volume, per unit mass, or per unit length.

Here, 
we use the following definitions:

 \begin{itemize}
     \item monochromatic extinction coefficient per gram of material, also referred to as \textit{opacity} $\kappa_{\nu}$, given in cm$^2/$g;
     
     \item monochromatic extinction coefficient per unit volume $\chi_{\nu}$ is defined as $\chi_{\nu} = \kappa_{\nu} \times \rho$, with $\chi$ given in cm$^{-1}$;

     \item monochromatic extinction coefficient per unit length $\chi$, equivalent to the one above;
 \end{itemize}
 
Definitions equivalent to the above can also be given for emissivities. We note that the notation of the geometric mean free path of the photon $l$ follows from the assumption of the photon propagation in a homogeneous gas without emission, and it represents the mean optical depth of unity ($<\tau_{\nu}> =1$), hence $l = 1/\chi_{\nu} = 1/(\kappa_{\nu} \rho)$.
Opacities and emissivities (as well as, of course, the source function defined as their ratio, that is, $S_{\nu} = \eta_{\nu}/\chi_{\nu}$), depend on local kinetic temperature, gas density (and pressure), detailed chemical abundances, atomic and molecular data, and, in the case of NLTE, also on the radiation field. 

The coupling between the atomic/molecular parameters and the properties of the gas are rather simple, for their derivation of this quantity, we refer to the afore-mentioned book by \citet{Rutten2003}. Assuming the Maxwell velocity distribution for particles involved, the following relations for \textit{opacity and emissivity due to bound-bound radiative transitions} between a lower energy state $i$ and a higher energy state $j$ hold:  
\begin{align}
    \eta_{ij,\nu} &= \frac{h \nu}{4 \pi} A_{jl} n_{j} \psi \label{eq:eta} \\ 
    \chi_{ij,\nu} &= \frac{h\nu}{4 \pi} (n_{i} B_{ij} \varphi - n_{j} B_{ji} \zeta)  \label{eq:chi} \\
    S_{ij, \nu}  &= \eta_{\nu}/\chi_{\nu} = \frac{A_{jl} n_{j} \psi}{(n_{i} B_{ij} \varphi - n_{j} B_{ji} \zeta)} = \frac{2 h \nu^3}{c^2} \frac{\psi/\varphi}{(g_j n_i/g_i n_j) - \zeta / \varphi}, \label{eq:source}
\end{align}
where $A_{ji}, B_{ij}, B_{ji}$ are the Einstein coefficients and $\varphi, \zeta, \psi$ are the profile functions for absorption, stimulated emission and spontaneous emission, respectively. The latter equation for the source function reduces to the Planck function ($S_{\nu} = B_{\nu}$), if the level populations $n_i$ and $n_j$ are LTE. 

If the assumption of complete redistribution (CRD) is used, all profile functions are assumed to be equal $\varphi=\zeta=\psi$. With the application of relationships between Einstein coefficients ($g_i B_{ij} = g_j B_{ji}$ and a similar expression for $A_{ji}$), the Eqs.~\eqref{eq:eta} and \eqref{eq:chi} in NLTE then simplify to the commonly used form in CRD:
\begin{align}
    \chi_{ij,\nu} &= \frac{h\nu}{4 \pi} \varphi B_{ij} n_{i} (1  - g_i n_j /g_j n_{i}) \\
     &= \sigma_{\nu}^l n_{i} (1  - g_i n_j /g_j n_{i}) \\
     &= \sigma_{\nu}^l n_{i}^{\rm LTE} (b_i - \frac{g_i}{g_j} \frac{n_{j}^{\rm LTE}}{n_{i}^{\rm LTE}} b_j) \\
     &= \sigma_{\nu}^l n_{i}^{\rm LTE} (b_i - b_j \exp(-h\nu/k_{B}T)) \\
     &= \chi_{ij,\nu}^{\rm LTE} b_i (1 - \frac{b_j}{b_i} \exp(-h\nu/k_{B}T))/(1 - \exp(-h\nu/k_{B}T)) \label{eq:alpha_NLTE} 
\end{align}
where, the NLTE departure coefficient is defined as $b_i=n_i /n_i^{\rm LTE}$, that is the ratio of atomic number densities of an atomic (or molecular) energy state $i$ as computed in NLTE, $n_i$, and its counterpart computed in LTE using the Saha-Boltzmann statistics $n_i^{\rm LTE}$ \citep{Rutten2003}. The LTE extinction coefficient was defined as $\chi_{ij,\nu}^{\rm LTE}$. 
For a bound-bound radiative transition, its numerical value at a given frequency $\nu$ (note that in codes wavelength units are often used) is simply:
\begin{align}
    \chi_{ij,\nu}^{\rm LTE} &= \sigma_{\nu}^l n_{i}^{\rm LTE}(1 - \exp(-h\nu/k_{B}T)) \\
    & = \frac{\pi e^2}{mc} f_{ij} \varphi(\nu - \nu_{0}) n_{i}^{\rm LTE} (1 - \exp(-h\nu_{0}/k_{B}T))
\end{align}
Correspondingly, the emissivity and the source function are written as:
\begin{align}
    \eta_{ij,\nu} &= \frac{h\nu}{4 \pi} \varphi A_{ji} n_j \\ 
     &= \sigma_{\nu}^l n_{j} \frac{2 h \nu^3}{c^2} \frac{g_i}{g_j} \\
    S_{ij, \nu}  &= \frac{2 h \nu^3}{c^2} \frac{1}{(g_j n_i/g_i n_j) - 1} \\ 
     &= \frac{2 h \nu^3}{c^2} \frac{1}{\exp(h\nu/k_{B}T) b_i/b_j - 1} \\
     &= B_{\nu} \frac{(\exp(h\nu/k_{B}T) - 1)}{\exp(h\nu/k_{B}T) b_i/b_j - 1} \label{eq:src_NLTE}
\end{align}
where $B_{\nu}$ is the Planck function and $\sigma_{\nu}^l$ is the already introduced monochromatic extinction coefficient per particle for a frequency point within the line ($\sigma^l =\frac{h \nu}{4 \pi} B_{ij} = \frac{\pi e^2}{mc} f_{ij}$, with $f_{ij}$ being the oscillator strength). The relationships are very powerful, as they can be conveniently employed to convert an LTE spectrum synthesis code into a NLTE code, as long as the departure coefficients $b_i$ and $b_j$ are available from NLTE statistical equilibrium codes. These relations, or numerically simplified re-formulations thereof (as only $\chi$ and $S$ are used), are therefore used in post-processing 1D NLTE spectrum synthesis codes, such as SIU \citep{Reetz1999}, Turbospectrum v20 \citep{Gerber2023}, and PySME \citep{Piskunov2017}. A comparison of NLTE synthetic profiles computed with such scaling relations and with direct NLTE RT codes can be found, e.g. in \citet{Wehrhahn2023}. NLTE departure coefficients computed with different codes are compared, e.g. in \citep{Bergemann2019}. These relationships in principle allow the application of NLTE departure coefficients in 3D NLTE codes, but we are currently not aware of such implementations in any of the codes described in Sect.~\ref{sec:codes}.

For the interpretation of NLTE effects for the near-UV and optical diagnostic lines, it has become convenient to re-write the equations \ref{eq:alpha_NLTE} and \ref{eq:src_NLTE} as follows (since $h \nu > k_{\rm B}T$, which holds in the optical):
\begin{align}
    \chi_{ij,\nu}^{\rm NLTE} &\approx \chi_{ij,\nu}^{\rm LTE} b_i \\
    S_{ij}  &\approx  \frac{b_j}{b_i} B_{\nu}
\end{align}
The assumption of complete redistribution physically means that the probability to emit a photon somewhere within a spectral line is not correlated with the probability to absorb there, so the source function is defined for a line and is assumed to be independent of frequency. The CRD is a reliable assumption if the collisions are frequent enough to ensure a random redistribution of the frequency of the scattered photon. However, this approximation may break down for the strongest chromospheric spectral features, such as the Na D lines,  Mg II H \& K,  Ca II H \& K, the core of H$_{\alpha}$, which require partial re-distribution \citep{Leenaarts2009a, Leenaarts2010, Leenaarts2013a, Leenaarts2013b, Pereira2015, CruzRodriguez2019, Moe2024} and we refer the reader to \citet{Linsky2017} for a review of PRD-sensitive diagnostic lines in stellar atmospheres.

The distinction between continuum and line opacities has classically relied on the underlying physical process at hand. The continuum opacities as typically defined as due to bound-free or free-free transitions in interactions  between the radiation field and different types of gas particles. These can be of various types \citep[][their table 1]{Gustafsson2008}, including photo-ionization reactions in atoms  \citep{Bautista1997, Nahar1997, Nahar2015, Bautista2017,Smyth2019}, photo-dissociation and photo-ionization reactions in molecules \citep{Heays2017,Hrodmarsson2023}, and their reverse reactions (recombination and photo-induced attachment), but also free-free transitions in various species, e.g. negative ions of atoms, such as H$^-$, C$^-$, N$^-$, O$^-$ \citep[e.g.][]{John1975a, John1975b, Bell1988, Ramsbottom1992}, negative ions of molecules CO$^-$, H$_2$O$^-$. For stellar atmospheres of FGK-type stars, the most important source of continuum opacity are the H$^-$ bound-free transitions, for which new cross-sections were recently published by \citet{McLaughlin2017}. Also other physical processes, such as CIA (collisionally-induced absorption) between H I atoms or H$_2$ molecules \citep{Doyle1968, Borysow2001} may lead to a quasi-continuum type absorption. Finally, isotropic scattering representing Rayleigh scattering on H I, He I, H$_2$, and on free electrons, is typically included. The latter is coherent isotropic scattering, and it is only relevant in the UV spectrum of extremely metal-poor RGB stars \citep{Gustafsson1975, Cayrel2004, Sobeck2011}, but in this wavelength regime there are additional complexities due to the chromosphere \citep{Avrett2008,Dupree2016} and these effects make it hardly possible (and barely reasonable) to study such types of coherent isotropic scattering in any detail. 

\begin{figure}[ht]
\centering
\includegraphics[width=0.9\textwidth]{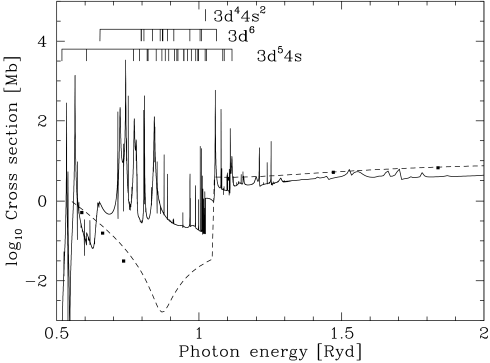}
\caption{Detailed photo-ionization cross-section for the ground state of neutral manganese. The data shown with the solid line were computed using the R-matrix. We also show the data computed using the central-field approximation by \citet{Verner1995} and \citet{Reilman1979}. Image reproduced with permission from \citet{Bergemann2019}, copyright by the author(s).}
\label{fig:photoMn}
\end{figure}

The continuum absorption is usually wide-band, covering a huge range of frequencies and, for most of the afore-mentioned reactions can be triggered by any photon, as long as its frequency is greater then the threshold (minimum) energy value $\nu_{\rm threshold}$. However, for practical computational reasons, especially in 3D NLTE, it is normally not possible, neither feasible, to include the semi-infinite range of corresponding cross-sections. Hence for the background continuum, a fixed grid is typically adopted, which covers fully the dominant range of stellar fluxes for a given type of stars. For example the DETAIL code for FGK-type ($\teff \sim 4000$ to $7000$ K) stars covers the frequency range from $6 \times 10^{15}$ Hz (500 \AA) to $10^{11}$ Hz (3000 $\mu$m = 3 mm) for the continuum with a mesh, which is non-equidistant in the wavelength space and has a dense sampling of 3 to 5 \AA\ in the optical and IR, but very coarse sampling at higher wavelengths. MULTI, in contrast, defines the continuum mesh directly by the properties of individual bound-free transitions provided in the model atom, which are usually taken from published datasets of atomic physics calculations. For example, for Mn I the photo-ionization cross-sections are sampled at 5000 evenly-spaced energy points between zero and 0.8 Ryd above the first ionisation threshold, followed by 250 points from 0.8 Ryd to 2.0 Ryd \citep[][]{Bergemann2019}. Figure~\ref{fig:photoMn} shows examples of cross-sections for two Mn I states obtained from theoretical quantum mechanical calculations. In other codes, such as \Cloudy which is mostly used for planetary atmospheres or very hot environments, the wavelength coverage as offered by linelists and continuum opacities extends to much greater ranges from $\sim 1$~\AA\ to over 3 meters \citep{Ferland2013}. However, for the vast majority of neutral and singly-ionized atoms of astrophysical interest (except a few core species like H I, He I, and recent update to Fe II in \citealt{Chatzikos2023}), no detailed quantum-mechanical photo-ionization cross-sections are available in \Cloudy \citep{Ferland2017}.

In bound-bound transitions, it is only the transition probability $f_{ij}$ that matters, which enters the equations together with the statistical weight $g_{ij}$ of the energy state, its degeneracy. The values of $f_{ij}$ for atoms and molecules can be calculated using theory or measured via experiment by combining the level lifetimes and branching fractions. Large-scale spectroscopic surveys like APOGEE or Gaia-ESO devote a substantial effort to compile accurate datasets to describe the stellar spectra taken within the survey programs \citep{Shetrone2015, Heiter2021}. For atoms, the statistical weight depends only on the orbital angular momentum quantum number of each energy level. For molecules, however, one has to take into account the nuclear spin degeneracy $g_{\rm nuc}$ \citep{Demtroeder2010}:
\begin{align}
  & \text{atoms}     & g_{ij} &= (2J + 1) \\
  & \text{molecules} & g_{ij} &= g_{\rm nuc} (2J + 1),
\end{align}
where $g_{\rm nuc}$ depends on several properties of molecules. One of them is the nuclear spin $I$ of the individual atoms within the molecule. The other is the molecular point symmetry group. For example, diatomic heteronuclear molecules, such as CN and CO, have the symmetry term of unity \citep{Ochkin2009}. The main isotopes of C and O ($^{12}$C, $^{16}$O) have $I=0$ and hence for the main isotope $^{12}$C$^{16}$O, $g_{\rm nuc} = 1$. For N ($^{14}$N), $I=1$, as a consequence for the $^{12}$C$^{14}$N isotopologue, $g_{\rm nuc} = 3$ (product of $g_{\rm nuc}$ for $^{12}$C and $^{14}$N). There are more complexities for non-linear poly-atomic molecules, such as H$_{2}$O. These differences are significant and shall be accounted for, when computing the molecular opacities for applications in stellar and planetary atmospheres.

\subsection{Broadening}
\label{sec:broadening}
\newcommand{\qD}{q_{\rm D}}
\newcommand{\vproj}{v_{\rm proj}}
The absorption and emission lines in spectra of astronomical objects are not infinitely narrow, but represent a characteristic shape reflecting different broadening processes. 
The broadening processes can be broadly classified into intrinsic (internal to the physical system) and extrinsic (external). The former represent the natural broadening, thermal broadening due to motion of particles (usually according to Maxwellian distribution), and pressure broadening. The pressure broadening arises due to elastic collisions between different kinds of particles. The natural broadening is the consequence of the Heisenberg uncertainty principle and the finite lifetime of energy states. External broadening mechanisms include, but are not limited to, stellar rotation and broadening of the intensity due to the finite slit width (instrumental).

The profile function $\phi$ is usually defined by the Voigt profile, i.e. the convolution of profiles due to different broadening processes: 
\begin{align}
\psi(\nu - \nu_0) = \phi(\nu - \nu_0) = \dfrac{H(a, \upsilon)}{\sqrt{\pi} \Delta \nu_D}
\end{align}
with
\begin{align}
a = \dfrac{\gamma_R + \gamma_3 + \gamma_4 + \gamma_6}{4\pi \Delta \nu_D} \hspace{0.5cm} \upsilon = \dfrac{\nu - \nu_0}{\Delta \nu_D} 
\end{align}

where $\gamma_R$ for natural damping and $\gamma_n$ for pressure broadening, $n$ being the impact parameter. The Doppler broadening is due to the thermal and non-thermal (microturbulence) motions of atoms:
\begin{equation} 
\Delta \nu_D \equiv \dfrac{\nu_0}{c}
\sqrt{\dfrac{2kT}{m} + v_{\rm mic}^2} \label{eq:DopplerWidth}
\end{equation}
with $m$ the mass of the atom under consideration. The parameter $v_{\rm mic}$ is the microturbulence velocity, which is defined as the parameter describing qualitatively random motions on scales \emph{smaller} than the photon mean free path. The concept of micro-turbulence will be described in more detail in the next section (Sect. \ref{sec:velocities}). 

Doppler broadening arises because the observer sees the integrated radiation from particles moving with different velocities along the \emph{line-of-sight}. It is described by a Gaussian profile: 
\begin{equation} \label{eq:gaussian} \psi(\nu - \nu_0) = \dfrac{1}{\Delta \nu_D \sqrt{\pi}} e^{[(\nu - \nu_0)/\Delta \nu_D]^2}
\end{equation}
The Voigt profile, i.e., the convolution of a Gaussian and a Lorentzian profile is described as:

\begin{align} 
    \phi(q,\qD,\gamma) \label{eq:voigt_profile}
    &= \frac{e^{-q^2/\qD^2}}{\sqrt{\pi}\qD} \ast
    \frac{\gamma}{\pi(q^2 + \gamma^2)} \\
    \phi(u, a) 
    &= \frac{e^{-u^2}}{\sqrt{\pi}} \ast
    \frac{a}{\pi(u^2 + a^2)},
    \label{eq:voigt_function}
\end{align}
where $q = \nu/\nu_0 - 1$, $q_D = \Delta\nu_D/\nu_0$, and (in 1D static model atmospheres, for 3D RT in convective models see below) $u = q/q_D$, and $a = \gamma/q_D$.

In 3D RT calculations, a critical difference in the calculations of profile and hence, opacity and emissivity, arises because of the 3D velocity fields and hence the requirement to project the real velocity vectors (gas velocities) on the direction of photon propagation \citep[see also][section 8.3]{Ibgui2013}. In this case, the $u$ parameter that enters the exponent in the numerator and the denominator in Eq.~\eqref{eq:voigt_function} changes its form to a slightly more sophisticated expression:
\begin{align}
  u= \frac{(\nu/\nu_0 - 1 - v_{\rm proj}/c)}{\sqrt{2kT/m c^2}} = \frac{(\nu - \nu_0 (1 + v_{\rm proj}/c))}{\frac{\nu_0}{c}\sqrt{2 kT/m}} = \frac{q - v_{\rm proj}/c}{\qD}
\end{align}
where $v_{\rm proj}$ is the projection of the velocity vector on the ray. \textit{It is exactly this exponential dependence of the opacity - through the profile function, Eq.~\eqref{eq:voigt_function} - on gas temperature and gas velocity that leads to drastic differences in the line formation between 1D and 3D radiation transfer in 1D hydrostatic and 3D radiation-hydrodynamics simulations of atmospheres, respectively.}

The Voigt function can be computed using different recipes. For example, in the \Cloudy code \citep{Ferland2017}, the expression based on \citet{Wells1999} is used. Other techniques include \citet{Kuntz1997,Letchworth2007,Zaghloul2007,Ngo2013, Tran2013}. We are not aware of dedicated tests of the profile implementation within the astrophysical domain of the objects we are dealing with in this review. However, the following study deserves to be mentioned. \citet{vonClarmann2002} present a comparative analysis of LTE and NLTE line profiles computed with different RT codes under conditions of the Earth's middle atmosphere. Their results suggest that despite the vast differences in the underlying micro-physics and numerical recipes, the agreement between the synthetic observables is very good and is usually better than $0.5$ percent, in some cases up to a few percent. For exoplanet or stellar applications, such effects are of no relevance, except when extremely small astrophysical signatures are sought for. One special example is, however, the case when arbitrary fits using a Voigt profile are needed to determine \textit{the line shifts} in the observed spectra for the purpose of studies of the variability of the fine structure constant $\Delta \alpha/\alpha$ \citep{Murphy2001, Griest2010, Murphy2022}.

Regarding the damping coefficient, the general expression for the damping constant $\gamma$ is as follows:
\begin{equation}\label{eq:damp_const} 
\gamma_n = N \int_{\infty}^{\infty} \upsilon f(\upsilon) \sigma_n(\upsilon)
d\upsilon 
\end{equation}
where $N$ and $f(\upsilon)$ are the number density and velocity distribution of perturbing particles, $\sigma_n$ is the broadening cross-section and this physical quantity can be obtained from theoretical calculations for transitions between different energy states of different ions. Depending on the type of perturbers, the impact parameter $n$ may take different values. Resonance broadening, $n = 3$, represents interaction between two hydrogen atoms. The Quadratic Stark effect, $n = 4$, causes broadening due to the interaction of an atom with a charged particle, e.g. free electrons or protons. A rather special case is the van der Waals broadening, $n = 6$, describing the interactions in collisions of atoms or molecules with neutral H atoms. The latter is formulated using the ABO theory \citep{Anstee1991, Anstee1995, Barklem1998, Barklem2000}, which is based on the full interaction potential between an atom and a hydrogen particle, avoiding series expansion in $1/r$. Line broadening cross-sections $\sigma_6(\upsilon_0)$ for a relative collision speed $\upsilon_0 = 10$ \kms\ are tabulated as a function of effective principal quantum numbers for the upper and lower level of a transition. Damping constants are then calculated as:
\begin{equation} 
\gamma_6 = N_H \left(\dfrac{4}{\pi}\right)^{\alpha/2} \Gamma\left(\dfrac{4 -
\alpha}{2}\right)  \left(\dfrac{\upsilon}{\upsilon_0}\right)^{-\alpha} \upsilon
\sigma_6(\upsilon_0)
\end{equation}
with $\Gamma$ the gamma function, $\upsilon$ the relative velocity of an atom and a perturber, and $\alpha$ the velocity exponent. For a comprehensive review of the broadening cross-sections for different types of perturbers, we refer to the reader to \citet{Barklem2016}.

Finally, high-resolution synthetic spectra are then  convolved with a the instrumental profile (see \cite{Gray1992}) in order to compare with observations with a given spectral resolution of $R$. The instrumental broadening is commonly treated like macroturbulence, by a Gaussian with the full width at half maximum $c/2\sqrt{\ln 2}R_{\rm sp}$. Here, $R_{\rm sp} = \lambda/\delta \lambda$ is the resolution of a spectrograph.
\subsection{Velocities} \label{sec:velocities}

Microscopic thermal motion of particles as well as macroscopic convective motions of the gas lead to observable broadening of spectral features. Typically, at conditions of atmospheres of FGKM-type stars, the Doppler shifts are of the order a $\sim$ few $100$ meter to a few km/s \citep{Sheminova2020}. For the solar spectral lines, for example, Doppler shifts of up to 800 m/s are observed \citep{AllendePrieto1998, AllendePrieto2002b, Reiners2016}. This implies a wavelength shift of a 0.1 \AA\ to few milli-\AA. Therefore continuum opacities can be considered as fairly isotropic, i.e., independent of the apparent motion of the gas. However, the velocity shifts may be significant in spectral lines, and these may interfere with other effects, like hyperfine splitting, isotopic structure, or planetary motion.

In RT calculations in 1D plane-parallel (or spherically-symmetric) homogeneous atmospheres, the only source of microscopic broadening is the thermal broadening due to random isotropic motions of particles. However, this thermal broadening is not sufficient, and it is common in all 1D RT calculations \citep{Kurucz1970, Gustafsson2008} to adopt an additional non-thermal component to 'correct' the Doppler width for the \textit{absence of turbulence and convection} (see eq. \ref{eq:DopplerWidth}). This component is known as microturbulence $v_{\rm mic}$.
Typically, $v_{\rm mic}$ is either a free parameters, or it solved iteratively together with the determination of stellar abundances and metallicities \citep{Bergemann2012a}. In some cases, e.g., in large spectroscopic surveys like Gaia-ESO \citep{Smiljanic2014} and GALAH \citep{BuderSharma2021}, functional forms for $v_{\rm mic}$ are assumed. For the Gaia-ESO survey, in particular, we (M. Bergemann and V. Hill; ref. v2, 2013) developed the following $v_{\rm mic}$ velocity parametrisation using the fits to the spectroscopic measurements of $v_{\rm mic}$ obtained from the literature. 

The following relations for microturbulence velocity v$_{\rm mic}$ (in km\,s\(^{-1}\)) are valid for:
\[
4000 < T_{\text{eff}} < 7000,\quad 0 < \log g < 5,\quad -4.5 < \mathrm{[Fe/H]} < +1
\]

Let:
\[
t_0 = 5500\ \text{K},\quad g_0 = 4.0,\quad t = T_{\text{eff}},\quad g = \log g,\quad m = \mathrm{[Fe/H]}.
\]

\begin{itemize}
    \item \textbf{Main sequence and subgiants} (\(T_{\text{eff}} \geq 5000\,\text{K},\ \log g \geq 3.5\)):
    \begin{align*}
    v_{\rm mic} = 1.05 + 2.51 \times 10^{-4}(t - t_0) + 1.5 \times 10^{-7}(t - t_0)^2 - \\ 0.14(g - g_0) - 0.005(g - g_0)^2 + 0.05m + 0.01m^2
    \end{align*}

    \item \textbf{Main sequence stars} (\(T_{\text{eff}} < 5000\,\text{K}\,\ \log g \geq 3.5\)):
    \begin{align*}
    v_{\rm mic} = 1.05 + 2.51 \times 10^{-4}(5000 - t_0) + 1.5 \times 10^{-7}(5000 - t_0)^2 - \\ 0.14(g - g_0) - 0.005(g - g_0)^2 + 0.05m + 0.01m^2
    \end{align*}

    \item \textbf{Giants} (\(\log g < 3.5\)):
    \begin{align*}
    v_{\rm mic} = 1.25 + 4.01 \times 10^{-4}(t - t_0) + 3.1 \times 10^{-7}(t - t_0)^2 - \\ 0.14(g - g_0) - 0.005(g - g_0)^2 + 0.05m + 0.01m^2
    \end{align*}
\end{itemize}
This formula is based on fitting the following datasets: \citet{Bensby2003,Bensby2005, Bruntt2012, Barklem2005, Bonifacio2009, Francois2007, AllendePrieto2004a, Takeda2005, Johnson2002, Ghezzi2010, Yong2008}, and \citet{Ruchti2013}.

We remind that $v_{\rm mic}$ is a free parameter, which does not have a physical basis in 1D calculations and cannot be derived from first principles. Since $v_{\rm mic}$ enters the equation for the line opacity, the resulting line profile is strongly correlated with the assumed $v_{\rm mic}$ \citep{Bergemann2012c, Takeda2022}. Such that for the spectral lines with EWs $> 80$ mA, a 0.1 \kms\ increase in $v_{\rm mic}$ leads to a 0.05 dex lower abundances, and vice versa \citep{Takeda2022}. Hence adopting a $v_{\rm mic}$ of 0.2 \kms\ only may change abundances by 0.1 dex (see also the discussion in \citealt{Bergemann2019}.
%

%


%
%
%
%
%
\subsection{Frequency quadrature} \label{sec:FreqQuad}
The primary objective of 3D NLTE radiative transfer is to output detailed synthetic spectra for comparison with spectroscopic observations. The accuracy of the resulting spectra and the computational cost of the calculation are directly impacted by the number and choice of frequencies for which the monochromatic radiative transfer equation is solved. One has to choose a frequency grid, which captures all the relevant features in the spectral region to be modelled, while keeping the computational cost at a reasonable level. The outputs of radiative transfer codes have infinite spectral resolution. When comparing to observations with low spectral resolution, it is therefore not enough to use the same frequency discretisation as employed in the observations.

In LTE one can compute a spectrum in a wavelength range of choice, while neglecting all other wavelengths without a change in precision. Even when taking coherent scattering into account (such as Thomson and Rayleigh scattering), photons are only scattered into different directions, but never change wavelength.
In NLTE this is not the case, since photons can be absorbed in one spectral line and be emitted in another at a different wavelength. In principle one has to model the entire spectrum even if one is only interested in the intensity of a single spectral line. This makes NLTE line formation significantly more complex than line formation under the assumption of LTE and is the main reason why it is computationally a lot more costly. In NLTE one should differentiate between the frequencies used when integrating the radiative rates and when printing the emerging flux. The latter has the same frequency requirements as LTE spectrum synthesis, while the former requires an adequate discretisation of the $\sigma_{ij} J_\nu$ term. Here the $J_\nu$ factor usually truncates the spectrum which has to be modelled in a certain astronomical environment.

A popular way to setup a frequency grid is to define a wavelength range between $\lambda_{\rm min}$ and $\lambda_{\rm max}$ and to distribute NF points between them with a constant wavelength step \citep[e.g.][]{deLaverny2012} or constant spectral resolution $\lambda/\Delta\lambda$ \citep[e.g.][]{Plez2008, Chiavassa2018}. For codes, which work on the frequency scale, one then has to convert wavelengths into frequencies via $\nu = c/\lambda$. This scheme is reasonable for LTE as well as the NLTE flux calculations, since one usually wants to compare to observations with a given $\lambda_{\rm min}$ and $\lambda_{\rm max}$. In principle one can construct an equally accurate frequency grid with fewer points, by placing the points exactly where strong gradients in the intensity with wavelength are expected. The problem is that the frequency grid has to be defined before any intensities are calculated and one can only guess where $dI/d\nu$ is strong before having a look at the actual spectrum.

One attempt to improve the efficiency is to distribute the frequencies on a line by line basis. This idea was adopted from older RT codes, which were not able to deal with overlapping lines. Therefore there was no  single frequency grid, but each line had its own frequency grid. If one assumes a nearly Gaussian line profile, it is clear that more frequency points are needed near the core of the line, while fewer points are needed where the wings fade into the continuum. Furthermore one has to specify  $\Delta\lambda_{\rm max} = \lambda_{\rm max} - \lambda_0$ for each line. Choosing a reasonable $\Delta\lambda_{\rm max}$ alone is not trivial as it depends on the modelled environment and is often adjusted manually in practice. One can combine this multitude of grids into a single frequency set by concatenation and subsequent sorting of frequencies. This approach is used in many NLTE applications when defining a frequency grid for the integration of the radiative rates. The spectral resolution is secondary when performing this integration, since the uncertainty in atomic data, such as oscillator strength, for example, easily surpasses the uncertainty that results from using a frequency quadrature that is relatively coarse compared to the spectrum synthesis resolution.

As a result, the frequency resolution of the lambda-iteration may be much lower than the final spectrum synthesis resolution. Nevertheless, this does not imply that the total number of frequencies utilised is lower. Bound-free transitions play a crucial role in establishing statistical equilibrium, even though their absorption features may hardly be visible in a spectrum. Since bound-free transitions are ``active'' over a wide frequency range, we must include a sufficient number of frequencies from these ranges to ensure the $\sigma_{ij} J_\nu$ term is probed sufficiently.

Photo-ionisation cross-sections, either computed \citep{Cunto1993, Badnell2005} or measured \citep{Shafique2022}, are typically on a very detailed frequency grid, since $\sigma_{ij}$ has a very complex dependence on energy owing to strong resonances in the cross-sections \citep{Nahar1997, Bautista2017}. Some of which are so insignificant to the statistical equilibrium that they can be omitted entirely. 

A rather coarse frequency discretisation of a spectral line is usually enough to obtain a reasonable estimate of the radiative rates. Weak spectral lines with low $\log (gf)$ are more likely to be insignificant, but this cannot be guaranteed in advance, and it must be validated by repeated calculation with and without them. In what follows we list the choices of the number of points per radiative transition. But, we note that the total number of frequency points is a product of the latter with the number of lines. For large atoms like Fe and Mn with their large number of radiative b-b transitions, e.g. $3000$ for Fe \citep{Lind2017} versus $1681$ for Mn \citep{Bergemann2019}, the total number of frequency points far exceeds 10$^4$.  \citet{Bergemann2019} employ 13 frequency points per line in the rate integrals, which was sufficient to obtain reliable level populations of Mn. \cite{Gallagher2020} use only 9 frequency points per line, arguing that a refined frequency resolution did not affect their diagnostic Ba II lines. \cite{Lind2017} use less than 5 frequency points per line on average to make calculations of their Fe NLTE model feasible in 3D. Employing the same Fe NLTE model, \cite{Asplund2021}, use roughly 10 frequency points per line. The references above dealt with FGK-type stars, where velocity gradients are small enough that convective broadening only has a minor impact on $\lambda_{\rm min}$ and $\lambda_{\rm max}$ chosen for each line.
For applications in hot stars with winds, where radiative transfer is usually carried out in the comoving frame, the situation is slightly more complicated. Strong velocity gradients shift the line profiles in the observer's frame far away from their rest position. Therefore $\Delta\lambda_{\rm max}\gg\rm{FWHM}$ of the line if we define $\lambda_{\rm max}$ as the wavelength where the line profile ``ends". The radiative rates can be integrated in the comoving frame, however, and only has to be evaluated where the intrinsic line profile without any Doppler shifts is non-zero.  Even though the P Cygni line profiles are represented by 100 frequency points in \cite{Lobel2008}, the rate integral is only performed over 3 frequency points. \cite{Hennicker2018} state that their frequency discretisation is comparatively much finer and the rate integral covers 15 frequency points per line. \texttt{CMFGEN} typically uses 13 points for the frequency quadrature. The default is to assume the same Doppler width for all species (a typical assumption in hot star atmospheres), but this can be relaxed. The frequency grid is based on the location of bound-free edges and lines.
\section{Non-Local Thermodynamic Equilibrium}\label{sec:nlte}

\subsection{Kinetic equilibrium or rate equations} \label{sec:rates}
The assumption of radiation in local equilibrium with the gas violates any object which has an outer transmitting boundary. 

The most computationally expensive part of NLTE is the calculation of the number density $n_i$ of the particles which occupy different internal energy states, $i$. In scientific literature, $n_i$ is also known as the level population or occupation number of the level $i$. Depending on the physical context, the concept of the level varies, and may refer to, e.g. the excitation state of an atom or a molecule, which may either refer to a hyperfine structure level, a fine structure level, a term (in the notation of NIST handbook), or even a ``hyper-state'' which is a statistical representation of hundreds of terms.

In NLTE, the level populations directly depend on the radiation field, that is, on the mean specific intensity of radiation $J_\nu$ at different frequencies $\nu$. Extensive  summaries on the determination of occupation numbers in NLTE have been presented elsewhere \citep[e.g.][]{Auer1969, Mihalas1988, RybickiHummer1991, RybickiHummer1992, Kubat2014}, and we refer the reader to these papers. Below we briefly outline the basic aspects of the coupled radiation transfer and SE solution.

Following \citet{Mihalas1973} and \citet{Rutten2003}, we define the probability $P_{ij}$ of a particle to transition from the energy state $i$ to $j$ as given by the sum of all collision-induced transition rates $C_{ij}$ and the radiation-induced transition rates $R_{ij}$: 
\begin{equation} \label{eq:Pmatrix}
    P_{ij} = C_{ij} + R_{ij}, 
\end{equation}
where the individual rate coefficients are defined as:
\begin{align}
    \label{eq:Rij}
    R_{ij} &= \int_0^\infty \frac{4\pi}{h\nu}\sigma_{ij}J_\nu {\,\rm d}\nu \\
    \label{eq:Rji}
    R_{ji} &= \int_0^\infty \frac{4\pi}{h\nu}\sigma_{ij}G_{ij} \left(J_\nu + \frac{2h\nu^3}{c^2}\right) {\rm d}\nu
\end{align}
with
\begin{equation}
    G_{ij} = \left[\frac{n_i}{n_j}\right]_{\rm LTE} e^{-h\nu/kT}
\end{equation}
where the first equation represents excitation or ionisation by photons, and the second equation represents the inverse process of radiative de-excitation or recombination. For molecular NLTE, these equations may also be used to represent, in a simplified expression, photo-dissociation processes \citep{Popa2023, Hoppe2026}. We note that in a molecule both kinds of processes, photo-ionizations and photo-dissociations may take place simultaneously. The cross-sections for all radiative transitions are typically computed with quantum-mechanical methods or are obtained via experimental methods (see Sect.~\ref{sec:models}).

For collision-induced reactions, including excitations and ionization, the rates of transitions are calculated as follows, with the detailed balance assumed for reverse reactions:
\begin{equation}\label{eq:col}
C_{ij} = n_e \int \limits_{\upsilon_0}^{\infty}
\sigma_{ij}(\upsilon)\,\upsilon\,f(\upsilon) d\upsilon
\end{equation}
where $\sigma_{ij}(\upsilon)$ is the electron collision cross-section, $f(\upsilon)$ is the Maxwellian velocity distribution, $\upsilon_0$ is the threshold velocity with $m \upsilon_0^2/2 = h \nu_0$. 

\begin{equation}
    C_{ij} = \left[\frac{n_i}{n_j}\right]_{\rm LTE} C_{ji}
\end{equation}

Statistical equilibrium, that is, the time dependence and advection terms are neglected, is then formulated as:
\begin{equation} 
\label{eq:stateq} n_{i}\sum_{j\neq i}P_{ij} = \sum_{j\neq i}n_{j}P_{ji}
\end{equation}

In contrast, the rate equations including the time dependence and advection:

\begin{equation}
\frac{\delta N_i}{\delta t} + \frac{\delta \upsilon N_i}{\delta z} = \sum N_j
P_{ji} - N_i \sum P_{ij}
\end{equation}

The equilibrium populations $n$ can be computed once the (K$\times$ K) transition matrix P is known. The equilibrium distribution is defined as the eigenvector of the transition matrix with eigenvalue unity.

\begin{equation} \label{eq:matrix}
    \mathrm{P}(n - 1) = 0
\end{equation}

The set of equations is closed by the particle conservation equation: 

\begin{equation}
\label{eq:numcons} \sum_{i,c}n_{i,c} = \frac{\alpha_{\rm el}}{\alpha_{\rm H}}\left(\sum_{i}n_{i,{\rm H}} + n_p\right)
\end{equation}
where $\alpha_{\rm el}/\alpha_{\rm H}$ is the fraction of all atoms and ions of an element relative to that of H atoms, $n_{i,\rm H}$ are the hydrogen level populations and $n_p$ is the number of ionised hydrogen atoms (free protons).
\subsection{Lambda iterations} \label{sec:lambdaIterations}
For a given $J_{\nu}$ we can fill the entries of the transition matrix P (Eq.~\eqref{eq:Pmatrix}) and predict how many atoms occupy each  energy state of a species i.e. the level populations numbers $n$. Unfortunately, $J_{\nu}$, the total incident radiation on a local point arriving from all directions, depends on the population numbers in the entire physical system.

\begin{figure}[ht]
\includegraphics[width=\columnwidth]{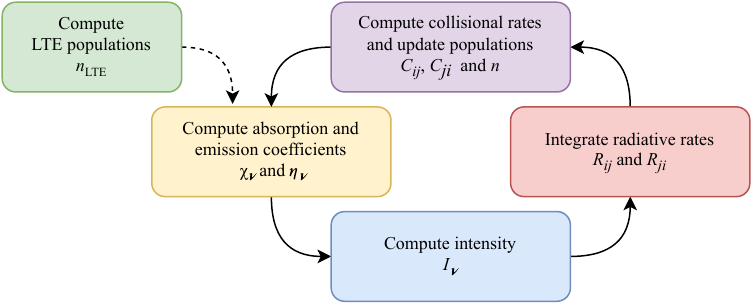}
\caption{The four conceptual steps of lambda iterations.}
\label{fig:LambdaIteraion}
\end{figure}

Solving radiation transfer and the equations of statistical equilibrium simultaneously is an overwhelming task, which is extremely time-consuming in a multi-D system with correlated motions within. It is more reasonable to estimate $J_{\nu}$ based on some initial level populations $n^0$ in order to predict new level populations $n$, which allow us to recompute $J_{\nu}$ in turn. This method results in iteratively updating $J_{\nu}$ and $n$ and has been shown to converge to the correct solution after a sufficiently large number of iterations. This is the method of $\Lambda$-iterations, which is schematically illustrated in Fig.~\ref{fig:LambdaIteraion}.

Here Lambda, $\bm\Lambda_{\nu}$, is the so-called \textit{operator}, or - more precisely - a method used to predict $J_{\nu}$ from the source function $S_{\nu}$. $\bm\Lambda_{\nu}$ can be expressed as a NP $\times$ NP matrix, where NP is the number of discrete points in the atmosphere model. The matrix element $\bm\Lambda_{\nu,ij}$ then contains the contribution weight of $S_{\nu}$ from point $j$ to $J_{\nu}$ at point $i$.

\begin{equation} \label{eq:Lambda}
 J_{\nu} = \bm\Lambda_{\nu}S_{\nu}
\end{equation}

Theoretically there is a non-zero chance of a photon arriving at point $i$ from any point in the atmosphere including $i$ itself, therefore $\bm\Lambda_{\nu}$ is a dense matrix with no zeros. The value of $\bm\Lambda_{\nu,ij}$ mainly depends on the optical depth $\Delta\tau_\nu$ between points $i$ and $j$, but one should not forget about the angle-dependence of $\Delta\tau_\nu$ introduced by velocity fields. 

In practice, this matrix is never constructed in 3D NLTE RT codes, not only because NP $\times$ NP is a very large number, but also because the matrix elements are not completely uncorrelated. This becomes more obvious when splitting $\bm\Lambda_{\nu}$ into its contributions from a discrete set of $\mu$-angles $\bm\Lambda_{\mu\nu}$.

\begin{equation} \label{eq:LambdaMu}
 J_{\nu} = \sum_{\mu}w_{\mu}\bm\Lambda_{\mu\nu}S_{\mu\nu},
\end{equation}
where $w_{\mu}$ is the angle integration weight and $S_{\mu\nu}$ is the angle dependent source function.
Consider a vertical ray travelling from the bottom of the atmosphere to the top. Firstly, all matrix elements $\Lambda_{\mu\nu,ij}$ for $\mu=1$ are zero if point $j$ is above point $i$ in the atmosphere. Secondly, if point $i$ and $j$ are not directly connected by a vertical line, $\Lambda_{\mu\nu,ij}$ should also be zero. Here it also becomes clear how important the angular quadrature is during the lambda-iterations.

How lambda iterations are carried out in practice without the explicit construction of the Lambda-operator is discussed in Sect.~\ref{sec:lambdaIterations}. 
\subsection{Accelerated Lambda Iteration} \label{sec:ALI}

A very common method to solve the statistical equilibrium and radiative transfer equations is the method of Accelerated Lambda Iteration) \citet{RybickiHummer1991, RybickiHummer1992}, which is employed in vast majority of modern astronomy codes, including \Cloudy, \Multi, \MultiD, \Detail, \Magritte, etc. This method builds upon the fundamental realisation of unsatisfactory convergence properties of $\Lambda$-iteration that seeded the development of  operator-splitting techniques \citep{ Scharmer1981, Scharmer1985, Olson1987, RybickiHummer1991}.

In this, the radiative transfer equation is solved approximately with a simplified operator \citep[see][for an overview]{Hubeny2003}. The exact operator is written as 
$\hat{\Lambda}_{\nu} = \hat{\Lambda}_{\nu}^* +
(\hat{\Lambda}_{\nu} - \hat{\Lambda}_{\nu}^*)$, where
$\hat{\Lambda}_\nu^*[S_\nu]$ is the \textit{approximate} operator.
The formula \eqref{eq:Lambda} is expanded to: 
\begin{equation} \label{eq:ali1} J^{(n)}_{\nu} =
\hat{\Lambda}_{\nu}^*[S^{(n)}_{\nu}] + (\hat{\Lambda}_{\nu} - 
\hat{\Lambda}_{\nu}^*)[S^{(n-1)}_{\nu}] \end{equation}
where $S^{(n-1)}$ is the source function from the previous iteration. The approximate operator acts on the new estimate of the source function, whereas the difference between the exact and approximate operator, $\hat{\Lambda}_{\nu} - \hat{\Lambda}_{\nu}^*$, operates on the previous known source function. The latter can be found from the formal solution. Then the new estimate is found from the inversion of $\hat{\Lambda}_{\nu}^*$. This inverted approximate operator acts as an accelerating factor that ensures fast convergence.

Equation \eqref{eq:ali1} must be solved for all frequencies $\nu$ that are related to the transitions in the atom under consideration and for all depth points $\tau_{d}$ in the model atmosphere. However, because the lambda-operator is integrated over all depths, the mean intensity for a certain depth point requires knowledge of  source functions at all other depths. One could take into account the depth-dependency for both exact and approximate operators, but this is costly in the computational sense. The solution is to choose the special form of the approximate operator $\hat{\Lambda}_{\nu}^*$, so that it is local for the first term in the right-hand side of the Eq.~\eqref{eq:ali1}. In this case, only the source function at one depth point $k$ is considered. This term can be written as $\hat{\Lambda}_{\nu}^*[S^{(n)}_{\nu,k}]$, whereas the dependency on other depth points $k'$ is taken into account only in the second term of the Eq.~\eqref{eq:ali1}. This is called the diagonal operator, because the matrix $\Lambda^*$ contains terms corresponding to the specified depth point $d$ diagonally. Using the diagonal operator results in a Jacobi iterative scheme \citep{Olson1986}. In the context of radiative transfer this means the following execution of tasks:

\begin{enumerate}
    \item Start from an initial guess of the source function at each point in the atmosphere $S^{(0)}_{\nu}$
    \item Solve the RTE by propagating the intensities $I$ from one boundary to the other. This can be done in arbitrary order or even in parallel.
    \item Integrate $I$ over all angles to get $J$. Compute an effective mean intensity $J_{\rm eff}$, which excludes the contribution of the local source function at each point.
    \item Plug this $J_{\rm eff}$ into the rate equations to determine the radiative rates of each transition.
    \item Solve the statistical equilibrium and obtain a new set of source functions $S^{(n+1)}_{\nu}$. 
    \item If the global update has not met the convergence criterion then jump to step 2 and repeat.
\end{enumerate}

\begin{figure}[ht]
    \centering
    \includegraphics[width=\columnwidth]{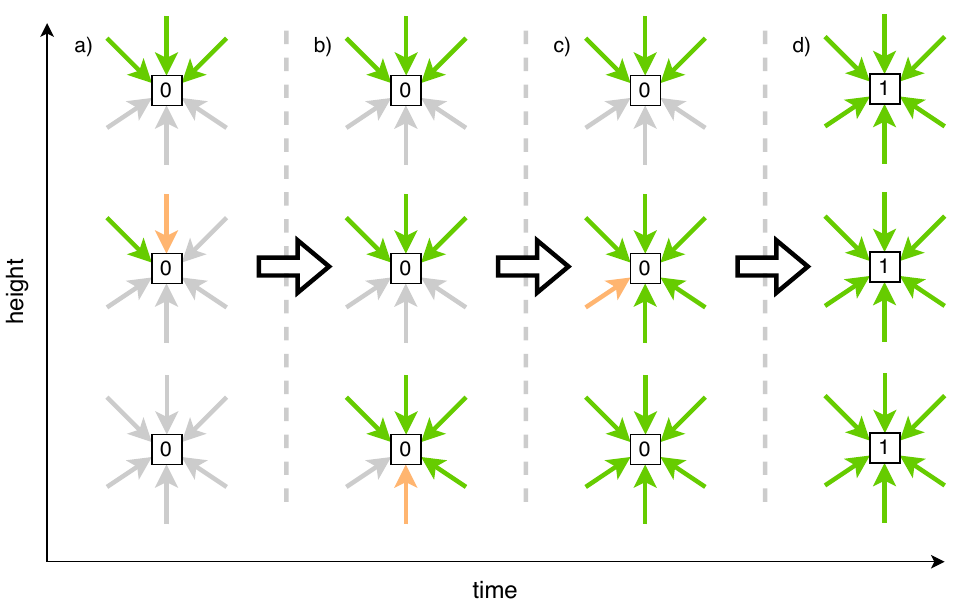}
    \caption{Jacobi iterations visualised in 4 steps (a, b, c and d) for a one-dimensional atmosphere consisting of 3 points.  The squares mark the grid points with the number showing the number of times the source function has been updated. The grey arrows mark the intensities that have yet to be computed, the green arrows mark those that have already been computed and the orange arrow marks the intensity that is computed in the current step. Most notably, the source function of all grid points is updated in the last step.}
    \label{fig:Jacobi}
\end{figure}

\begin{figure}[ht]
    \centering
    \includegraphics[width=\columnwidth]{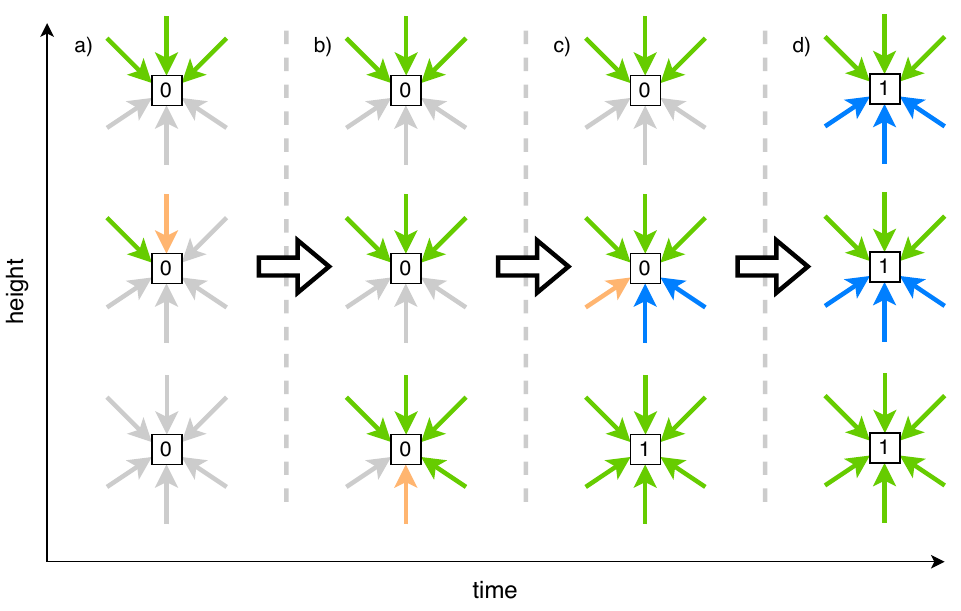}
    \caption{Same as Fig.~\ref{fig:Jacobi}, but for Gauss--Seidel iterations. While steps a) and b) are the same, the source function at the bottom of the atmosphere is already updated in step c). The intensities including the updated source function from this point already are coloured in blue.}
    \label{fig:Gauss-Seidel}
\end{figure}

The Jacobi method ensures the fastest execution of a single iteration step. A common alternative to this are the Gauss--Seidel (GS) method and its extension the successive over-relaxation (SOR) method. When iteratively solving a system of equations, these methods update the vector $S^{(n)}_{\nu}$ in-situ, meaning that one overwrites the source function results from the previous iteration as soon as new values can be obtained. Since the vector $S^{(n)}_{\nu}$ is updated element by element, one takes $i-1$ already updated elements into account, when updating the element $i$. The GS and SOR methods represent a lower triangular operator as an alternative to the diagonal operator of the Jacobi method. In order to implement such a scheme for radiative transfer think of $J$ as being made of an up and down part $J = J^+ + J^-$, where $J^+$ contains contributions from all angles with $\mu>0$. Now lets update the scheme from above in the following way, where we have dropped the frequency index for clarity:

\begin{enumerate}
    \item Start from an initial guess of the source function at each point in the atmosphere $S^{(0)}$
    \item Sweep the atmosphere from the top boundary to the bottom while calculating and storing all values of $J^-$ on the way. Now we start from layer k=1 (counting from the bottom) where we can make use of the bottom boundary condition.
    \item Compute $J^+_k$ and $J_k$ . Compute $J_{{\rm eff}, k}$ and plug it into the rate equations. Update element $S^{(n+1)}_k$.
    \item In case of a 3-point formal solver, which makes use of the downwind source function, one then has to correct $J^-_{k+1}$ as well as $J^+_k$  using $S^{(n+1)}_k$, since both originally had contributions from $S^{(n)}_k$.
    \item Increment k by 1 and repeat step 3 until reaching the other boundary.
    \item If the global update has not met the convergence criterion then jump to step 2 and repeat.
\end{enumerate}

For a more detailed description see \cite{Trujillo1995}, who were the first to propose this scheme in the context of NLTE radiative transfer. See \cite{TrujilloBueno1999} for an application to polarized line transfer. The SOR method differs from the GS method in the way that it calculates the source function update as
\begin{align}
    S^{(n+1)}_k = S^{(n)}_k + \omega \Delta S^{\rm GS}_k,  && \Delta S^{\rm GS}_k = \left[S^{(n+1)}_k - S^{(n)}_k\right]_{\rm GS}
\end{align}
where $1 <\omega < 2$. One has some freedom in choosing the parameter $\omega$, but the optimum value is the one which leads to the highest rate of convergence. Mathematically, this corresponds to the value of $\omega$ which minimises the spectral radius (the largest absolute eigenvalue) of the approximate operator $\hat{\Lambda}_{\nu}^*$.

The original GS and SOR methods proposed in \cite{Trujillo1995} only consider using the partially updated source functions during the upward sweep. \cite{TrujilloBueno2003} introduces the possibility to apply the procedure stated in step 3 also to the downward sweep, which the author called symmetric-SOR (SSOR). 

The Jacobi method requires twice as many iterations as the one-sided GS method \citep{Trujillo1995}. The symmetric-GS method improves convergence by another factor of 2, so it requires only a quarter of the iterations needed by the Jacobi method \citep{Fabiani2003}. The SOR and SSOR methods accelerate convergence even further with the factor depending on the choice of $\omega$. The SSOR method was found to be generally less sensitive to the exact choice of $\omega$ \citep{TrujilloBueno2003}. Theoretically, an implementation of a GS or SOR iteration does not need to perform many more operations than a Jacobi iteration. In practice, this acceleration comes at a cost, however, as the in-situ updates of the source functions requires alternating solutions of the SE and the RTE for each point in the domain. This requires a more complex implementation which prohibits efficient parallelisation and vectorisation \cite{StepanTrujillo2013}. This is a well known drawback of GS like schemes \citep{Bertsekas1988} and for this reason the Jacobi method, i.e. the diagonal operator, remains the method of choice for parallelised NLTE applications such as \Porta, \Nicole, \MultiD and \Balder. 

Otherwise, one can take into account the interdependence of $S^{(n)}_{\nu}$ between nearby grid points. For example, in 1D calculations one can account for changes in the upstream and downstream points, $S_{k-1}$ and $S_{k+1}$, yielding a tridiagonal operator. In multidimensional setups the contributions from the neighbouring points of a cell result in a sparse matrix operator, which is no longer tridiagonal. The inversion of so-called multi-band or non-local operators requires sparse matrix inversions at each grid point instead of a single simple division in the case of a diagonal operator. The multi-band operators can improve the performance of the ALI despite this additional computational effort, since they converge in less iterations. \cite{Hauschildt1994} define the optimum bandwidth as the one, which minimises the total CPU time needed for the solution of a particular radiative transfer problem. They find that the number of iterations required for convergence drops rapidly with increased bandwidth, but the optimum bandwidth strongly depends on the vectorising capabilities of the CPU and the number of frequency points $N_\nu$ and angles $N_\mu$ considered. The formal radiative transfer solver vectorises well, but its cost scales linearly with $N_\nu$ and  $N_\mu$. The inversion of the matrix operator vectorises poorly, but this is done after the radiative rates have been integrated over angle and frequency, so its cost does not increase with an increase in $N_\nu$ or  $N_\mu$. For multidimensional applications \cite{Hauschildt2006} point out that the memory requirements of non-local operators prevent employment of large bandwidths. Considering the interaction of a point with its direct neighbours requires the storage of $3^3$ matrix elements times the number of grid points. Large bandwidth setups, going beyond the nearest neighbour interactions, already require much more memory and the authors only consider them useful for highly complex problems. Interestingly, it was also observed that Ng acceleration works better for narrower bandwidths and deteriorates when increasing the bandwidth. \cite{Hennicker2020} developed a nearest neighbours non-local operator for spherical geometries and confirm its fast convergence behaviour.

The convergence rate of operator splitting methods generally depends on the resolution of the mesh and convergence is obtained much faster when working on relatively coarse grids. \cite{Auer1994b} give an intuitive explanation: \textit{By taking larger optical depth steps we reduce the influence of the off-diagonal terms, make the diagonal approximate operator more accurate, and accordingly speed the convergence.} More precisely, the low-frequency oscillations around the solution are suppressed very effectively on coarser meshes. This leads to rapid initial convergence, but the converged solution might still be far from the true, analytic solution. The analytic solution can be defined as the converged solution on a mesh with infinite resolution. Simulations with a higher resolution therefore get closer to the true result, but take more iterations to converge which is multiplied with the increased computational cost needed to solve each single iteration. Multi-grid methods have been developed in order to achieve quick convergence independent of the mesh resolution \citep{Steiner1991, Auer1994b, FabianiBendicho1997, BjorgenLeenaarts2017}. 

The basic idea is to to solve the NLTE problem using multiple mesh resolutions. One interesting result of changing mesh resolution is that one can estimate the uncertainty in the calculation through the change in convergence speed between the different grid resolutions. \cite{Auer1994b} developed a method which refines the mesh resolution once the relative change between iterations $R_e$ falls below the approximated true error $T_e$. They are defined as
\begin{align}
    R_e({\rm I,L}) &= {\rm Max}\left[\frac{|n({\rm I,L}) - n({\rm I-1, L})|}{n({\rm I, L})}\right] \\
    T_e({\rm I,L}) &= {\rm Max}\left[\frac{|n({\rm I,L}) - n({\rm \infty,\infty})|}{n({\rm \infty,\infty})}\right]
\end{align}
where $n({\rm I,L})$ is the set of level populations at iteration I and refinement level L (higher L meaning finer resolution). $T_e$ basically describes how far the populations are from the analytical solution $n({\rm \infty,\infty})$. There is no point in iterating further than $R_e < T_e$, since there is no more gain in accuracy. The analytical solution is of course unknown, but $T_e$ can still be roughly approximated, which allows the definition of a convergence criterion resulting in a solution with the desired $T_e$. For how $T_e$ is approximated in detail, we refer the reader to \cite{Auer1994a}.

\cite{FabianiBendicho1997} generalise the grid refinement method further and develop a nested multi-grid method. Their method is based on iterating on grids of different refinement levels, where the refinement is always a factor of 2 in the grid spacing. The solutions are interpolated back and forth between the multiple levels ensuring rapid decay of the low-frequency error oscillations as well as the high-frequency error oscillations. In contrast to the method by \cite{Auer1994b}, they do not only go from low to high-resolution, but they also go from high to low-resolution iterations. This is called a V-cycle (see figure 1. of \cite{BjorgenLeenaarts2017}). The complex setup comes with an overhead, however, which only pays off over the classic Jacobi method when interested in particularly fine resolutions. For 1D applications the break-even point is reached approximately at $\Delta z\approx$\SI{10}{\km}. For multidimensional applications the break-even point is reached at much coarser resolution, since the computational cost is dominated by solving the formal solution on the finest grids only \citep{Fabiani2003}. Assuming the resolution is doubled between refinement levels, the finest grid already has 8 times as many points as the next coarser grid.

\cite{StepanTrujillo2013} implemented a multi-grid method for radiative transfer in 3D and tested its efficiency compared to the standard Jacobi method. From their figure 15 it is clear that the multi-grid method is at least an order of magnitude faster, with the exact numbers depending on the desired convergence criterion. Their test bed, however, was not a true 3D hydrodynamical simulation, but based on a plane-parallel 1D atmosphere with smooth, artificial perturbations in the horizontal directions.

\cite{BjorgenLeenaarts2017} showed that when using real MHD snapshots of stellar atmospheres, the multi-grid method does not converge as quickly as previously thought, but still gives a speed up with a factor of 3 to 6 compared to the Jacobi method. Furthermore there was no single setup of V-cycle parameters that provided optimal convergence in all tested atmospheres. One for example has to decide how many iterations are performed at each refinement level. This number is largest for the finest mesh, while only few iterations are needed on the other meshes, but finding the optimal values required manual fine tuning of each simulation. When applying coarse grid corrections to the populations numbers on the finer grids, unfortunately, negative population numbers may arise, which is not possible in the Jacobi method. This demands additional alertness from the side of the user.

In a proof-of-concept paper \cite{Arramy2024} recently presented a NLTE solver, which is not based on the operator splitting method. Alternatively they explore a Jacobian-Free Newton--Krylov (JFNK) method.
Another radically different approach was explored by \citet{PanosMilic2025}, who tackle the 2-level atom NLTE problem using Soft Actor-Critic Reinforcement Learning. This "Approximate-Lambda-Operator-Free" method requires a pre-trained agent to predict the source function (or potentially level populations) to be used in the next lambda iteration. While it is not surprising, that a well-trained agent could potentially achieve rapid convergence, it remains to be shown how such an agent may be trained in order to generalise to unseen environments. Also the computational cost of the training phase should not be neglecting when benchmarking performances of pre-trained operators against classical methods. 

%
%
%
%
\subsection{The diagonal operator}
Here we would like to look at the diagonal approximate lambda iterator in more detail. The diagonal elements are determined by the choice of formal solver (see section \ref{sec:Formal}. Defining $\Lambda^*_{kk}$ means defining the impact of the local source function $S_k$ on the local mean intensity $J_k$ (here we have dropped the frequency indices). The angle-dependent intensity at point $k$ can be expressed in a multitude of ways, but here we list the most commonly used expressions:

\newcommand{\RTE}{I_k &= I_{k-1} e^{\Delta\tau}}
\newcommand{\ui}[1]{_{\rm #1}}
\begin{flalign}
    \RTE + \alpha\ui{L} S_{k-1} + \beta\ui{L} S_{k},                                 & \text{linear} \label{eq:diagLin} \\
    \RTE + \alpha\ui{Q} S_{k-1} + \beta\ui{Q} S_{k} + \gamma\ui{Q} S_{k+1},                & \text{quadratic} \label{eq:diagQuad} \\
    \RTE + \alpha\ui{B} S_{k-1} + \beta\ui{B} S_{k} + \gamma'\ui{B} C\ui{B},             & \text{quadratic Bezier} \label{eq:diagQuadB} \\
    \RTE + \alpha\ui{H} S_{k-1} + \beta\ui{H} S_{k} + \beta'\ui{H}  \frac{dS_{k}}{d\tau},  & \text{quadratic Hermite} \label{eq:diagCube} \\
    \RTE + \alpha\ui{C} S_{k-1} + \beta\ui{C} S_{k} + \alpha'\ui{C} \frac{dS_{k-1}}{d\tau}
                                        + \beta'\ui{C}  \frac{dS_{k}}{d\tau},  & \text{cubic Hermite}
\end{flalign}
where we would also like to remind that for quadratic B\'ezier polynomials $C\ui{B}$ is commonly defined as

\begin{align}
    C\ui{B} = S_k - \frac{\Delta\tau}{2}\frac{dS_{k}}{d\tau} \,.
\end{align}
The $\alpha$ and $\beta$ factors are unique for each formal solver and depend on the exact details of the interpolation of the source function between $k$ and the upstream point $k-1$.  Let us define the local angular contribution $\Lambda^*_{\mu}$ to $\Lambda^*_{kk}$ such that
\begin{equation}
    \Lambda^*_{kk} = \sum^{N_\mu} w_\mu \Lambda^*_{\mu}.
\end{equation}
$\Lambda^*_{\mu}$ is the contribution of $S_k$ to $I_k$. For the linear case (Eq.~\eqref{eq:diagLin}) it is clear that $\Lambda^*_{\mu} = \beta\ui{L}$. For higher order interpolation the choice of $\Lambda^*_{\mu}$ is more ambiguous and depends on the way $dS/d\tau$ is defined.

\cite{Olson1987} suggest to determine $\Lambda^*_{\mu}$ by using the equation for $I_k$ but setting $I_{k-1}$ and $S_{k\pm1}$ to zero and setting $S_k=1$. This eliminates contributions from the $\alpha$ and $\gamma$ terms, but not necessarily from the terms including $\alpha', \beta'$ or $\gamma'$. Monotonicity preserving solvers enforce $dS_k/d\tau=0$ if $S_k$ is a local maximum, as is the case when $S_k=1$ and $S_{k\pm1}=0$. When choosing the Hermite methods, this results in no contributions from the $\beta'$ term, but in the B\'ezier formulation $\Lambda^*_{\mu} = \beta\ui{B} + \gamma'\ui{B}$, since $C\ui{B}=S_k=1$. \cite{Hayek2010} decide to reformulate their quadratic B\'ezier solver to resemble Eq.~\eqref{eq:diagQuad}, i.e. using  $\Lambda^*_{\mu} = \beta\ui{Q}$ in the case when there are no overshoots in the source function interpolant. \cite{SocasNavarro2015} claim that there is no reason to split the gradients $dS/d\tau$ into the individual contributions from $S_k$ and $S_{k\pm1}$. By direct comparison, they find that $\Lambda^*_{\mu} = \beta\ui{B} + \gamma'\ui{B}$ achieves convergence in fewer iterations than $\Lambda^*_{\mu} = \beta\ui{Q}$. The cubic Hermite implementation in \MultiD employs $\Lambda^*_{\mu} = \beta\ui{C}$.
\subsection{Preconditioning the radiative rates}
The discretised radiative rates are given by:
\begin{align}
    R_{ij} &= \sum^{N_\mu}\sum^{N_\nu} w_\mu w_\nu \frac{4\pi}{h\nu}\sigma_{ij}I_\nu, \\
    R_{ji} &= \sum^{N_\mu}\sum^{N_\nu} w_\mu w_\nu \frac{4\pi}{h\nu} G_{ij}\, \sigma_{ij} \left(I_\nu + \frac{2h\nu^3}{c^2}\right),
\end{align}
where we have replaced the integrals by weighted sums using the angle and frequency quadrature weights $w_\mu$ and $w_\nu$. 

These expressions can be modified to reduce the feedback between the local source function and the local mean intensity. This is also known as preconditioning the rate equations. Here the approximate lambda operator comes into play. 

\cite{RybickiHummer1992} derive the \textit{full-preconditioning} strategy for multilevel atoms, which can handle overlaps between multiple transitions. Their recipe was implemented in the \RH code \citep{Uitenbroek2001}, which got its name from this seminal paper. Its implementation was subsequently adopted by \MultiD and its off-spring \Balder. However, \cite{RybickiHummer1992} state that in many situations the full treatment is most likely an overkill and preconditioning the rate equations by only considering interactions of a transition with itself is probably more efficient. This scheme, we call \textit{auto-preconditioning}, appears to gain popularity in radiative transfer codes \cite{Hubeny2017, Amarsi2020a, Osborne2021}. In the auto-preconditioning scheme the preconditioned radiative rates are given by

\begin{align}
    R_{ij}' &= \sum^{N_\mu}\sum^{N_\nu} w_\mu w_\nu \frac{4\pi}{h\nu}\frac{2h\nu^3}{c^2}\sigma_{ij} \left(\mathcal{I}_\nu - \frac{\Lambda_{\mu\nu}^*}{\chi_{\rm tot}} G_{ij} \, \sigma_{ij} \, n_j\right) \\
    R_{ji}' &= \sum^{N_\mu}\sum^{N_\nu} w_\mu w_\nu \frac{4\pi}{h\nu}\frac{2h\nu^3}{c^2} G_{ij} \, \sigma_{ij} \left(\mathcal{I}_\nu - \frac{\Lambda_{\mu\nu}^*}{\chi_{\rm tot}} \sigma_{ij} \, n_i + 1\right)
\end{align}
where we have introduced the dimensionless intensity $\mathcal{I}_\nu$, also known as the photon occupation number
\begin{equation}
    \mathcal{I}_\nu \equiv \left[\frac{2h\nu^3}{c^2}\right]^{-1} I_\nu
\end{equation}
Using $\mathcal{I}_\nu$ has the advantage that the factors $4\pi/h\nu$ and $2h\nu^3/c^2$ can be included in $w_\nu$ which is usually computed and stored during initialisation and does not change between iterations.  In contrast to the full-preconditioning, the auto-preconditioning scheme allows for computing the rates on a line-by-line basis, rather than on a frequency-by-frequency basis, since no information has to be shared between transitions. The $\Lambda_{\mu\nu}^*/\chi_{tot}$ factor is also referred to as the $\Psi_{\mu\nu}^*$ operator, where the Psi-operator acts on the emissivity $\eta$ rather than the source function, i.e., $J=\Psi[\eta]$. 
\subsection{Complete linearisation}

Another method to solve the NLTE problem is known as complete linearization. In this method, the statistical equilibrium equations and the radiative transfer equation are solved simultaneously by expanding them in a Taylor series, dropping 2nd and higher order terms, and solving the resulting linear system. However, the system is coupled over all depths (through radiative transfer) and all frequencies (statistical equilibrium). The system is huge and only small atomic models can be handled with this approach. Complete linearization still can be used for numerical testing, because the method is very efficient and stable, and it can deal with very complicated multilevel cases on line formation problems. A combination of ALI and CL is used, for example, in the \texttt{Cloudy} NLTE RT code \citep{Ferland2017}.
\subsection{Scattering and two-level (coronal) approximation}
\label{sec:twolevel}

Following \citet{Ferland2017}, we distinguish between the full NLTE RT solution (the collisional-radiative model) and simplified NLTE approaches, which, for example, neglect either the internal level structure of the system (atoms or molecules) or neglect everything, except the ground and first excited level of the ion. The former approach is commonly used in studies of the ISM \citep{Spitzer1978, Osterbrock2006}. The latter approach is used, e.g in the calculations of the solar corona \citep[e.g.][]{Gudiksen2011} and modelling winds of massive stars \citep[e.g.][]{Hennicker2020}.

In the two-level approximation \citep[][their Chapter 1 for derivation and references]{Crivellari2019}, the radiation intensities are computed using the line or continuum source function of the following form:
\begin{equation}\label{eq:scat}
    S_{\rm l,c} = \epsilon B(T) + (1-\epsilon) J_{\Phi}, 
\end{equation}
where $J_{\Phi}$ is the integral of mean intensity over the frequencies sampled by the line and $\epsilon$ is the so-called \textit{photon destruction probability (by inelastic collisions)} and (1-$\epsilon$) is photon creation probability (here by spontaneous de-excitation in the transition between two states). The parameter is unity in LTE, but can range from $0.1$ to express the ``moderate scattering'' to $10^{-6}$ to express ``strong scattering'' \citep{Gudiksen2011}. 

Physically, the assumption is that the NLTE effects are driven by  radiation and collisions in the transition between two energy states only, and all other effects (that may normally change the populations) play a negligible role. This assumption, for example, quantitatively holds in very strong resonance lines of species that have a rather special atomic structure (such as Li). Yet, generally this approach holds primarily the academic value allowing a convenient method to explore qualitatively the influence of scattering on the structure of the model (e.g., for 3D applications). In stellar atmospheres, the value of $\epsilon$ is typically about $10^{-7}$. For example, the form of Eq.~\eqref{eq:scat} for the continuum was used in \citet{Hayek2011} to explore the influence of coherent isotropic scattering due to free e$^-$ scattering and Rayleigh scattering on H in the atmospheres of metal-poor red giants. A similar formalism, assuming either the coherent continuum scattering or the two-level approach for a spectral line, is used in studies of 3D RT in expanding atmospheres of hot stars with winds \citep{Hennicker2020}.
 
Another form of a two-level approximation, as used e.g in Cloudy  \citep{Ferland2017} is the representation of the number densities of ions through the ionization-recombination balance. In this approximation, it is assumed that the populations of ions are determined solely by photo-ionisations and/or recombinations followed by (instantaneous) suction to the ground state, hence the following expressions are used:
 \begin{equation}
     \dfrac{n_{i+1}}{n_{i}} = \dfrac{\int F_{\nu} \times \sigma_{\nu} d \nu}{\alpha(i+1) n_{e}},
 \end{equation}
 
and 
 \begin{equation}
     \dfrac{n_{i+1}}{n_{i}} = \dfrac{q(i)}{\alpha(i+1)} 
 \end{equation}
where alpha is the total recombination rate coefficient for an ion (i$+$1) and F$_{\nu}$ the radiative flux, and sigma the photoionization cross-sections. The former equation is used for systems with photo-ionization equilibrium, and the latter for systems with recombination equilibrium (e.g., for the solar corona).

\subsection{Timing of RT calculations}\label{sec:timing}
The computational time $t$ for LTE and NLTE RT in pre-computed atmospheres (stars, planets, etc.) depends on many variables. But the three most important quantities are: the complexity of the grid (dimensionality, geometry, size, number of points, step between them), the way the radiation field propagates (discretisation in wavelength/frequency, number of rays), and the underlying physics used to compute the matter parameters (opacities, source function). The scaling relations are defined as follows:

We begin with the definition of variables to be used in subsequent formulae:
\begin{itemize}
    \item $N_{\text{xyz}}$: Total spatial grid points (for 1D, $N_{\text{xyz}} = N_z$) in the input atmosphere (typically down-sampled for 3D models).
    \item $N_{\mu}$: Number of angles considered in the radiative transfer.
    \item $N_{\lambda}$: Number of wavelength points. 
    \item $N_{\text{trans}}$: the average number of active transitions, that is, opacity contributors (spectral lines and continua) per wavelength point.
    \item $N_{\text{level}}$: Number of energy levels (atoms, molecules) in the NLTE atomic models.  
    \item $N_{\text{iter}}$: Number of ALI iterations.
    \item $N_{\text{time}}$: Number of 3D RHD cubes (each representing a point in stellar time, also referred to as a "snapshot")
    \item $c_{\text{RT}}, c_{\text{SE}}, c_{\text{MI}}$: Implementation and machine-specific constants for the formal (RT) solver and the NLTE statistical equilibrium (SE) solver as well as the matrix inversion (MI) needed to update the NLTE populations. 
\end{itemize}

The following equation holds for 1D LTE:
\begin{equation}
t_{\text{1D, LTE}} \approx c_{\text{RT}} N_{\text{z}} N_{\mu} N_{\lambda} N_{\text{trans}}
\end{equation}

The following equation holds for 3D LTE:
\begin{equation}
t_{\text{3D, LTE}} \approx c_{\text{RT}} N_{\text{time}} N_{\text{xyz}} N_{\mu} N_{\lambda}  N_{\text{trans}}
\end{equation}

The following equation holds for 3D NLTE. The starred variables indicate that the chosen quadratures may differ during the NLTE iterations and the final flux synthesis:
\begin{align}
t_{\text{3D, NLTE}} \approx N_{\text{time}} N_{\text{xyz}} [ N_{\text{iter}}\{c_{\text{SE}} N_{\mu}^* N_{\lambda}^* N_{\text{trans}} + c_{\text{MI}} N_{\text{level}}^3\} + c_{\text{RT}} N_{\mu} N_{\lambda} N_{\text{trans}}]
\end{align}

One has to keep in mind the following additional dependencies, most of them leading to declining performance in 3D compared to 1D:

\begin{itemize}
    \item \textbf{Opacity Bottleneck:} The value $N_{\text{trans}} \cdot N_{\lambda}$ is the primary driver, in 1D as well as 3D, rather than $N_{\lambda}$ alone. $N_{\text{trans}}$ can be vastly greater than unity, because continua typically overlap with each other and line lists contain many overlapping lines, especially in molecular bands (see sect. \ref{sec:FreqQuad}). This issue is usually tackled by pre-computing opacity tables, resulting in a single interpolation per wavelength.
    \item \textbf{Extra angles in 3D:} $N_{\mu}$ is typically larger by a factor $\textgreater\,4$ in 3D geometries compared to 1D. 
    \item \textbf{NLTE Iterations:} We note that $N_{\text{iter}}$ is affected by the grid spacing, as denser grids typically require more iterations. This is due to ALI approaching the convergence speed of regular lambda iterations as grids become denser. 3D grids typically have finer grid spacing than their 1D counterparts, leading to slower convergence and higher total costs (see sect. \ref{sec:ALI}).
    \item \textbf{RT Solver:} A linear formal solver yields the lowest value for $c_{\text{RT}}$, while quadratic and cubic solvers are more expensive. Solvers using monotonicity enforcing interpolation typically have higher computational cost, since the interpolation weights depend on the interpolated data and cannot be precomputed (see sect. \ref{sec:Formal}). Additionally, extra interpolations are needed in 3D to map the relevant quantities to the photon path before the formal solution (see sect. \ref{sec:interp2D}).
    \item \textbf{SE Solver:} While $N_{\text{level}}^3$ is theoretically significant, it only competes with the RT term in cases of extreme atomic complexity (e.g., $N_{\text{level}} > 10^3$).
    \item \textbf{Memory proximity:} 3D codes rarely scale perfectly according to the scaling relations above compared to their 1D counterparts. As $N_{\text{xyz}}$ gets larger the reduced memory proximity leads to declining CPU efficiency compared to smaller arrays in 1D codes.
\end{itemize}
Let us estimate roughly the computational cost of 3D compared to 1D. Hydrostatic model atmospheres from the public \texttt{MARCS} grid have $N_z=56$, whereas the public \texttt{STAGGER2.0} grid has already down-sampled 3D models with $N_{\text{xyz}}=80\times80\times240 \approx 1.5\cdot10^6$. Typically one uses 5-10 snapshots so we may approximate $N_{\text{time}}\cdot N_{\text{xyz}}\approx10^7$. Assuming $N_{\mu,\text{3D}}/N_{\mu,\text{1D}}\approx4$, while the atomic parameters $N_{\text{level}}$ and the resulting $N_\lambda$ and $N_{\text{trans}}$ cancel as they are the same in 1D and 3D. Therefore the 3D LTE vs. 1D LTE cost is $\approx10^6$. In NLTE we have to account for $N_{\text{iter}}$ which typically lies between 10-100, resulting in an approximately $10^7-10^8$ times higher computational cost of 3D NLTE compared to 1D LTE. We note that this is assuming one models all transitions of a given element in both cases. In practice one cannot model a narrow spectral region or even a single spectral line in NLTE as is routinely done in LTE, unless the NLTE populations have been precomputed.
\section{3D and/or NLTE radiative transfer codes}\label{sec:codes}

In this section, we provide a brief description of 3D LTE, 1D NLTE, and 3D NLTE codes that are used for spectral modelling and diagnostics of exoplanets, hot stars, and cool stars. For more details and specific applications, such as kilonovae, which share much of the numerical approach with SNe modelling, we refer the reader to the next section.  Table~\ref{tab:codes} presents a compact overview of some of the commonly-used codes in different communities.

\subsection{Exoplanets}
Different physical codes are available for exoplanet atmosphere and synthetic spectral modelling. \texttt{petitRADTRANS} is among the most widely used codes for 1D LTE modelling of exoplanet spectra, both in transmission and in emission \citep{Molliere2019}. For NLTE RT, the most commonly-used code is Cloudy (\url{https://trac.nublado.org/}). 

Although originally intended for modelling and diagnostics of emission spectra from externally irradiated molecular clouds \citep{Ferland1983, Ferland1998} and referred to as \texttt{HAZY}, this iterative NLTE RT code has over decades become a general-purpose software and it is presently broadly applicable to modelling diverse astrophysical environments with temperatures in the range from $\sim 3$ K to 10$^{10}$ K \citep{Ferland2013, Ferland2017}. The code is more restrictive in terms of density owing to prescriptions used to compute the heating and cooling rates, hence the applicability is limited to densities $\rho \lesssim 10^{15}$ cm$^{-3}$. For more details, we refer the reader to \cite{Chatzikos2023}, and details on microphysics and numerical relevant to exoplanet atmospheres are further provided in \citet{Turner2016}. 

In the latest public version \citep{Gunasekera2023}, the code includes available atomic data from diverse databases (NIST, LAMDA, STOUT, CHIANTI) for chemical elements with atomic numbers up to zinc (Zn), as well as selected molecular data permitting chemical equilibrium calculations for some of the key species for planet atmospheres like carbon monoxide CO, water H$_{2}$O, molecular hydrogen H$_{2}$, methane CH$_4$, and carbon dioxide \citep{Madhusudhan2012}. For the H and He-like iso-electronic sequence atoms (e.g. He I, Li II, Be III, etc) and diatomic molecules, full coupled equations of SE and RT are solved. For other species, however, simpler approaches, e.g within the framework of the two-level (also known as 'coronal' approximation, see Sect. \ref{sec:twolevel}) are used. Comparative examples of H number densities computed the two limiting cases (LTE and the coronal approximation), compared to the full comprehensive NLTE solution, can be found in \citet{Ferland2013}.
\subsection{Hot stars}

In contrast to cool stars, LTE approximation in the analysis of spectra of hot OBA-type stars was abandoned very early \citep{Auer1969,Kudritzki1990} and powerful NLTE RT methods were employed not only for spectrum synthesis, but also for atmospheric structure calculations (see Sect.~\ref{sec:atmospheres}). The specific development progressively implied that most groups have directly relied on NLTE solvers for various diagnostic problems, circumventing the phase of individual analysis and determination of NLTE abundance corrections. Here we provide only a brief summary of recent work in modelling NLTE atmospheres of hot stars.

Among several powerful numerical codes, there are the \texttt{Kiel} code \citep{Dreizler1996} and \Tlusty \citep{Hubeny1992, Hubeny1995, Hubeny2017}, which are commonly-used iterative codes to compute NLTE model atmospheres in 1D and hydrostatic equilibrium with detailed handling of line blanketing due to metals and multi-element NLTE calculations. \Tlusty operates in a hybrid framework both using the complete linearisation \citep{Auer1969} and ALI methods and it includes detailed atomic and molecular linelists permitting physically comprehensive treatment of physical conditions from a few tens of K to $\sim 10^8$ \citep{Hubeny2017}. The lower temperature limit reflects the availability of molecular data, whereas at $T \gtrsim 10^8$ the need to include more complex atomic processes, such as relativistic Compton scattering and inner-shell electronic structure restricts the applicability of the code. Similar restrictions apply for densities in excess of 10$^{20}$ cm$^{-3}$, requiring a more accurate handling of line broadening and non-ideal effects \citep{Hubeny2017}. 

Another widely-used code for modelling OB-type expanding stellar atmospheres is \texttt{Fastwind} based on the ALI scheme. The code has similar physical capabilities to \Tlusty, however, it also permits comprehensive treatment of stellar winds \citep{Santolaya-Rey1997, Puls2005, Herrero2002}. The code is suitable for calculations of hot stars with \teff\ greater than roughly 8500 K and for all \logg\ and mass loss values. 

Also the code \texttt{CMFGEN} \citep{Hillier2012} is used to determine the atmospheric structure and the atomic level populations primarily for massive stars with winds. These include Wolf-Rayet (WR), OBA-type stars, luminous blue variables and central stars of planetary nebulae. It also has a 1D time dependent radiative transfer mode for SNe \citep{HillierDessart2012}. The usual operating mode assumes spherical geometry and solves the transfer code in the comoving-frame but it also has a 1D plane-parallel and 2D spherical modes that assume static atmospheres. It uses a linearisation technique with either purely local (i.e., a diagonal operator) or adjacent grid points (tridiagonal operator) \citep{Hillier1990}.

The 1D codes \texttt{PoWR} \citep{SanderShenar2015} and \texttt{METUJE} \citep{Krticka2018} also solve the comoving-frame radiative transfer in spherical geometry for O and WR type stars but use different techniques to facilitate the simultaneous solution of the rate equations. They are similar to \texttt{CMFGEN}, but unlike the former they have implemented techniques to solve for the velocity law.

More recently, a multi-dimensional framework based on \texttt{MPI-AMRVAC} \citep{XiaTeunissen2018} utilises RHD models to reveal the highly turbulent and variable surfaces of O-type and WR stars \citep{MoensPoniatowski2022, DebnathSundqvist2024, MoensDebnath2025}. While radiation is treated as fluid in the 2D and 3D RHD calculations \citep{MoensSundqvist2022}, these models have been post-processed for spectrum synthesis using the radiative transfer code by \citet{Hennicker2018, Hennicker2020}. Departures from LTE are accounted for in the RHD step as well as in the post-processing by an approximate NLTE formalism (aNLTE) following \citet{LucyAbbott1993, PulsSpringmann2000}. This approach is based on utilising potentially different radiation and gas temperatures for the ionisation and level excitation fractions. Furthermore, these multi-dimensional results are used to calibrate the treatment of turbulent pressure and hydrodynamics in the \texttt{PoWR} code \citep{Gonzalez-ToraSander2025b, Gonzalez-ToraSander2025a}.
\subsection{Cool stars}

The 1D NLTE and 3D NLTE codes used in RT calculations for cool stars were extensively described in Sect. 4. Here we provide only a small selection of studies that were seminal either by introducing a new mathematical or numerical framework, and/or introducing a new code, which found applications in many subsequent studies. We caution that this list is not complete.

\begin{itemize}
\item \citet{RybickiHummer1992}: This paper is the first to derive the preconditioned statistical equilibrium equations for overlapping transitions. Including overlapping transitions requires the use of the Psi operator which operates on the emissivity rather than on the source function. A recipe for full-preconditioning is given as well as a recipe for self-preconditioning of single transitions is given. Preconditioning  accelerates convergence, because it removes the self-interaction of transitions. 

\item \citet{Trujillo1995, FabianiBendichoTrujilloBueno1999} introduce the MUGA code and propose Gauss--Seidel iterations to accelerate convergence in lambda iterations.

\item \citet{MansoSainzTrujillo-Bueno1999} and \citet{Dittmann1999} address the problem of scattering line polarization and the Hanle effect in 3D.

\item \citet{Auer2003}: This paper describes the translation from observer's frame to comoving frame quantities and also describes the monotonic cubic Hermite interpolation scheme, which was later implemented in the IRIS code and after that in the MULTI3D code.

\item \cite{Carlsson2009} and \cite{Leenaarts2009b}: The MULTI3D codes applies a linearization of the statistical equilibrium equations following \citep{Scharmer1985} and the approximate operator of \cite{RybickiHummer1991}. The code was adapted to a full preconditioning scheme from \cite{RybickiHummer1992} similar to RH. 

\item \citet{Ibgui2013}: This paper showcases the 3D RT code IRIS, which was purely LTE at the time and not MPI parallelised. The main contribution of this paper was a very efficient piecewise cubic, locally monotonic, interpolation technique, that considerably reduces the numerical diffusion effects of the short- characteristics method. This technique was also adapted in the MULTI3D code as the ``hermite'' Sinterpol option. It is mentioned that IRIS is to be fused with the NLTE code TLUSTY in order to provide a 3D NLTE RT code. 

\item \citet{HauschildtBaron2014} The 3D version of Phoenix is using a long-characteristics scheme and has been parallelised using the MPI library making it the most scalable code at the time. The conclusion in this paper is that there are many mature 3D NLTE RT codes, but too few realistic 3D RHD simulations and high resolution observations to test them on. 

\item \citet{Janett2021}: A novel fourth-order WENO interpolation technique. A possible new tool designed for radiative transfer; discussion of advantages r.t. Hermite and other more sophisticated approaches.

\end{itemize}
\section{Main results and applications}\label{sec:results}

%
\begin{table}[htbp]
\caption{3D and/or NLTE studies of elements in different astrophysical regimes. For stellar conditions, only papers, where NLTE model atoms or comprehensive multi-element NLTE modelling is presented are listed below. We do not distinguish here between 1D NLTE, 1.5D NLTE, or 3D NLTE, but when all kinds of results are available, we give a preference to the latter (3D NLTE) obtained with comprehensive micro-physics and atomic/molecular models. The listed applications and references are merely examples, and there are many more detailed studies on NLTE analyses.}
\label{tab:nlte}
\begin{tabular}{l p{6cm} p{4cm}}
\toprule
 Element, ion  & Applications &  NLTE (ref)  \\
\midrule     
 \textit{Stars}  &  &  \\

   Li I     &  cosmology, Spite plateau      &  \citet{Wang2021} \\
   C (C  I) &  Sun, star-planet connection   &   
   \citet{Amarsi2019a}  \\
   C (CH)   &  Sun, red giants        & \citet{Popa2023}  \\
   N I      & [C/N] ratio, CNO cycle and the RGB & \citet{Amarsi2020a} \\
   O I      &  standard solar model, interior &  \RaggedRight{\citet{Amarsi2018, Bergemann2021}}     \\
   Na I     &  globular clusters, multiple populations  &   \\
   Mg I,II  &  $\alpha$-enhancement proxy for stars   & \citet{Bergemann2017}  \\
   Al  I    &  globular clusters, anomalies  & \citet{Nordlander2017}   \\
   Fe I,II  &  metallicity proxy for stars &    \citet{Amarsi2018}  \\
   Mn I, II &  Fe-peak, SN Ia progenitors  & \citet{Bergemann2019}  \\
   Ba  II   &  s-process, 2nd-peak    & \citet{Gallagher2020}  \\
   Y  I, II &  s-process, 1st-peak    & \citet{Storm2024}  \\
   Eu II    &  r-process              & \citet{Storm2024}  \\
  &  &  \\
\textit{Exoplanets}  &  &  \\
   H$\alpha$  &  dynamics &    \citet{Fossati2023} \\
   He I  & 1083 nm, atmospheric escape & \citet{Oklopcic2018, Young2020} \\
   Na I  &  Na D (589 nm), winds, clouds &  \citet{Barman2002, Fischer2019, Canocchi2024} \\
   K I   &  atmospheres &  \citet{Canocchi2024} \\
   Ca II &  850 nm, heating, dynamics  &   \citet{Deibert2021} \\
   Mg I,II &  285,280 nm, opacity  &  \citet{Young2020} \\
   Mg I, Fe I, II &  broad-band, atmospheric heating  & \citet{Fossati2021, Young2024}\\
 &  &  \\
\textit{Kilonovae}  &  &  \\
   He I     &  1.08 $\mu$m, spectral modelling, LC evolution &   \citet{Tarumi2023} \\
   Sr II    &  1.037 $\mu$m, r-process nucleosynthesis &   \citet{Perego2022} \\
   Ce II, Pt III    &  opacity &   \citet{Pognan2022a} \\%
\bottomrule
\end{tabular}  
\end{table}
%

\subsection{Physical domains and 3D NLTE codes}

\begin{figure*}[ht]
\hbox{
	\includegraphics[width=0.34\textwidth]{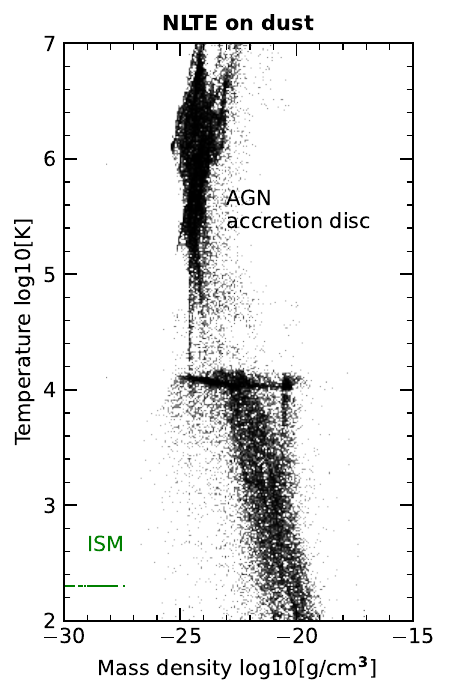}
	\includegraphics[width=0.32\textwidth]{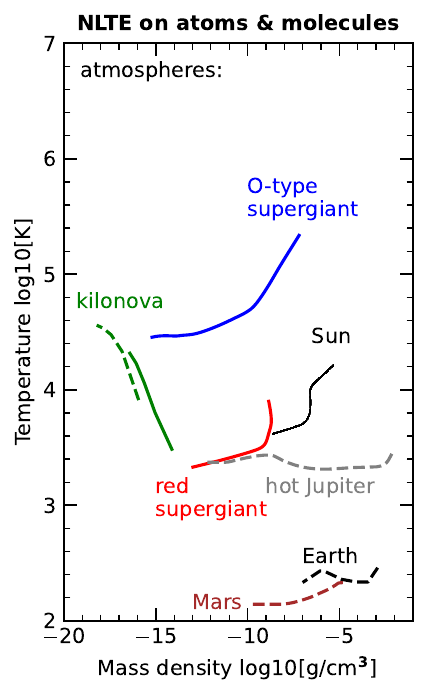}
	\includegraphics[width=0.37\textwidth]{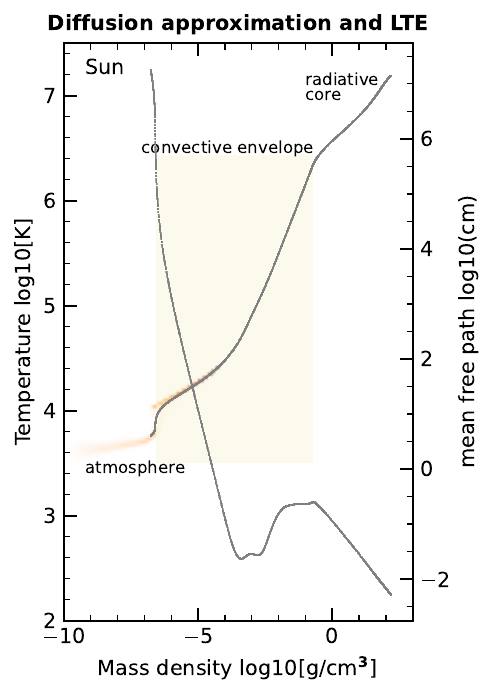}
    }
    \caption{Gas density versus gas temperature in different astrophysical domains. See text.}
    \label{fig:maindiagram}
\end{figure*}

We begin with summarising the physical conditions, that correspond to atmospheres of different astronomical objects, in which 3D geometry and/or NLTE radiation effects become relevant in Fig.~\ref{fig:maindiagram}. With this figure, we are responding (indirectly) to the critical comment by \citet{Ivanov1979} made in response to the textbook on Stellar Atmospheres by \citet{Mihalas1978}. Precisely, the comment by \citet{Ivanov1979} is:
 \begin{quote}Even the most diligent student will, after working through Mihalas's book, remain fully in the dark about the chemical composition of stellar atmospheres and the accuracy with which it has been determined, as well as the diverse range of problems which astrophysics today links with differences in the composition of stellar atmospheres. Moreover, he (NB: she) will not even gain an idea of what order of magnitude the density is in the atmospheres of various types of stars: [\dots] Mihalas [\dots] has not thought it necessary to provide his readers with numerical data on how temperature and density are distributed with depth in the atmospheres of various stars.
 \end{quote}
Therefore in this paper, we provide such an overview of physical conditions, but we include also other systems, which are better aligned with the landscape of modern astronomy and astrophysics research.

This figure shows the gas mass density versus gas temperature in different domains, including the atmospheres of hot O-type (\texttt{Tlusty} OSTAR2002 model with \teff $= 55\,000$ K, $\log g =4.0$, and solar $Z$ from \citealt{Lanz2003}) and cool G- and M-type (the Sun, red supergiant Betelgeuse) stars, the expanding atmosphere of a high-energy lanthanide-rich kilonova for two different epochs (5 and 20 days after the merger of two neutron stars, from \citealt{Pognan2023}, Y$_{\rm e} =0.15$ and  $M_{\rm ej} = 0.05\, M_{\odot}$), the atmospheres of an exoplanet (hot Jupiter, (c) Paul Molliere, \teff $= 2200$ K\footnote{Here, the $\teff$ of the planet refers to the following definition $T_{\rm eff} = (T_{\rm int}^4 + T_{\rm equilibrium}^4)^{1/4} $, where $T_{\rm equilibrium}$ is the temperature of a body (here the exoplanet), which is in a radiative equilibrium given the external radiation (stellar radiation) and its internal heat (planet emission).}, $\log g =3.00$, and solar metallicity $Z$), of the Earth (the standard ISA 1976 model) and Mars, and --- for comparison ---  the parameter space probed by an AGN torus from \citet{Matsumoto2023} based on \citep{Wada2016}, where NLTE effects are common and have to be included in line transfer. For guidance, we also show the T-$\rho$ parameter space covered by the interior structure of the Sun and and the solar 3D RHD atmosphere from \citet{Eitner2024}. The solar interior model is adopted from \citet{Herrera2023} based on the calculations of \citet{Vinyoles2017} and adopting the solar composition from \citet{Magg2022}. The distinction between different NLTE domains is based on reported published evidence of the sensitivity of different  species (ions, atoms, molecules, dust) to NLTE effects. 

The physical basis of 3D and NLTE effects on radiation transfer is described in detail in subsequent sessions, which discuss the recent literature addressing the topic.
\subsection{Overview of NLTE effects}
In the astronomical literature, it has become common to refer to the qualitative influence of NLTE on line profiles as NLTE effects, whereas the quantitative influence that can be quantified via the effect of the departures from LTE on the line EW or the chemical abundance of an element derived from a given line is referred to as ``NLTE abundance correction". In what follows, we first briefly address the physical mechanisms behind NLTE effects in stellar atmosphere. An overview of NLTE effects in spectral lines can be found in \citet{Asplund2005}, \citet{Bergemann2014}, and \citet{Lind2024} for cool stars. We note that such classification has not been adopted in other areas of research, e.g., in exoplanet research or kilonovae. Spectra of the latter objects are are vastly more complex, as all absorption and emission features overlap and hence individual de-compositions into atomic and molecular diagnostic features caused by transitions in individual chemical species are difficult.

Generally, since the NLTE conditions are driven by the radiation field, the effects are generally larger in physical conditions that lead to stronger, that is, more intense, radiation fields. These are the conditions in hotter stars (F-type), more metal-poor stars, or stars with lower density rarefied atmospheres (red giants or red supergiants). These conditions are illustrated in Fig.~\ref{fig:maindiagram}. The NLTE effects can be divided into two classes, radiation-driven and collision-driven \citep{Gehren2001, Bergemann2014}. In case of radiation-driven species, the NLTE effects are primarily caused by the over-ionisation due to super-thermal radiation, $J_\nu > B_{\nu}$ (exceeding the Planck function) in the UV \citep{Rutten2003, Bergemann2012a}. These conditions are, physically, the consequence of radiation propagation in an RE model atmospheres with no incoming radiation from outside, that is the weighting of the source function by the E1 kernel, which represents the exponential behaviour of the opacity with depth. In the Rayleigh--Jeans regime, where the depth gradient of the source function is steep (for a given radiative equilibrium T gradient), equation \ref{eq:src_NLTE} leads to J$_\nu > S_\nu$, and thus super-thermal radiation field. In the Wien regime (IR), the gradient of the source function is much shallower (flatter) and J$_\nu < S_\nu$, so the radiation field is sub-thermal. The latter situation favours recombination over over-ionisation. 

The type of NLTE effect also depends on the structure of the atom or molecule and on its abundance in the atmosphere. Atoms like, e.g. Mg I, Ca I, Ti I, Mn I, Fe I, Cr I, and Sr I  have large ionization cross-sections in the near-UV and blue wavelength regime (a few electron-Volt), and therefore in physical conditions of cool stars are typically over-ionised \citep[e.g.][]{Korn2003, Gehren2004, Bergemann2010a, Mashonkina2011, Bergemann2011, Bergemann2012a, Bergemann2012c, Lind2012, Amarsi2018, Mallinson2022}. The atomic level populations of these  neutral ions are thus all systematically smaller compared to LTE populations. A similar process, but acting in bound-bound transitions, is referred to as a super-thermal over-excitation, and this process may also contribute to the depletion of the number densities of low-excitation states in the system. 

In contrast, the energy levels system of species like Na I and K I is much less complex \citep{Baumueller1997, Zhao1998, Baumueller1998, Mashonkina2000b, Gehren2004, Zhao2006, Reggiani2019, Lind2022}. The NLTE effects in these species are driven by over-recombination coupled with photon losses in the wings of strong spectral lines. Photon loss describes the following microscopic physical process. As energy is redistributed in the spectral line, a fraction of energy is lost through the line wings, which become optically thin in deeper atmospheric layers. As a consequence of this redistribution, some photons from the line core leak and escape through the wings, which leads to brighter (weaker) wings and fainter (stronger) cores in NLTE conditions \citep[see][for example]{Mashonkina2000b, Gehren2004}. Similar NLTE effects are found for singly-ionised species of trans-Fe-group elements, such as Nd II, Ba II, Sr II, and Eu II \citep[e.g.][]{Mashonkina2000a,Mashonkina2005,Bergemann2012c,Gallagher2020,Storm2024}. Finally, certain ions, such as Li I and O I, are mixed-type ions that may behave similar to either collision- or to radiation-driven species, depending on physical conditions or the diagnostic lines under consideration \citep[e.g.][]{Carlsson1994, Lind2009, Amarsi2019b}.

In 3D NLTE radiative transfer, the effects are qualitatively similar. Yet, so far, only selected species have been subject to a thorough analysis. In 3D NLTE modelling, overall, the tendency of certain species to be over-ionized is amplified, so that positive NLTE effects on line profiles become even larger. This is due to the prevalence of up-flows over down-flows in 3D convective models. The up-flows are characterised by steeper temperature gradients and thereby enhance the inequality between $J_\nu$ and the local value of $B_{\nu}$. For a more detailed discussion of the differences between 3D NLTE and other approaches, we refer the reader to Sect.~\ref{sec:coolabu}.

\subsection{OBA and FGK model structures}\label{sec:atmospheres}
Standard calculations of NLTE radiation transfer in the context of stellar atmospheres are currently routinely performed for OB-type stars. Physical conditions of these hot systems with $\teff \gtrsim 10\,000$ K are such that strong departures from NLTE arise owing to extreme radiation field in the UV, and radiation pressure on lines is also the main driver of stellar winds  \citep[e.g.][]{Mihalas1970, Auer1972, Kudritzki1978, Kudritzki1990, Kudritzki2000}. Whereas both self-consistent NLTE calculations for the model structure and spectrum synthesis have been explored, more recent work progressively relies on a hybrid approach, where LTE and 1D are adopted for structure of the atmospheres, where multi-element NLTE RT is performed for the key diagnostic elements, such as H I, He I and He I, O II, C II and C III, Si III \citep[][]{Nieva2007, Nieva2012, Nieva2014, Aschenbrenner2023}.

Multiple grids of NLTE model atmospheres and NLTE synthetic spectra for OBA-type stars exist. They are publicly available to the community, including but not limited to the following databases:

\begin{itemize}
 
 \item OB-type and early A-type stars: \texttt{Fastwind} / WM grid  \citep{Puls2005, Bestenlehner2024}. Following the definition in \citet{Puls2005}, the code is to be written \textit{not} in capital letters. This grid covers the domain of \teff in the range from 17.8 to 56.2 kK and \logg\ from 2 to 4.5, as well as variable He abundance and metallicity;

 \item early B-type stars: the \texttt{Tlusty} grid \citep{Lanz2007}. This grid covers \teff in the range from 15 to 30 kK, \logg\ from 1.75 to 4.75, and $v_{\rm mic}$ of 2 \kms;

 \item O-type \texttt{METUJE} \citep{Krticka2010,Krticka2018};
 
 \item hot B- and O-type sub-dwarfs (\teff from 30 to 50 kK, \logg\ from 4 to 8) \citep{Kudritzki1976, Przybilla2006}, see also \citet{Heber2009} for an overview of the relevance of NLTE on the surface structure of these stars;

 \item \texttt{Phoenix} Nextgen grid \citep{Hauschildt1999a, Hauschildt1999b} for \teff from 2\,000 to 10\,000 and \logg\ from 0.0 to 5.5 (separate grids for RGB and MS stars, the latter computed using spherically-symmetric radiation transfer);

 \item A-type - there are no A-type NLTE or 3D grids. Perhaps because there has not been enough motivation for this work, or due to evidence that NLTE effects in A-type stars are not critical and can be ignored \citep{Przybilla2011}.

\end{itemize}
A comprehensive comparison of OBA type model atmospheres, synthetic energy distributions, and individual spectral lines in LTE and NLTE can be found in \citet{Przybilla2011}. This study also found that the structures of LTE and NLTE model atmospheres are very similar for stars with $15\,000 \lesssim T_{\rm eff} \lesssim 35\,000$, including both dwarfs and giants. This justifies the use of the hybrid approach (LTE atmospheres, NLTE line profiles) in spectral diagnostics of B to late-O type stars. A rather novel approach in massive star atmosphere modeling is to use insights from full 3D simulations to improve the structure of 1D HE atmosphere models. This was done, for example, in \citet{Gonzalez-Tora2025}.

NLTE calculations of stellar model atmospheres for cool late-type (FGKM-type) stars are considerably more complex, owing to a gigantic number (over 500 millions, \citealt{Hauschildt1999b}) of absorption lines due to line transitions in neutral atoms and molecules across the entire spectrum from the UV to the IR.  So far, these calculations have been carried out using the 1D hydrostatic version of the \texttt{Phoenix} model atmosphere code, yet with treating the line blanketing with comprehensive atomic and molecular databases, and including the effect of NLTE on the atmospheric structure through changes in the radiative heating and cooling \citep{Hauschildt1997}. For the Sun, the effects of NLTE the model atmosphere structure are marginal and do not exceed about 20 K in the outer layers relevant for line formation \citep{Short2005}. The study by \citet{Haberreiter2008} took a step further and explored the impact of NLTE on the SEDs and fluxes of the Sun. They showed that using NLTE opacity distribution functions in the calculations of the solar model atmosphere has a noticeable effect at only far-UV wavelengths shorter than 260 nm (their Fig. 7), although, they did not compare the fluxes computed self-consistently with LTE and NLTE opacity distribution functions. Their calculations also included NLTE effects in H$^{-}$ continuum opacity. Other studies of NLTE effects on the H$^{-}$ include \citet{Shapiro2010} and in \citet{Barklem2024}. These analyses suggest that the effects for FGKM-type stars are small, at a sub-percent level, although may become more important in the outer layers of the atmospheres of warm and low gravity stars, $\log g \lesssim 2$ and \teff $\gtrsim 6000$ K \citep[][their Fig. 7]{Barklem2024}. \citet{Young2014} investigated NLTE model atmospheres for red giants. Also, their work shows that NLTE effects influence the spectral energy distribution in stars with \teff~~greater than 4000 K only below 400 nm.
\subsection{Cool stars: fundamental parameters}\label{sec:coolparam}

The analysis of stellar parameters has so far been primarily limited to 1D models with NLTE included for selected chemical elements, such as Fe \citep[e.g.][]{Bergemann2012a,Ruchti2013,Kovalev2019}, Mg \citep{Kovalev2019}, and H \citep[e.g.]{Mashonkina2008, Amarsi2018}. Recent studies also addressed the influence of 3D NLTE in the formation of the optical H lines, specifically, $H_{\alpha}$ and $H_{\beta}$ finding non-negligible effects in the line wings and in the cores. This may influence the determination of $\teff$ from the Balmer line wings, e.g. \citet{Giribaldi2021} showed that the effective temperatures derived using 3D NLTE line profiles of H lines tend to be larger, by up to $300$ K at low metallicity, [Fe/H] $\approx -2$, compared to 1D LTE values. Similar effects are found in 1D NLTE calculations \citep{Kovalev2019}. 

\begin{figure}[ht]
    \centering
    \includegraphics[width=1\linewidth]{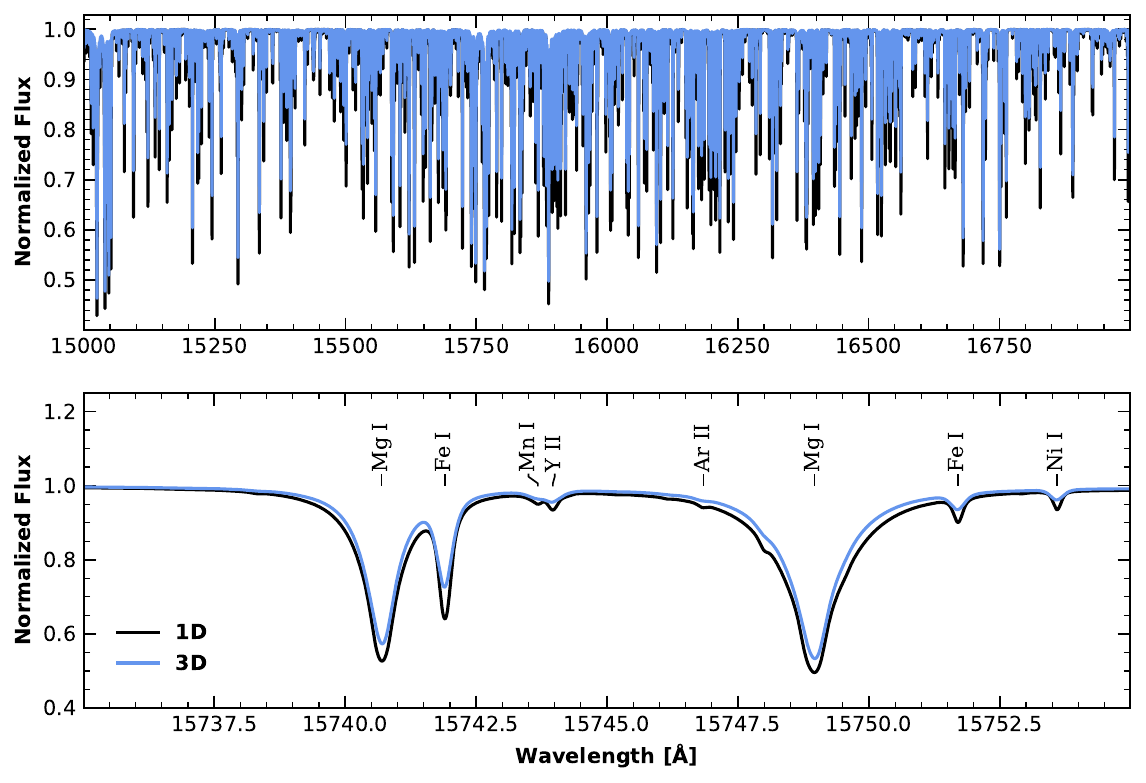}
    \caption{APOGEE spectral range modelled using a 1D solar \texttt{MARCS} model and a 3D \texttt{STAGGER} snapshot.}
    \label{fig:APOGEE}
\end{figure}

\begin{figure}[ht]
    \centering
    \includegraphics[width=1\linewidth]{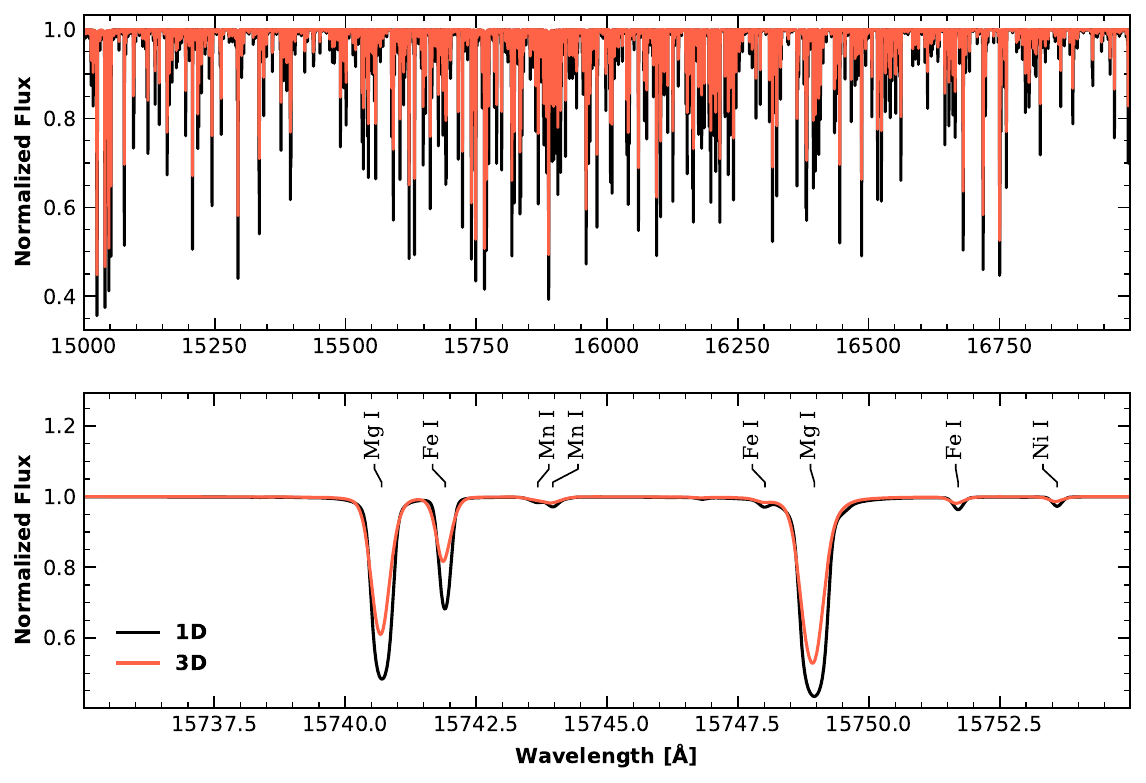}
    \caption{APOGEE spectral range modelled using a 1D red giant \texttt{MARCS} model (\teff=4500, log(g)=2.0, [Fe/H]=$-1.0$) and a 3D \texttt{STAGGER} snapshot.}
    \label{fig:APOGEE_RG}
\end{figure}

Comparative analyses of the bulk influence of NLTE on synthetic stellar spectra were presented, e.g. in \citet{Bialek2020}, who reported a substantial improvement of the estimates of $\teff$, $\logg$, and metallicity values obtained in NLTE compared to model-independent estimates obtained using the methods of interferometry and/or asteroseismology. Overall, however, substantial differences between 1D and 3D synthetic spectra are expected for different types of stars, and for different wavelength regimes. In Figs.~\ref{fig:APOGEE} and \ref{fig:APOGEE_RG} we compare synthetic spectra computed using 1D hydrostatic and 3D hydrodynamic model atmospheres in the APOGEE H-band window (1.5 to 1.7 $\mu$m). The difference in the spectral lines are systematic and ubiquitous across the entire 200 nm wavelength range, necessitating 3D RT calculations with RHD models from first principles.

Estimates of 1D and 3D NLTE abundance corrections for the diagnostic lines of Fe I and Fe II are available in different studies and a comprehensive review of those can be found in \cite{Lind2024}. Here we only briefly summarise the main effect and demonstrate how these influence stellar metallicities derived from stars of different Galactic populations. For main-sequence stars, the 1D NLTE  corrections for Fe I and Fe II lines typically do not exceed $+0.02$ to $0.15$ dex, at least within the [Fe/H] range from $-3$ to solar \citep{Lind2012, Ruchti2013, Kovalev2019, Larsen2022}. In comparison, 3D NLTE abundance corrections for Fe I lines are of the order $-0.2$ dex to $0.15$ dex for unevolved (main-sequence) stars \citep[][their Fig. 4, here inverted sign as they adopted the convention of 1D LTE-3D NLTE]{Amarsi2022} at solar metallicity. However, for metal-poor ([Fe/H] $= -3$) main-sequence and TO stars, the 3D NLTE corrections may exceed $0.5$ dex \citep{Amarsi2022}. The lines of Fe II, although not particularly affected by NLTE \citep{Lind2012}, are rather sensitive to the structure of the atmospheric models, especially the horizontal inhomegeneities of temperature and density. Recent detailed estimates, suggest effects of the order $0.05$ to $0.15$ for weaker Fe II lines (with reduced $\log_{10} \rm{EW}/{\lambda}$ of $\lesssim -5$), but for stronger Fe II features the 3D NLTE may be both positive and negative. The biases depend, however, on the atomic properties of the line, primarily on the excitation potential and transition probability, and on stellar parameters. For red giants, the effects of 3D NLTE on the Fe I lines are more substantial, at the level of $\sim 0.5$ dex, at least at very low metallicities [Fe/H] $\lesssim -4$ \citep{Lagae2023}. For such stars, however, also 1D NLTE calculations yield comprehensive values of abundances: at [Fe/H] of $\sim -2.5$, for example, the NLTE abundance corrections for RGB stars reach up to $0.3$ dex \citep{Kovalev2019}. 
\subsection{Cool stars: abundances}\label{sec:coolabu}
\begin{figure*}[htbp]
\centering
\includegraphics[width=0.86\columnwidth]{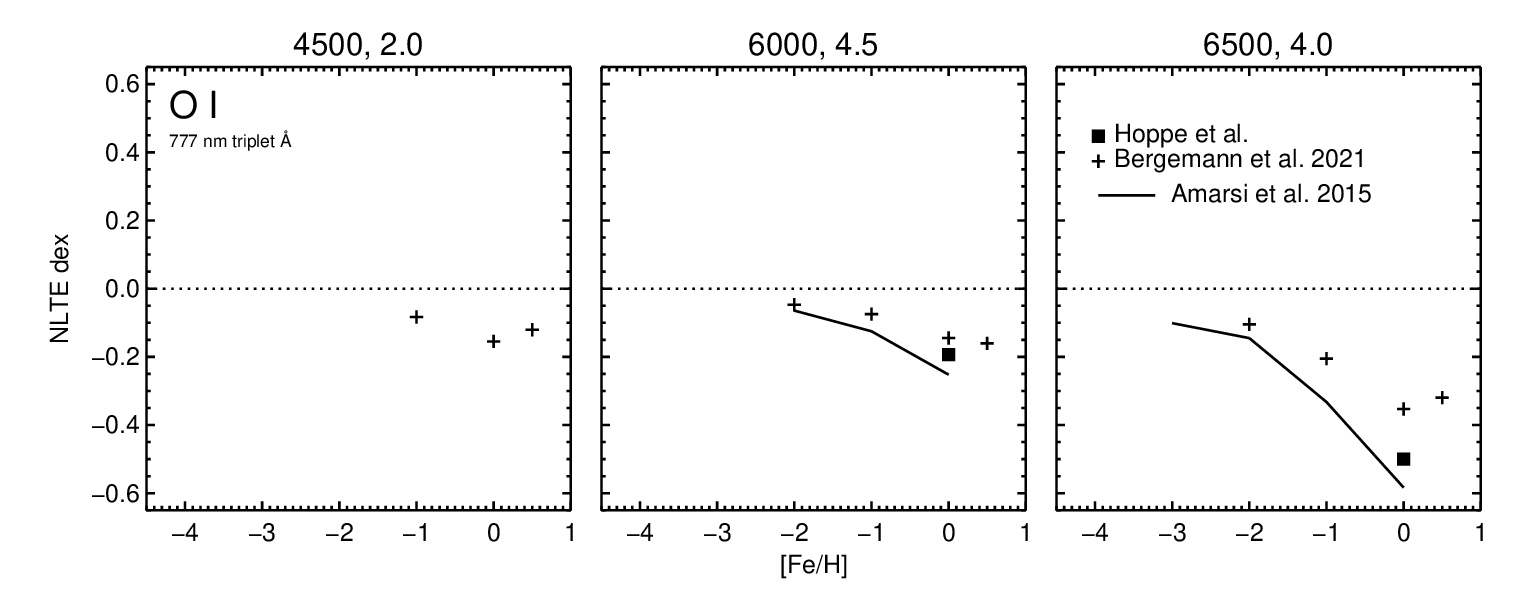}
\includegraphics[width=0.86\columnwidth]{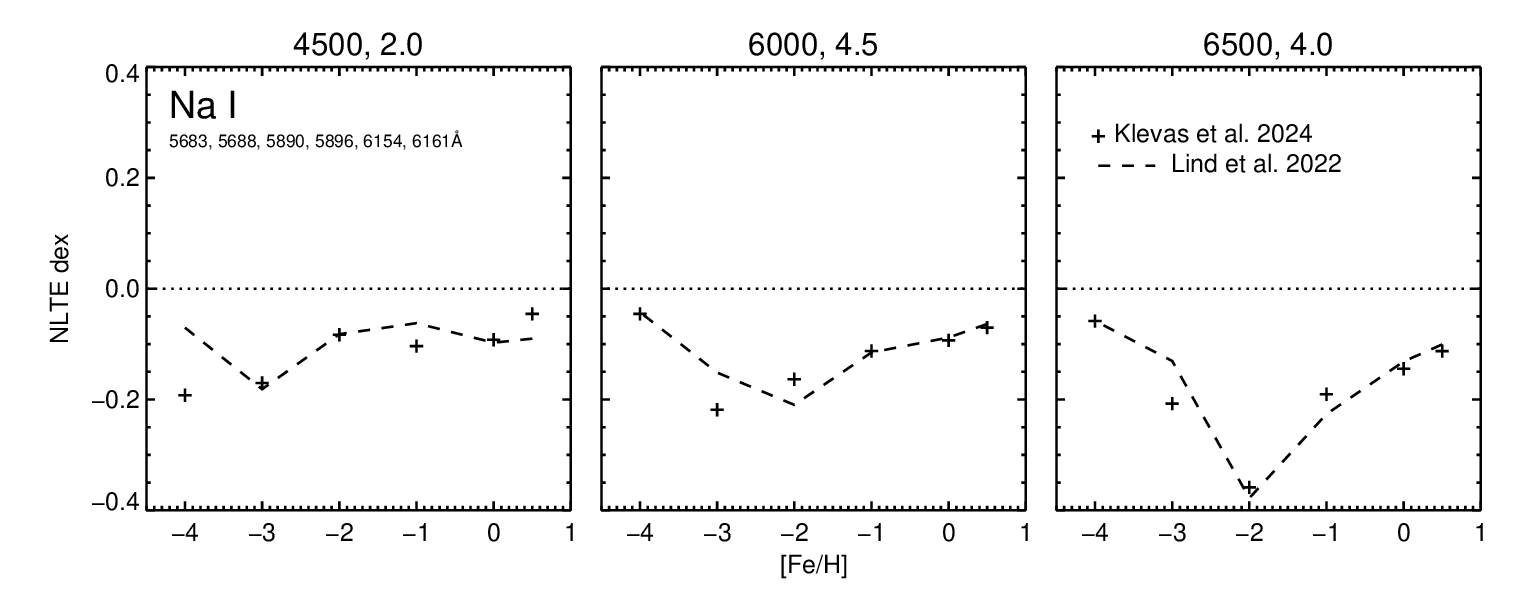}
\includegraphics[width=0.86\columnwidth]{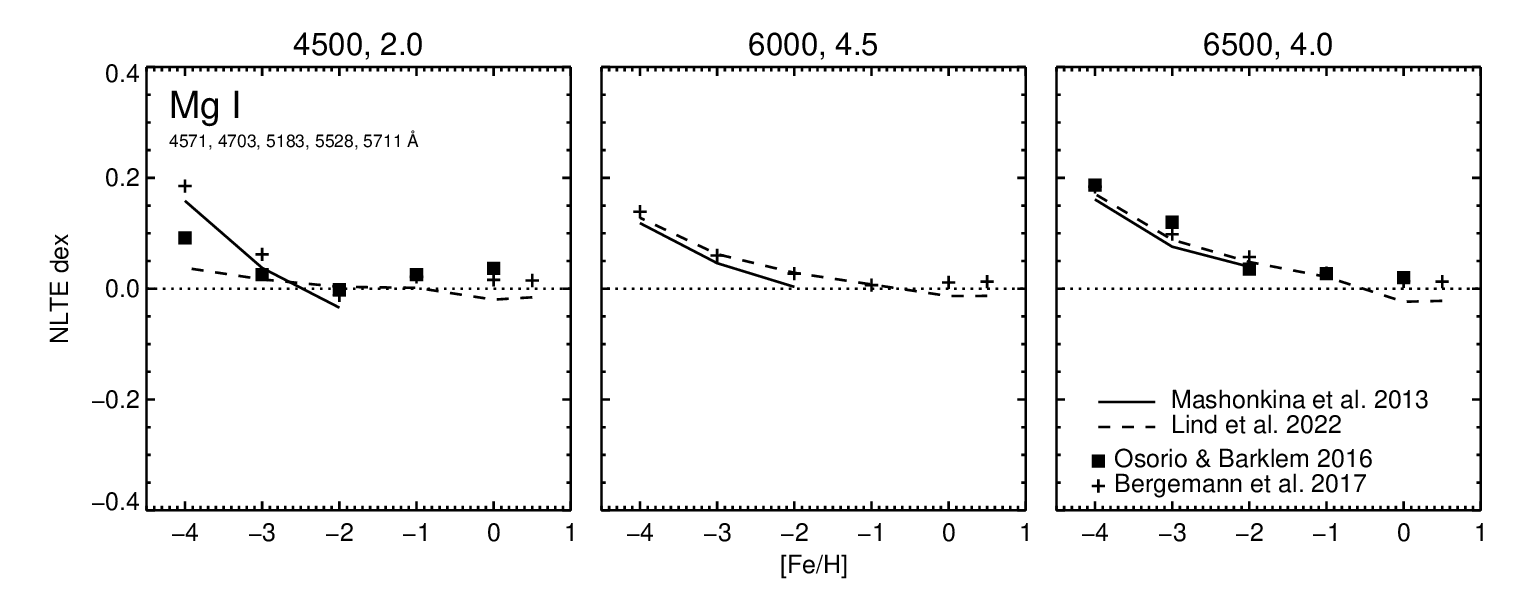}
\includegraphics[width=0.86\columnwidth]{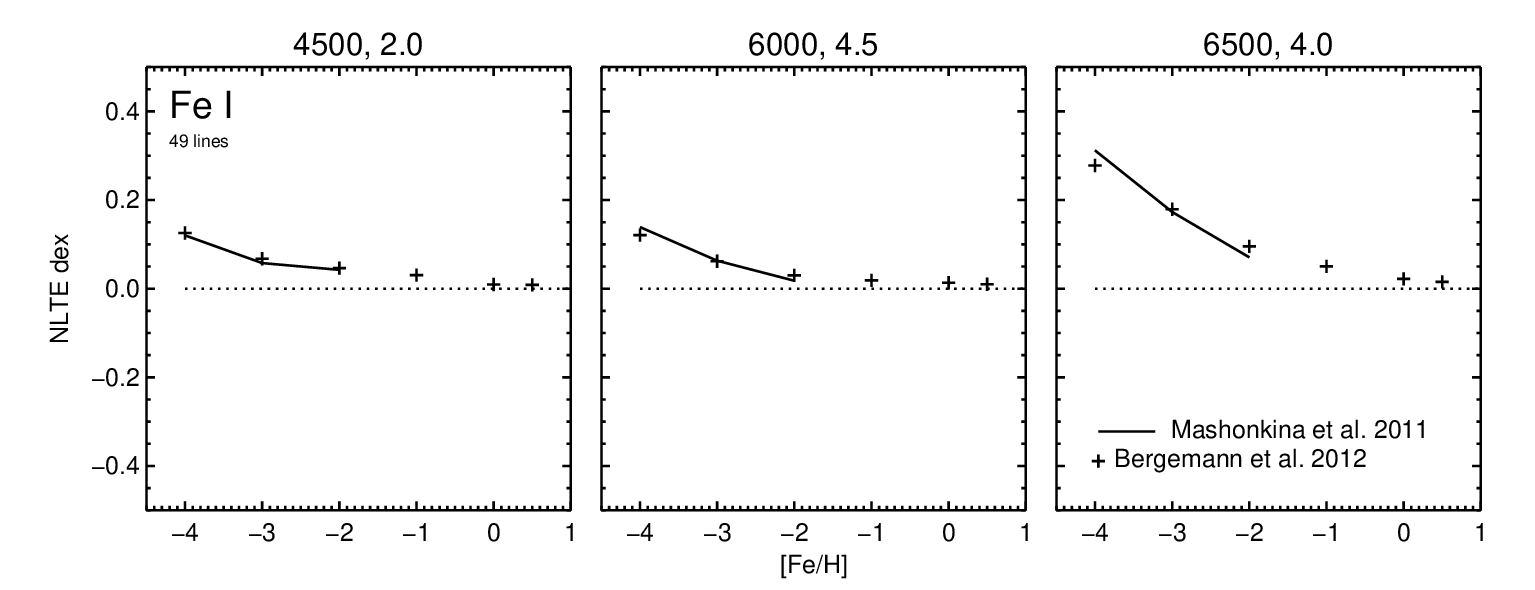}
\caption{NLTE effects on abundances of different chemical elements, including O, Mg, Fe, Mn, Sr, Y, and Eu, computed with 1D and 3D hydro-dynamical model atmospheres. See text. The figure highlights that overall the agreement between different independent calculations, based on different model atmospheres, atomic models, and NLTE RT codes (1D and 3D), is excellent. Hence the NLTE effects can be trusted.}
\label{fig:nltecorr1}
\end{figure*}

\begin{figure*}[ht]
\centering
\includegraphics[width=\columnwidth]{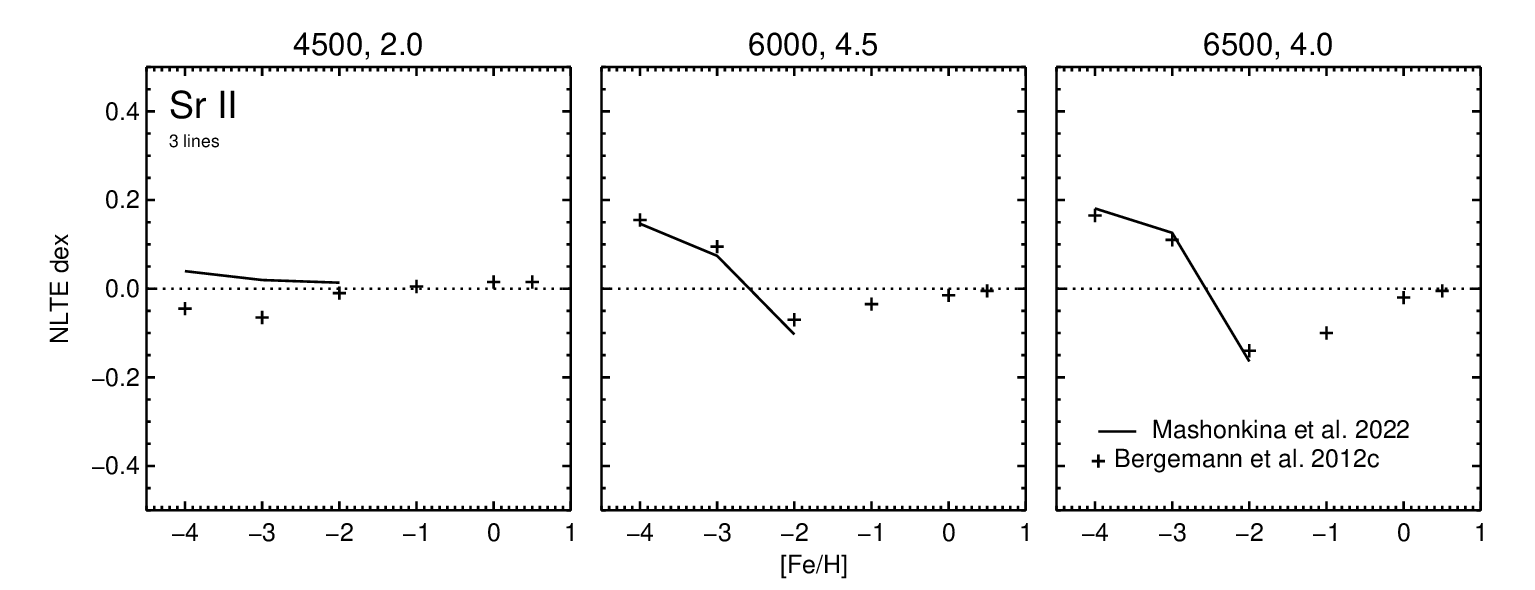}
\includegraphics[width=\columnwidth]{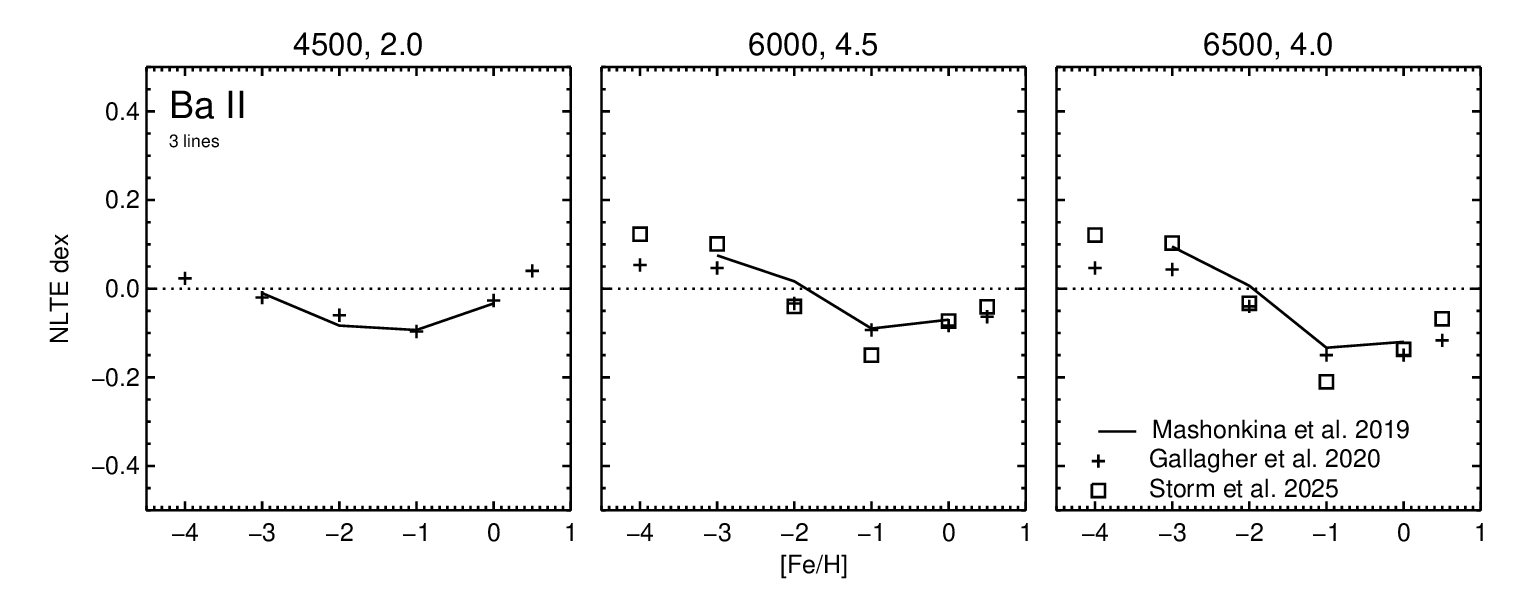}
\caption{NLTE effects on abundances of different chemical elements, including Sr, Ba, and Eu, computed with 1D and 3D hydro-dynamical model atmospheres. The label insets are as follows. Crosses: Sr II results based on the model atom from \citep{Bergemann2012c} updated to QM collisional data in \citet{Gerber2023}, computed using MARCS and MULTI2.3. Solid line: NLTE Sr II results from \citet{Mashonkina2022} computed using MAFAGS-OS models and DETAIL. For Ba, the NLTE results computed using the model atom from \citet{Gallagher2020} (MULTI and Turbospectrum) and \citet{Mashonkina2019} (DETAIL and SIU) are presented.}
\label{fig:nltecorr2}
\end{figure*}

Very substantial and tedious work has been carried out already regarding 1D NLTE calculations for vast majority of elements up to and including the post-Fe-group elements \citep[see also][for an overview]{Mashonkina2014}. This includes the light species (H, He, Li), CNO, $\alpha$-elements, and Fe-group. We collate these studies in Table~\ref{tab:nlte}. Some NLTE abundance corrections computed with 1D and 3D NLTE codes are further presented in Figs.~\ref{fig:nltecorr1} and \ref{fig:nltecorr2}. 

For trans-Fe group elements, selected estimates of NLTE effects are available for the 1-st peak s-process elements Rb, Sr, Y, Zr \citep{Korotin2020, Mashonkina2007b, Bergemannn2012b, Storm2023, Storm2024, Velichko2010}, 2nd-peak elements Ba and Pr \citep{Mashonkina2000a, Mashonkina2009}, and selected r-process species Eu, Gd, and Nd \citep{Mashonkina2000c, Mashonkina2005, Storm2024, Storm2025}. Also estimates of NLTE effects for the heaviest elements of the actinide series (Pb, Th) are available \citep{Mashonkina2012}. 

\begin{figure*}[ht]
    \centering
    \includegraphics[width=1\textwidth]{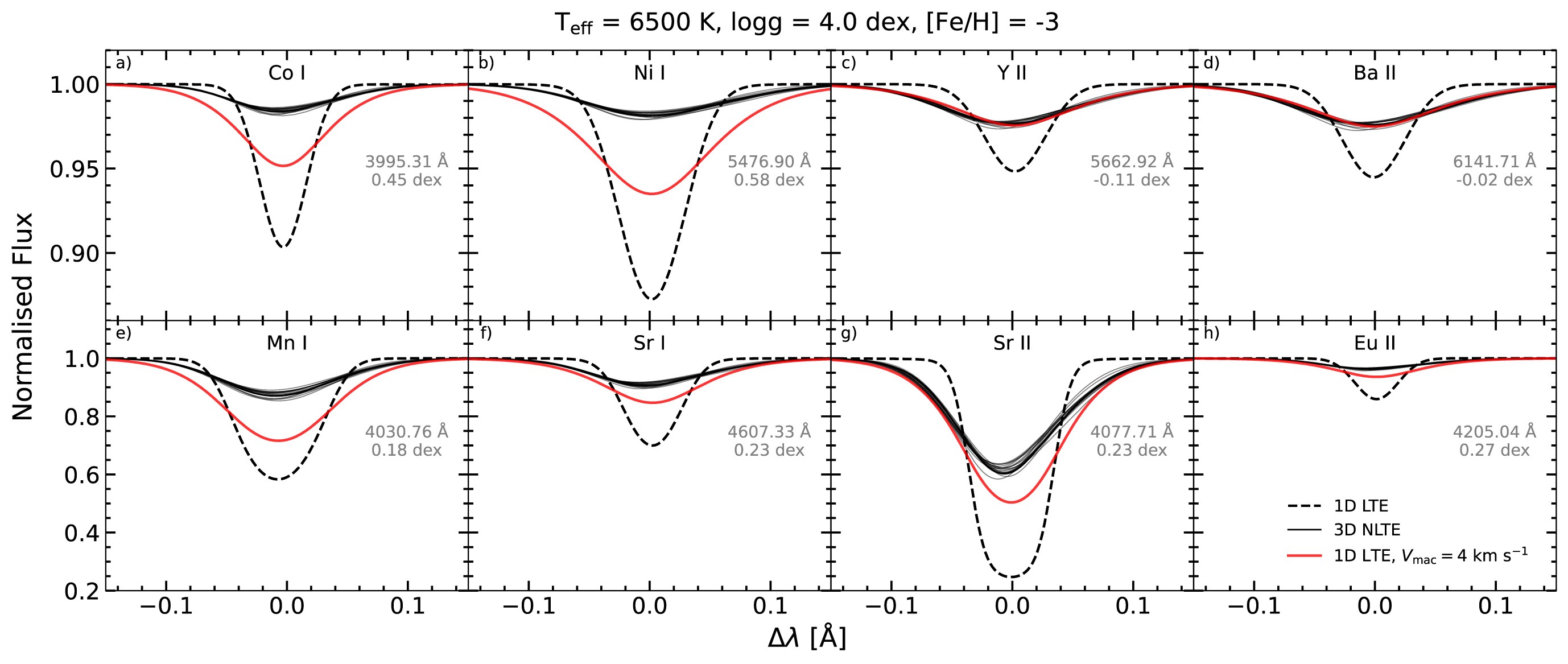}
    \caption{Figure illustrates the theoretical line profiles computed using 1D LTE and 3D NLTE models using the  parameters of a very metal-poor turn-off star with metallicity [Fe/H]$=-3$. The same abundances were used. Note that major differences between 3D NLTE and 1D in line profiles for iron-group elements, Co I, Ni I, and Mn I, with 1D LTE over-estimating the line strength and EW by over a factor of two. As a result, the abundances of Co and Ni ([Co/Fe] and [Ni/Fe]) may be erroneous by over 0.5 dex (a factor of 3, panels a and b).  Image reproduced with permission from \cite{Storm2025}, copyright by ESO.}
    \label{fig:st2025a}
\end{figure*}

\begin{figure}[ht]
    \centering
    \includegraphics[width=0.7\textwidth]{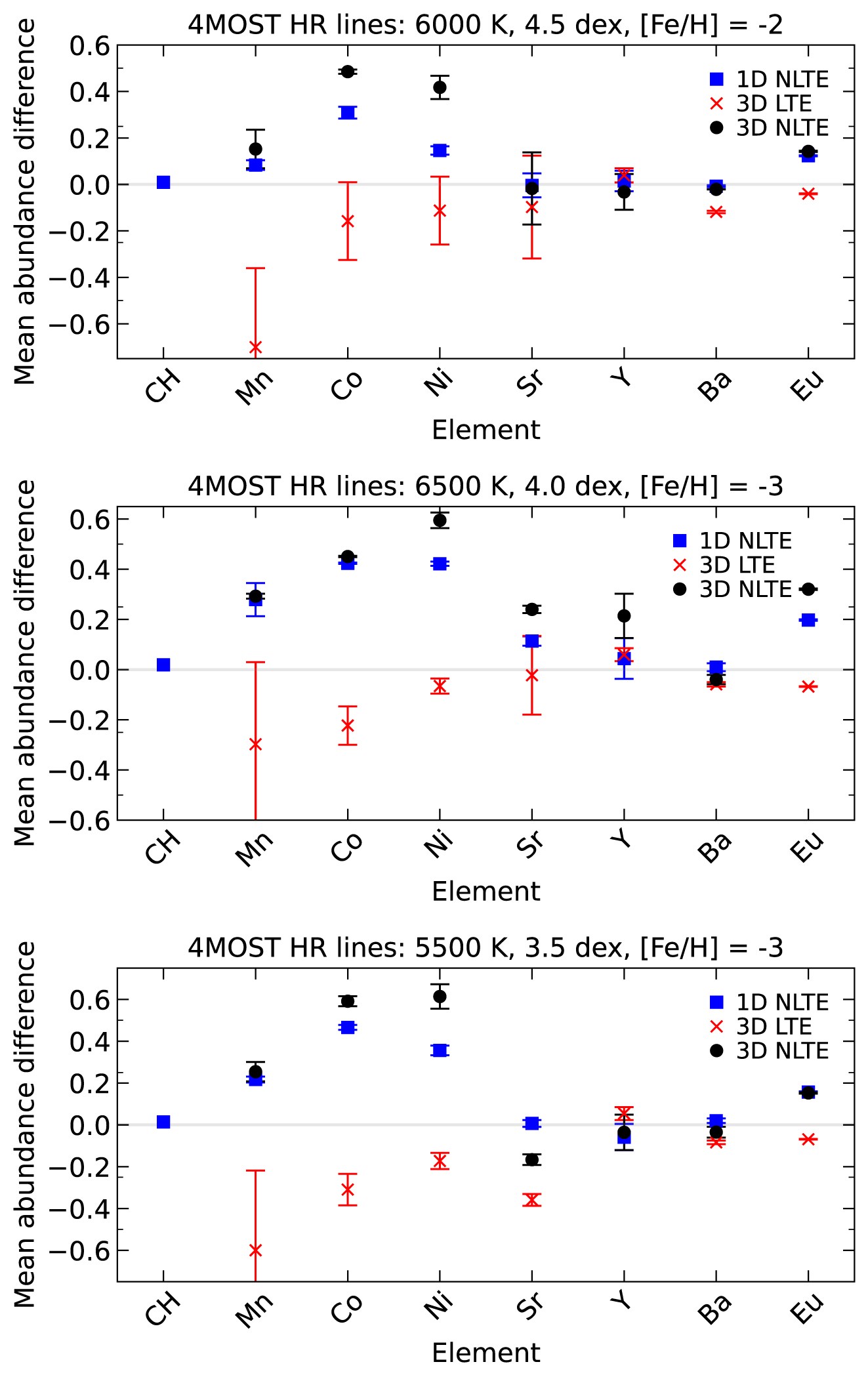}
    \caption{3D NLTE abundance corrections for different chemical elements computed using very metal-poor 3D model atmospheres, as indicated in the figure titles. Image reproduced with permission from \cite{Storm2025}, copyright by ESO.}
    \label{fig:st2025b}
\end{figure}

Regarding full 3D NLTE, much of the early work focussed on chemical elements, for which rather simple model atoms (a few tens of energy states) can be used to derive astrophysically relevant results. These, specifically, addressed line formation of Li I \citep{Barklem2003, Cayrel2007}, O I \citep{Asplund2004}, and H I \citep{Pereira2013}. More recently more complex 3D NLTE calculations for astrophysically relevant atomic lines of Fe-group and trans-Fe neutral elements became possible. As a consequence, estimates of 3D NLTE effects on abundances are now available for the following atoms: H \citep{Amarsi2018}, Li \citep{Wang2021}, C \citep{Amarsi2019a}, N \citep{Amarsi2020a}, O \citep[e.g.][]{Amarsi2016, Bergemann2021}, Na \citep{Asplund2021}, Mg \citep{Bergemann2017, Asplund2021}, Al \citep{Nordlander2017}, Si \citep{Amarsi2017,Asplund2021}, K \citep{Reggiani2019}, Ca \citep{Asplund2021, Lagae2023}, Fe \citep{Amarsi2016, Lind2017}, Mn \citep{Bergemann2019}, Ba \citep{Gallagher2020}, Y \citep{Storm2024}, and Eu \citep{Storm2024}. These studies are typically performed for a small sample of stars with high-quality astrophysical parameters inferred by independent techniques. All these calculations, with a few exceptions, were carried out under the trace element assumption. Also grids of 3D NLTE abundance corrections for selected elements were computed by some groups \citep[e.g.][]{Amarsi2020b, Storm2025}. In Fig. \ref{fig:st2025a} we show the resulting line profiles for selected chemical elements computed using 1D LTE, 1D NLTE, and 3D NLTE. The corresponding differences in abundance diagnostics are gigantic, and may exceed a factor of 3. The resulting difference in chemical abundances inferred using 1D and 3D NLTE models are furthermore demonstrated in Fig. \ref{fig:st2025b}.

\citet{Osorio2020} explored simultaneous multi-element NLTE RT using the \texttt{Tlusty} code. They find that the influence of cross-talk of radiation field is very modest, typically not exceeding $0.01$ dex in the abundance space, that is the effects are within (and even significantly smaller) than the physical quality of sub-grid microphysics, for example, cross-sections describing inelastic collisions with hydrogen \citep{Bergemann2019}. However, there are physical conditions, such as metal-poor red giants, where coupling between the radiation field in lines of \textit{different} chemical species becomes important. These are, for example the impact of NLTE on populations involved in formation of the key O I lines through the NLTE radiation in Ly $\beta$ \citep{Dupree2016}.
\subsection{Databases}

Public NLTE databases can be used to calculate 1D and 3D NLTE abundance corrections for the key diagnostic elements:
\begin{itemize}
     \item \url{https://www.inspect-stars.com/} hosted by University of Stockholm. NLTE corrections are available for Li I, O I, Na I, Mg I, Ti I, Ti II, Fe I, Fe II;
     \item \url{https://nlte.mpia.de/} hosted by the Max Planck Institute for Astronomy, Heidelberg. This database offers NLTE corrections for H I, O I, Mg I, Si I, Ca I, Ca II, Ti I, Ti II, Cr I, Mn I, Fe I, Fe II, and Co I;
     \item \url{https://spectrum.inasan.ru/nLTE2/} hosted by the Institute for Astronomy of the RAS. This database offers NLTE corrections for Ba II, Ca I, Ca II, Fe I, Mg I, Na I, Sr II, Ti II, Zn I and Zn II.
    \item NLTE corrections computed withing the ChETEC project \url{https://www.chetec-infra.eu/3dnlte/abundance-corrections/strontium/} and \url{https://www.chetec-infra.eu/3dnlte/abundance-corrections/barium/}
\end{itemize}
\subsection{M dwarfs and pulsating stars}
For M dwarfs and other types of stars, such as Cepheids and RR Lyr, only very limited knowledge of impact of 3D and NLTE effects on structure and diagnostics is available. 

\begin{figure}[ht]
    \centering
    \includegraphics[width=1\textwidth]{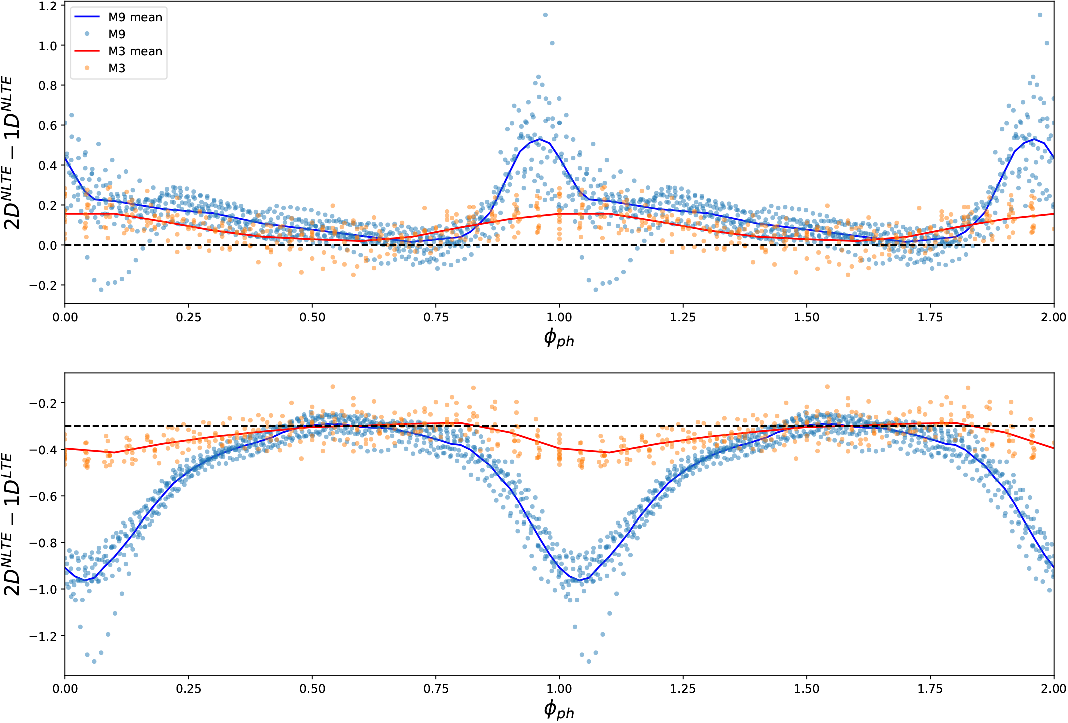}
    \caption{Figure illustrates the impact of 2D NLTE effects on the analysis of oxygen (O I) abundances from multi-phase spectra of Cepheids. Here two solar-metallicity pulsation-convective RHD models with the pulsation periods of 3 days (Teff $= 5600$ K, log g $= 2.0$) and 9 days  (Teff $= 5600$ K, log g $= 1.5$) are shown, respectively. Image reproduced with permission from \cite{Vasilyev2019}, copyright by ESO.}
    \label{fig:cepheid}
\end{figure}

Selected studies of statistical equilibrium were carried out for M dwarfs by \citet{Hauschildt1997} on Ti I and by \citet{Schweitzer2000} on CO. For Cepheids, preliminary work on NLTE abundances was carried out using the DETAIL code by \citet{Przybilla2021} in 1D, and in 2D NLTE by \citep{Vasilyev2019}. The latter study, in particular, found that using 2D pulsating models of Cepheids in combination with NLTE RT for oxygen leads to large impact on oxygen abundance. For the O I triplet lines (Fig. \ref{fig:cepheid}), in particular, the 2D RHD models with 3- and 9-day pulsation periods yield oxygen abundances lower by up to $-1$ to $-0.25$ dex depending on phase, compared to results obtained using standard 1D LTE hydrostatic models. The large order of magnitude difference is particularly prominent for the low-log(g) model (P9), at the phases of maximum contraction ($\phi = 0.8 ... 1.2$), when $\teff$ is the range $\sim$ 6200 -- 6700 K and $1.4 \lesssim \log g \lesssim 2.0$. The effect is primarily due to extremely strong over-population of the O I lower energy states due to photon losses in the lines. 

It remains to be explored, with 2D and 3D RHD convective-pulsating model atmospheres, whether similarly large NLTE effects influence the chemical abundance diagnostics of spectra of variable stars.
\subsection{Exoplanets}\label{sec:exoplanets}
This section is truly inspired by the following statement \citep{Ornes2017}
\begin{quote}
    ``One of the biggest surprises is that driving a system far from equilibrium doesn’t just lead to turbulence. It leads to structure, and the most fascinating one is life."
\end{quote} 

Physical conditions in exoplanet atmospheres favour substantial NLTE effects, owing to low gas densities of less than 10$^{-7}$ g cm$^{-3}$ (Fig.~\ref{fig:maindiagram}). Yet at lower gas temperatures, and with an addition source of irradiation from the host star, NLTE conditions are primarily limited to lines of neutral atoms and molecules. We note that in the context of exoplanet atmosphere work, NLTE may have entirely different meaning, depending on the literature source. Here we use the more inclusive definition, which covers different types of NLTE. On the one extreme, there is the very approximate semi-empirical NLTE modelling \citep{Treanor1968}, in which, e.g., population numbers of rotational and vibrational energy states are set by the product of Boltzmann distributions for several temperature values, one of them the kinetic temperature (set to be, $T_{\rm rot}$ for rotational states) and the other being the vibrational temperature, also known as the adjustable ``NLTE'' parameter, see e.g. \citep{Wright2022}. On the side of more comprehensive physically-base investigations, there are also models that solve coupled RT and SE equations and include a range of atomic and/or molecular processes, such as photo-ionization, ionization by free electrons, charge transfer with H and other atoms and ions \citep{Yelle2004, Koskinen2013}.

So far, the focus of dedicated NLTE studies \citep[e.g.,][]{Barman2002, Young2020, Deibert2021, Fossati2021, Fossati2023} has been primarily on hot exoplanet environments, with characteristic temperatures higher than 2000 K, and only limited knowledge of conditions in cooler atmospheres is available. In this context, perhaps the only planet that has been extensively studied by means of NLTE calculations is Earth atmosphere \citep[see][and references therein]{vonClarmann2002, Lopez-Puertas2001}. To a lesser extent, NLTE methods were applied in spectroscopic diagnostics of the atmosphere of Mars \citep[e.g.][and references therein]{Read2015} and Titan \citep{Yelle1991, Lopez-Puertas2013}.

The earliest NLTE studies can be traced back to \citet{Appleby1990} (CH4), \citet{Seager2000} (He), and \citet{Barman2002} (Na). The former study performed NLTE calculations for the He I triplet line at 1083 nm employing the methods developed for Cepheid atmospheres \citep{Sasselov1994}. The latter focussed on the effects of NLTE in the key diagnostic doublet lines of Na D at 598 nm in atmospheric conditions of the gas giant HD~209458~b. This study, in particular, went into a great detail what concerns the origin of NLTE effects, behaviour of the level departure coefficients, and the influence of collision rates on the NLTE results. The paper is also relevant, as the work that highlighted that even a qualitative interpretation of exoplanet spectra may change depending on physics of NLTE, e.g. misinterpreting the weakness of the Na D feature for low metallicity or presence of high-altitude clouds \citep{Barman2002}.

\begin{figure}[ht]
    \centering
    \includegraphics[width=0.9\textwidth]{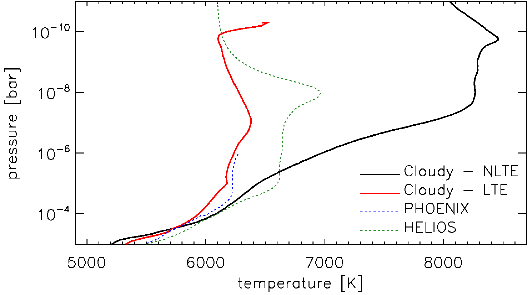}
    \caption{NLTE and LTE calculations of transmission spectra in the region of hydrogen H$_{\alpha}$ and H$_{\beta}$ lines in the atmosphere of a hot Jupiter Kelt-9b, compared with the observed data. Image reproduced with permission from \cite{Fossati2021}, copyright by the author(s).}
    \label{fig:exoNLTE}
\end{figure}

Over the past years, NLTE RT calculations have become more common and have since been applied to modelling the atmospheric structure and/or diagnostic spectroscopy of warm and hot Neptunes \citep{Kubyshkina2024}, hot and ultra-hot Jupiters Kelt-9~b, Kelt-20~b \citep{Fossati2021, Borsa2021, Fossati2023}, HD 209458~b \citep{Oklopcic2018, Lampon2020, Young2020}, WASP-12~b and WASP-76~b \citep{Wright2022}, WASP-33~b \citep{Wright2023}, WASP-121~b \citep{Young2024}, sub-Neptunes \citep{Garcia2024}, and Neptunes \citep{Oklopcic2018}. The findings cannot be yet generalised to all conditions relevant to planet atmospheres, as only targeted investigations for individuals models and/or ions are available, and no systematic study across the entire parameter space has been performed. Also, in terms of observations and detection of NLTE effects in exoplanet transmission spectra, one source of complexity is the degeneracy with amount of stellar light blocked by the planet, but also the still poorly understood effects of hydrodynamic outflows \citep[][]{Hoeijmakers2020}. Nonetheless, from the theoretical perspective, the evidence is as follows. Large and negative NLTE effects were reported for similar systems in many others species, including the O I 777 nm triplet \citep{Borsa2021}, H$_{\alpha}$ lines \citep{Fossati2021}, 1083 nm He I line \citep{Oklopcic2018, Young2020}, and in the strong UV Mg I and Mg II lines at $2852.12$ and $2802.7$ \AA\ \citep{Young2020}, respectively. For H I lines, we show the results obtained for Kelt-9b in Fig. \ref{fig:exoNLTE}.

\begin{figure}[ht]
    \centering
    \includegraphics[width=0.9\textwidth]{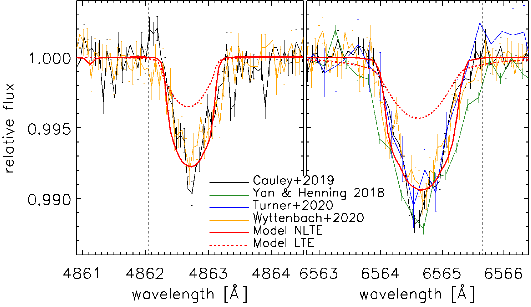}
    \caption{NLTE and LTE calculations of the atmospheric structure of hot Jupiter Kelt-9b. Image reproduced with permission from \cite{Fossati2021}, copyright by the author(s).}
    \label{fig:exoNLTEatm}
\end{figure}

The NLTE results for diagnostic spectral lines in exoplanet atmospheres are qualitatively similar to NLTE effects reported for FGK-type stars (see Sect.~\ref{sec:coolabu}). In this context, negative NLTE effects imply stronger NLTE absorption features, thus, lower abundances compared to LTE diagnostics. The Na D doublet at 598 nm shows weaker wings, but stronger cores in NLTE \citep{Fischer2019, Young2020}, which is, however, opposite to the findings by \citet{Barman2002}. The latter work explored irradiated and non-irradiated profiles of the atmospheric structure of HD 209458~b, taking into account collisions with free electrons and molecular hydrogen $H_2$, but in both cases their NLTE profiles of the resonance Na D lines were systematically weaker compared to LTE predictions, and even fully reverted to emission when collisions with $H_2$ were neglected. Hence according to \citet{Barman2002} work, Na abundance in exoplanet atmospheres is substantially underestimated.

Studies of NLTE or non-equilibrium effects in molecules are currently rather scarce. Among the first, \citet{Swain2010} proposed the possibility of NLTE emission from the methane $\nu_{3}$ band in the atmosphere of a hot Jupiter HD~189733~b. The relationship between this feature  and its possible NLTE origin were further explored in \citet{Waldmann2012}. We note, however, that several other studies \citep{Mandell2011, Birkby2013} did not confirm the emission using high-resolution observations of the planet with NIRSPEC and CRIRES, and instead link the signal at 3.25 $\mu$m to tellurics, specifically to the $\nu_{2}$ band of the Earth's water vapour. Non-equilibrium chemistry was investigated by means of detailed photo-chemistry calculations in different studies \citep[e.g.][]{Stevenson2010, Madhusudhan2011, Moses2011,Knutson2012, Moses2013, Baeyens2024}. Photo-chemistry is driven by the radiation field, but also advection (that is, the transport-induced processes) may further have a substantial impact on the population numbers of molecules and, hence, on the composition and structure of atmospheres. 

The question typically asked is what is the impact of NLTE on the structure of models. First estimates suggest that the outer thermodynamic structure of hot Jupiters is highly sensitive to the effect of photo-ionization by the stellar radiation. The classical effect is the \textit{re-heating of the outer layers}, leading to the characteristic inversion of the T($\tau$) and T-$\rho$ profiles, which was demonstrated in the early detailed calculations by \citet{Yelle2004} and further carefully explored by e.g., \citet{Koskinen2013}, \citet{Fossati2021} and \citet{Fossati2023}. The main physical process leading to \textit{heating} is the photo-effect reaction, that is, radiatively-driven ionization (by stellar X-UV field) \citep{Cecchi-Pestellini2009} and subsequent redistribution of energy of photo-electrons in inelastic collisions. Classical LTE atmospheric models that do not account for NLTE overionization of H (a continuum process) and NLTE pumping in metal lines are unable to reproduce this interesting physical effect (Fig. \ref{fig:exoNLTEatm}). Specifically, strong heating in Fe II lines, which constitute the vast majority of over 300$\,$000 transitions (priv. comm. Mitchell Young) and in Mg I lines was reported by \citet{Fossati2021}. Also, overionization of Fe I has been shown to influence the temperature-density structure  \citep{Young2024}. All these effects lead to significantly hotter upper atmospheres and thus, due to increased pressure scale height, greater radii of hot Jupiters. 

Over past years, some effort went into understanding the effects of hydrodynamics in planet atmospheres, especially in the context of mass loss, also known as \textit{hydrodynamic escape}. It is typically argued that the bulk planet atmospheres are nearly in HE, as from the analysis of the Jeans parameter \citep{Fossati2017} characteristic outflows in the line formation regions are subsonic  \citep{Koskinen2013, Fossati2023}. However, also 1D RHD simulations are now available that solve 1D equations of mass, momentum, and energy \citep[e.g.][]{Yelle2004, GarciaMunoz2007, Koskinen2013, Erkaev2016} and may take into account the effects of charge separation and viscosity. Available evidence suggests, however, that hydro-dynamical effects are not large, and the velocities are typically much smaller compared to thermal velocities. Such models be coupled with 1D NLTE codes, such as \Cloudy \citep{Salz2015, Kubyshkina2024}, therefore  NLTE effects can be taken into account by including photo-ionisation, recombination, charge transfer, and collision-driven transitions in various atomic species. 

With the impact of NLTE on the physical structure of exoplanet atmospheres \citep{Young2024},  we argue that a comprehensive treatment of NLTE is highly desirable for modelling these systems. Whereas the \textit{detectability (presence or absence)} of species may not be substantially affected by the assumption of LTE \citep[but see][for an interesting discussion]{Barman2002, Swain2009}, major NLTE effects may show in the spectroscopic diagnostics of the chemical composition. These may show when departures from atomic and especially molecular equilibrium are of a relevance in physics of exoplanet atmospheres.
\subsection{Cosmologically relevant elements}

Abundances of three elements, Li, Be, and B, are crucial in the context of observational cosmology. Their observed abundances right after the Big Bang Nucleosynthesis (BBN) are typically used to inform, among other constraints, the standard cosmological model \citep[e.g.][]{Boesgaard1985, Yang1984, Walker1991, Coc2012, Coc2017, Fields2020}.

The observed abundances of these elements are typically determined from the spectra of low-mass low-metallicity stars, which are thought to be born shortly after the BB \citep[e.g.][]{Spite1982, Charbonnel2005, Bonifacio2007, Aoki2009}. The measurements are model-dependent, and we are facing a growing body of work on the analysis of 3D and NLTE effects on the abundances of Li. Detailed studies \citep{Asplund2003, Klevas2016, Wang2021} suggest that in full 3D NLTE radiation transfer calculations, surprising cancellation effects occur at low metallicity, such as the abundances of Li in the so-called Spite plateau \citep{Spite1982} stars at [Fe/H] $\sim -2$ inferred using 1D LTE are nearly identical to those inferred using 3D NLTE modes (see \citealt{Klevas2016}, their Fig.3, right-handside panel, and \citealt{Wang2021}, their Fig. 11 bottom panel). Overall, 1D and 3D NLTE effects on Li abundances are also not large, and usually do not exceed 0.1 dex in modulus, except solar-metallicity red giants and cool M-dwarfs \citep{Klevas2016}. The latter study also emphasised the 1D NLTE and average 3D NLTE modelling approaches yield as reliable abundances of Li, as full 3D NLTE, for most of the parameter space, except low-metallicity red giants, where 3D NLTE effects are slightly larger in amplitude. 

Finally the effects of NLTE and 3D were quantified in the context of Li isotopic structure. 3D NLTE RT calculations are indeed very important for the analysis of ${^6}$Li/${^7}$Li isotopic ratios \citep{Harutyunyan2018}, because the spectral feature due to the ${^6}$Li isotope - although slightly offset form that of ${^7}$Li - is very weak, and its comprehensive treatment requires complete 3D calculations \citep{Cayrel2007, Steffen2012, Lind2013, Wang2022}.

\subsection{Kilonovae}\label{sec:kilonovae}
As summarised in Fig.~\ref{fig:maindiagram}, the physical conditions in which NLTE effects are important also include the so-called kilonova events. These events are thought to be, among other extreme systems like gamma-ray bursts (GRB), electromagnetically observed counterparts of compact binary mergers \citep[e.g.][]{Li1998,Metzger2010,Barnes2013, Tanaka2013, Pian2017}. With the development of gravitational wave astronomy following the detection of GWs from mergers of neutron stars (NS) or black hole--neutron star mergers (BH--NS)  \citep[e.g.][]{Abbott2017a,Abbott2017b,Abbott2017c}, it is becoming highly important to model accurately the broad-band fluxes and their time evolution (lightcurves) \citep{Villar2017,Watson2019,Domoto2021,Perego2022} including NLTE \citep[e.g.][]{Hotokezaka2021,Pognan2022b,Hotokezaka2022} in order to constrain physical conditions and nucleosynthesis in these enigmatic astrophysical events. 

Different from stellar conditions, the radiation of kilonovae is highly time-dependent. Following the merger, the subsequent time evolution of the optical emission is guided by an extremely rapid expansion of the envelope, at speeds of $\sim 0.1$ to $0.3$ of the light speed, synthesis of neutron-capture elements (specifically, in rapid neutron-capture process) on the timescales of a few seconds, and subsequent radioactive decays and photon losses on longer timescales of up to several months \citep{Villar2017, Waxman2018, Radice2018}. Fission cascades, $\alpha-$ and $\beta$-decays of radioactive nuclei produce energetic particles and photons \citep{Way1948}. Overall the evolution represents a transition from a hot and optically-thick system early (in full LTE) to a highly rarefied system owing to rapidly decreasing densities (in full NLTE), with corresponding effective temperatures $\teff$ of a few $\sim 10\,000$ at 0.5 days dropping to a few $1\,000$ K at 3--4 days \citep{Waxman2018}. While the bulk properties of the SED can be approximated by the Planck function, with heating driven by the $\beta$-decays \citep{Waxman2018, Hotokezaka2020}, the detailed SED structure is primarily set by high opacity in spectral lines due to, e.g., Ca II, Sr II, Y II, Zr II, and lanthanides Ce III, La III \citep[e.g.][]{Domoto2022,Pognan2023}, although the identifications are limited by the availability of atomic data \citep{Floers2023}. The corresponding statistical equilibrium is dominated by highly ionised species (up to 11th ionization stages) as early as 0.1 days after the merger at characteristic temperatures around $100\,000$ K, but transitions to singly-ionized and neutral species around $5\,000$ K \citep{Banerjee2022}. Subsequently approximate or more comprehensive (SE) NLTE treatment is used to compute NLTE populations. Recombination coefficients, accurate photo-ionization cross-sections, and comprehensive treatment of bound-bound radiative transitions are particularly important \citep[e.g.][]{Pognan2023}.

There are several RT codes that handle photospheric RT in expanding envelopes (with a steady-state system with a high outflow velocity). These include widely-used TARDIS \citep{Kerzendorf2014} and SUMO \citep{Jerkstrand2011,Pognan2023}, but also other custom numerical codes exist \citep{Tarumi2023, Sneppen2024}. TARDIS and SUMO are 1D spectrum synthesis codes employing the Monte Carlo RT approach and allowing for fluorescent emission. TARDIS is based on the  photospheric RT in the Schuster--Schwarzschild formalism, whereas SUMO carries out full-domain simulations, and NLTE RT is handled in significantly more comprehensive approach in the latter. The expansion is assumed to be homologous\footnote{Homologous expansion represents a constant shape of mathematical objects irrespective of time, albeit the size of the shape might scale with, e.g., time. This property is commonly used in astronomy, e.g. in the simulations of supernova explosions \citep{Noebauer2019}, cosmological expansion (cf Ed. Wright 1997, UCLA, \url{https://www.astro.ucla.edu/~wright/cosmoall.htm}), and dynamical ejecta during during mergers \citep{Rosswog2014,Wu2022}.} ($v = r/t$) and the underlying t$/\rho$ (density) structure follows from the velocity profile and solution of the 1st law of thermodynamics. Profiles like $\rho \sim v^{-4}$ or $\sim v^{-3}$ can be used, however, codes can handle other $\rho - v$ profiles. With various heating and cooling processes are included, such as heat exchange due to atomic line transitions, radioactive heating, and adiabatic cooling, with an assumed electron fraction $Y_e$. Radiation-hydrodynamic simulations predict diversity of velocity profiles, but overall the effects of dynamics are very small \citep{Wu2022}. Similar to SNe RT modelling, the concept of Sobolev expansion opacity\footnote{The total opacity accounting for the effect of the Doppler-shifted of the individual photon frequencies at a give time during the expansion.}(Sect. \ref{sec:Escape}) is usually adopted in the calculations \citep{Sobolev1957,Karp1977}. However, simplified 1D LTE spectrum synthesis calculations using the stellar atmosphere codes like \texttt{MOOG} \citep{Watson2019} have also found applications in the context of kilonovae spectral analysis and diagnostics. In the later paper, specifically the emission was represented by a stellar model atmosphere from Kurucz database with $T_{\rm eff} = 5\,500$ K, the surface gravity $\log g = 0$.

\begin{figure}[ht]
    \centering
    \includegraphics[width=1\textwidth]{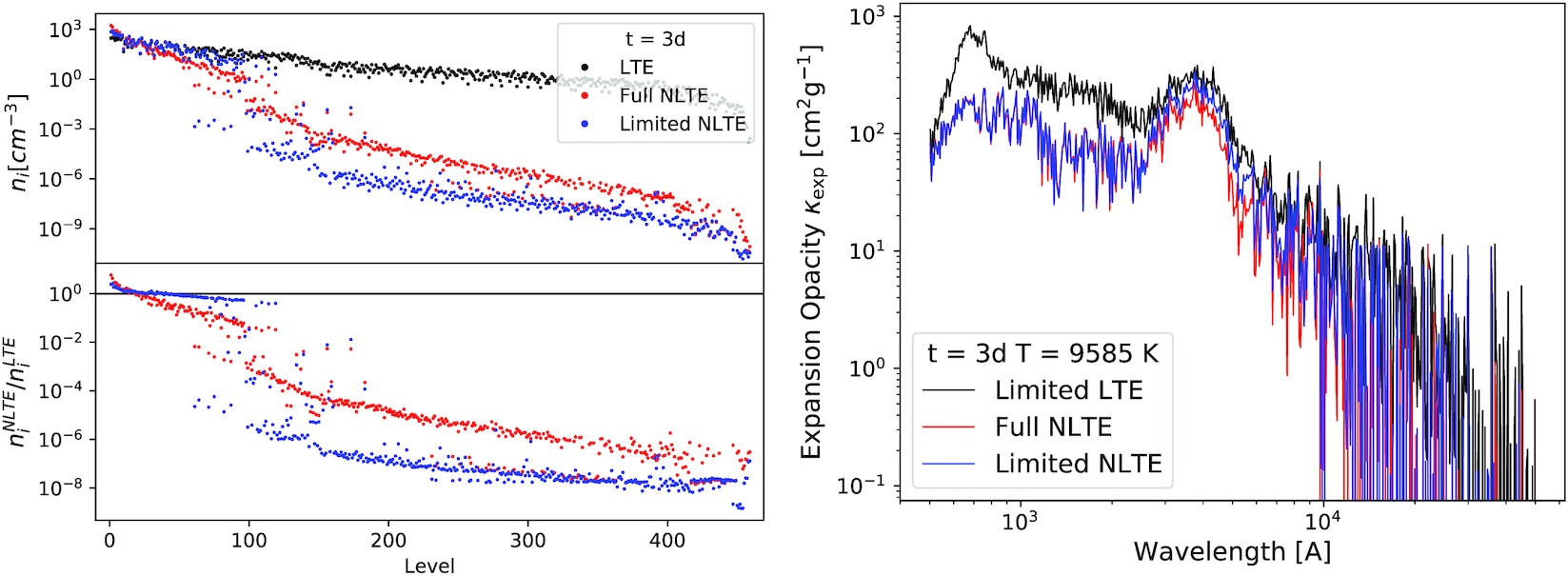}
    \caption{NLTE effect on the population of singly-ionized cerium (Ce II). The effect on NLTE populations on total opacity of kilonova at 3 d after the merger event, computed for a low-density model with the ejecta mass of $0.01$ M$_{\rm Sun}$ and expansion velocity of the ejected material of 0.2c (20\%  of the speed of light). Image reproduced with permission from \cite{Pognan2022b}, copyright by ESO.}
    \label{fig:Pognan2022}
\end{figure}

Recent studies suggest that NLTE effects become critical in the structure of KN after a few ($\sim $ 4 to 5 days) after the merger. In terms of thermodynamics, the general effect of NLTE is that it leads to progressively increasing kinetic temperature and ionization fractions with time, opposite to what simplified LTE calculations suggest \citep[e.g.][]{Hotokezaka2021, Pognan2022a, Pognan2023}. The temperature rise is thought to continue for months to a year, until adiabatic cooling sets in (priv. comm. A. Jerkstrand). Detailed calculations of NLTE opacities and number densities show that only at very early times (1 to 2 days) after the merger, level populations of the key species, such as Ce II and Pt III, are nearly thermalised and the LTE equilibrium is achieved \citep{Pognan2022b}. However, substantial (many orders of magnitude) de-population of upper energy states is achieved at later times ($\gtrsim 5$ days) after the merger due to very low densities, and hence inefficient collisional thermalisation and dominance of spontaneous radiative de-excitations. The LTE opacity can therefore be over-estimated by a factor of 2 (as for Pt III at early epochs) to a factor of $\sim 1\,000$ (as for Ce II 20 days after merger), which is due to a failure of the Saha ionization equilibrium to produce realistic ionization fractions \citep{Pognan2022b}. This major effect of NLTE on the population of main opacity sources, in this case Ce II, is demonstrated in Fig. \ref{fig:Pognan2022}, (c) \citet{Pognan2022b}.

For He I and Sr II, due to their particular atomic structure, the NLTE effects may arise in strong IR lines at 1.08 $\mu$m (He I: 10830 \AA, and Sr II: 10037, 10327, 10914 \AA). These NLTE effects are associated with over-ionization (relative to Saha equilibrium) caused by non-thermal electrons produced in radioactive $\beta$-decays \citep{Tarumi2023}, as also explored using an approximate NLTE model in \citet{Brethauer2026}. These spectral lines are critical for the interpretation of the PCyg like feature around 700--900 nm \citep{Watson2019}, which is thought to provide a strong evidence for r-process synthesis in NSM. We note that the He model of \citep{Tarumi2023} leads to a good fit to  these near-IR spectral features at early times, 1.4 days, if the photoionization is suppressed. Interestingly, another study by \citet{Sneppen2024} reach a different conclusion: that if photo-ionization in He is included, the observed feature at this epoch cannot be reproduced, unless the entire ejecta is composed of He. We refer the reader to \citet{Pognan2023}, who provide a parametric study of the evolution of thermodynamic structure and NLTE synthetic spectra of kilonovae. 

From the available body of evidence, it can be concluded that a more comprehensive understanding of NLTE effects in light elements, like He, and in trans-Fe-group species, is critical for the interpretation and analysis of nucleosynthesis in compact binary mergers \citep[][]{Perego2022, Tarumi2023}.
\subsection{Red supergiants}\label{sec:rsg}

RSGs are evolved core-He burning massive stars close (in the \teff-\logg\ space) to the Hyashi boundary \citep[][]{Maeder1981, Levesque2017}, with convective envelopes occupying $\sim$ 0.1 to 0.8 of their total mass \citep{Stothers1972}. Thus, 3D radiation-hydrodynamics is imperative for understanding the outer structure of these systems \citep[e.g.][]{Chiavassa2009, Chiavassa2011, Goldberg2022, Ahmad2023, Ma2024}. 

For RSGs, 3D radiation transfer and spectrum synthesis has been so far primarily exploited in the context of the atmospheric dynamics and convective flows \citep[see][for a review]{Chiavassa2024}. These calculations make use of 3D LTE radiation transfer simulations in selected spectral lines to invert the asymmetric observed profiles of selected photospheric lines \citep{Kravchenko2018}. These are, for example, the TiO band features that are sensitive to both temperature and velocity inhomogeneities due to large-scale convective motions \citep{Kravchenko2019}. The tomographic methods can be applied to different wavelength regimes, including the optical spectral lines, and they are also useful as diagnostics of relationships between photometric variability of red supergiants and sub-surface convection, as e.g., shown for Betelgeuse in the context of its Great Dimming event \citep{Kravchenko2021}. As \citet{Ma2024} show, 3D RT models are furthermore key to the interpretation of the widths of molecular lines, such as SiO and CO, that can be observed in the sub-mm with ALMA. These are relevant in the context of measurements of RSG rotation rates with ALMA \citep{Kervella2018} and the tantalising possibility that rapidly-rotating RSG may represent a product of merger in a massive binary system \citep{Chatzopoulos2020, Sullivan2020}. 

NLTE calculations for RSG stars have been carried out with 1D hydrostatic MARCS models \citep{Bergemann2012c,Bergemann2013,Bergemann2015} owing to their comprehensive treatment of molecular line blanketing. Since RSGs are brightest in the infra-red, abundance diagnostics typically rely in spectral lines in the J, H, or K bands \citep[e.g.][]{Davies2017}. The available evidence suggests that NLTE effects in the IR diagnostic features are typically driven by line scattering. Hence, for most diagnostic transitions, such as strong lines of Fe I \citep{Bergemann2012c}, Ti I \citep{Bergemann2011}, Mg I \citep{Bergemann2015}, and Si I \citep{Bergemann2013}, NLTE radiation transfer predicts significantly stronger spectral lines compared to LTE. Consequently, the chemical abundances inferred via LTE IR spectroscopy of RSG stars may be significantly over-estimated, and detailed NLTE calculations are needed to retrieve robust chemical composition of RSGs \citep[e.g.][]{Patrick2016, Davies2017, Asad2020} and young super-star clusters \citep{Gazak2014}.

\section{Summary written by AI}\label{sec:conAI}


Given the increasing relevance and integration of AI tools in research workflows, we have included in this section a summary of this document generated with the assistance of ChatGPT. Each point in the summary has been thoroughly reviewed and, where deemed necessary, revised by the authors to ensure accuracy, clarity, depth, and scientific precision. No other content in this document --- apart from this section --- was produced using ChatGPT.

\begin{figure}[ht]
    \centering
    \includegraphics[width=0.6\textwidth]{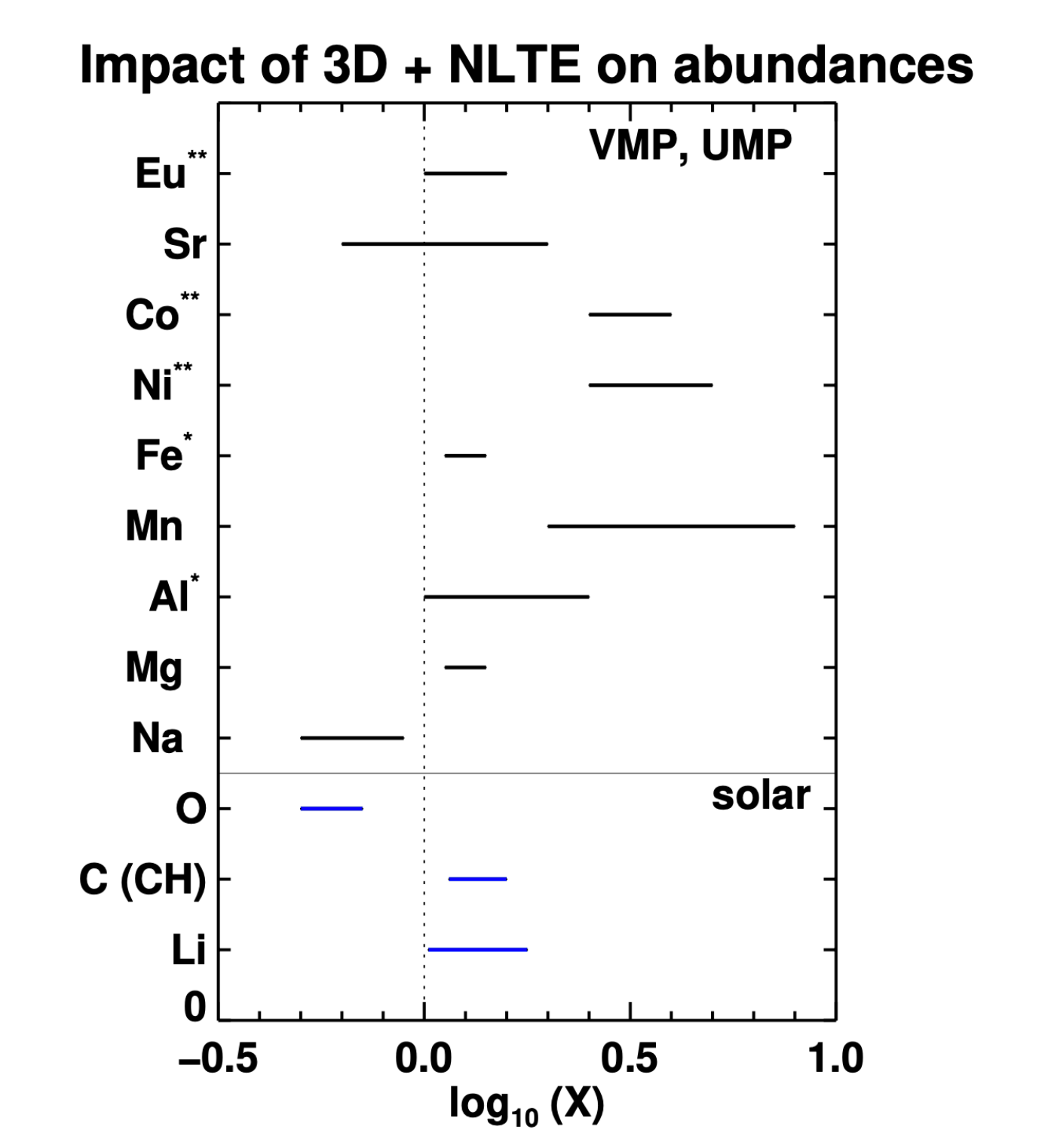}
    \caption{3D and/or NLTE effects on chemical abundances of  chemical elements in metal-poor (very metal-poor and ultra-metal-poor) and Sun-like stars. The figure only shows selected representative results and does not aim for completeness. The 3D NLTE and 1D NLTE abundance corrections are adopted from the  sources described in Sect.~\ref{sec:coolabu}. For Fe and Al, only 1D NLTE results are provided. For Co, Eu, and Ni, the data represent only turn-off and subgiant atmospheres. For all other elements, the bars represent the results for red giants and main-sequence stars.}
    \label{fig:sum3DNLTE}
\end{figure}

In Fig. \ref{fig:sum3DNLTE}, we also summarize the theoretical estimates of 3D NLTE abundance corrections for different types of stars and metallicities.

\medskip

\textbf{Motivation \& Scope}
\begin{itemize}[leftmargin=*]
    \item Radiative transfer (RT) modeling is essential for deriving physical parameters from astronomical spectra.
    \item 3D NLTE models are computed from first principles using hydrodynamics, radiation transfer, atomic and molecular physics, and statistical physics.
    \item 3D NLTE overcome limitations of traditional 1D LTE models that are highly-parameterized; the synthetic observables (spectra, SEDs, photometry) generated by 1D LTE models depend on input ad-hoc free parameters and hence any diagnostic based on 1D LTE is user-dependent.
    \item Applications span stellar types (OBA to FGKM), exoplanet atmospheres, and kilonovae.
\end{itemize}

\textbf{Conceptual Foundations}
\begin{itemize}[leftmargin=*]
    \item LTE assumes local equilibrium; NLTE solves statistical equilibrium (or rate) equations for greater realism.
    \item 3D RT accounts for spatial inhomogeneities and dynamics (e.g., convection, pulsations, magnetic fields), if multi-D RHD models are used as inputs.
\end{itemize}

\textbf{Model Inputs}
\begin{itemize}[leftmargin=*]
    \item Utilizes 3D radiation-hydrodynamics models (e.g., STAGGER, MURaM, CO$^5$BOLD, M3DIS) and atomic/molecular databases (e.g., NIST, VALD).
    \item Requires comprehensive equations of state (EoS) and partition functions and opacities for complex media.
    \item Full time-dependent 3D RHD model atmospheres available for FGKM-type stars of different metallicities [Fe/H] at \url{https://nlte.mpia.de/} (196 full 3D and average 3D stellar models) and at \url{https://staggergrid.wordpress.com/} (204 average 3D stellar models)
\end{itemize}

\textbf{Spectral Features \& Observational Phenomena}
\begin{itemize}[leftmargin=*]
    \item \textit{Convective Line Asymmetries}: Captures line bisectors and Doppler shifts from stellar surface granulation.
    \item \textit{Centre-to-Limb Variation (CLV)}: 3D NLTE models describe the center-to-limb variation of solar intensities.
    \item \textit{Limb Darkening}: Magnetic field effects accounted for in 3D SSD and 3D MHD simulations; near-UV discrepancies remain.
    \item \textit{Rossiter–McLaughlin Effect}: 1D NLTE and 3D NLTE  models outperform 3D LTE in simulating planet transits across stellar surfaces.
\end{itemize}

\textbf{Applications}
\begin{itemize}[leftmargin=*]
    \item \textbf{Cool Stars}: Accurate fundamental stellar parameters and chemical abundances;
    \item \textbf{Cool Stars}: 3D RT essential for spectral lines with strong asymmetries, e.g., due to isotopic structure, hyperfine splitting, or shifts due to variation of  fundamental physical constants. 
    \item \textbf{Exoplanets}: NLTE RT aids in atmospheric characterization; NLTE-driven strong heating in the outer atmospheres of exoplanets.
    \item \textbf{Kilonovae}: Adaptation of RT methods from supernovae for evolving spectra; transition from LTE early to a fully NLTE-dominated regime in late phases (several days) after the merger; fluorescence as a key NLTE effect.
    \item \textbf{Red Supergiants}: 3D models explain observed surface granulation and variability (e.g., Betelgeuse), as  revealed by interferometric data (VLT, ALMA).
\end{itemize}

\textbf{Computational Techniques}
\begin{itemize}[leftmargin=*]
    \item \textit{1.5D RT}: Approximation using independent vertical columns.
    \item \textit{Lambda Iteration \& Accelerated Methods}: Solve coupled RT and statistical equilibrium equations.
    \item \textit{Opacity Binning \& Mesh Refinement}: Optimize resolution vs. cost.
\end{itemize}

\textbf{Challenges \& Future Work}
\begin{itemize}[leftmargin=*]
    \item High computational demands limit broad application; grids of 3D atmosphere models and synthetic spectra are needed to enable physical analysis of data from next-generation  astronomical surveys, such as 4MOST, WEAVE, and SDSS-V.
    \item Need improved observational validation, especially for metal-poor stars (line shifts and asymmetries), exoplanet atmospheres (molecular lines), and kilonovae (opacities due to neutron-capture elements).
    \item Magnetic field treatment is needed; small-scale dynamo and magnetic fields to be explored over the entire parameter space of FGKM-type stars.
    \item Required for diagnostics of stellar physics with facilities, such as the ESO's Extremely Large Telescope (ELT), permits studies of time- and spatially-dependent phenomena (oscillations, variability) from first principles.
    \item Methods from stellar atmospheric conditions can be adapted to studies of kilonova and exoplanet atmospheres; A notable success in physical domain transfer is the \texttt{Cloudy} code, which has been widely adopted in astrophysical modeling. 
\end{itemize}

\section{Summary and conclusions}\label{sec:conHuman}
Over the past decade, the community has faced extraordinary advances in the area of radiation transfer - numerical simulations and applications - in the context of different astrophysical environments. This advance was only possible thanks to progressive uninterrupted developments of the theoretical basis for these methods over the past century. 

The two primary developments are systems in non-local thermodynamic equilibrium (NLTE) and inhomogeneous three-dimensional (3D) thermo-dynamical systems, which include convection and turbulence arising from first principles. The latter can be computed in the plane-parallel Cartesian geometry or in the full spherically symmetric geometry with depth-dependent gravitational potential. These advances have been mainly motivated by the development of instrumental facilities, atomic and molecular data, and the availability of new types of observational data for different kinds of astronomical objects, including stars, exoplanets, and compact binary mergers. The typical questions that can be asked in the framework of 3D NLTE calculations are as follows

\begin{enumerate}

 \item  How wrong are the astrophysical parameters of stars, planets, etc, obtained using simplified (1D LTE) models?

 \item Which qualitative advance [over 1D LTE] do 3D and/or NLTE models bring about, which cannot be learned from 1D LTE models? 

 \item How wrong are the 3D NLTE models and what are their observationally testable predictions?
 
 \item What is the effect of magnetic fields? 

\end{enumerate}

The codes developed by different groups to perform 3D NLTE calculations are very similar in their technical skeleton. However, the codes differ in the level of detail, micro-physics, and implementation of all sub-grid effects, such as interpolations, extrapolations, smoothing. At the NLTE level, various approaches are used, ranging from full comprehensive (and highly, many orders of magnitude, more time-consuming) solution of rate equations for all known energy states of a specie to various approximations and simplifications, such as the two-level atom coronal approach or even a double-temperature approximation in the Boltzmann equation for a mixed plasma in LTE. Regarding geometry or dynamics, or both, the approaches range from standard 1D system in hydrostatic equilibrium and analytical recipes to represent depth-dependent convective fluxes, through more sophisticated 1D hydrodynamics (e.g., ``Parker wind'' profiles used for stars and exoplanets, \citealt{Linssen2022, Kubyshkina2024}), to finally full 3D RHD hydrodynamics calculations \citep{Nordlund2009, Kupka2017, Chiavassa2024}. The latter, for the sake of maintaining the balance between relaxation and computational expense, have to assume LTE opacities or ignore them entirely adopting grey RT, or ignoring RT all together, as, e.g., in kilonovae for the early stages before the 1D homologous expansion can be assumed or in early phases of 3D RHD calculations of sub-surface stellar convection where Newton cooling is used \citep{Eitner2024}.

On the observational side, the most direct information comes from solar telescopes, which provide spatially-resolved high-resolution optical spectra for the solar surface. Such information is only barely accessible for other stars. However, first attempts have been made to reconstruct centre-to-limb intensity variation of a few nearby stars using exoplanet transit data \citep{Dravins2017a,Dravins2017b,Dravins2021}. More extensive and convincing observational arguments for 3D convective models come from imaging of stellar surfaces, such as, e.g., possible with interferometers. Although limited to large stars, the observed data reveal asymmetries and inhomogeneities on stellar surfaces. This sub-structure, driven by gigantic sub-photospheric convective cells, exhibits correlations between velocity and temperature. As a result, it gives rise not only to surface granulation, but also to hot spots, mass loss, stellar winds, and circumstellar shells. Finally, evidence for more complex 3D models of stellar atmospheres --- beyond the classical 1D LTE models --- is rooted in the failure of these models to describe key diagnostic features in observed spectra. This can be particularly well demonstrated for the spatially-resolved spectroscopy of the Sun. These do not require any larger or more powerful instruments, but are sufficiently well-known already from data acquired with small medium-resolution facilities. 

Overall, 3D RT modelling in the context of stars has been driven mainly by an interest in convection \citep{Stein1998, Vogler2004, Freytag2012}, atmospheric structure \citep{Gudiksen2002, Carlsson2016, Rempel2017}, and activity \citep{Vögler2007, Rempel2014, Chen2017}, the results of which could be compared to high precision observations. Atmospheric models of other stars were also developed for a handful of individual stars \citep{Caffau2011, Nordlander2017, Lagae2023} and grids of effective temperature and surface gravity were computed by \citet{Ludwig2009} and \citet{Magic2013a} and more recently by \citet{Rodriguez2024} and \citet{Eitner2024}.

For stars, the necessity for NLTE has been argued primarily from the theoretical standpoint. Recent calculations of cross-sections for reactions of collisional type suggest that classical formulations have to be used with caution, as these sometimes overestimate the cross-sections by orders of magnitude \citep{Barklem2016}. However, in general we conclude that the NLTE effects obtained by different authors \citep[][]{Amarsi2015, Osorio2016, Bergemann2017, Mashonkina2019, Gallagher2020, Mashonkina2022, Storm2024} are consistent, despite the differences in the atomic models, atmospheric physics, and numerical approaches in NLTE RT. Recent calculations for stellar conditions (Fig. 28, 29 in this review) confirm significant NLTE effects on lines in the UV, optical, and near-IR bands for most chemical elements in the periodic table, including the lightest (H, He, Li) through Fe-peak (Mn, Co, Ni) to heavy neutron-capture elements (e.g., Ba, Y, Sr, Eu). NLTE abundances strongly depend on stellar parameters (\teff, \logg, [Fe/H]) and the effects cannot be calibrated out, thus requiring NLTE modelling of stellar spectra from first principles. These effects can be included either directly (via spectrum synthesis) or indirectly, by correcting the inferred 1D LTE abundances for effects of 3D NLTE.

Coupled calculations of NLTE RT in 3D hydrodynamical atmospheric models of stars are still rare. However, the available evidence (for lighter species like H and Li, but also heavier like Na, Al, Mn) suggests that 3D NLTE modelling is crucial and may have a profound impact on the analysis of stellar chemical composition, especially so for low metallicity and/or low gravity stars \citep[e.g.][]{Amarsi2019b, Bergemann2019, Nordlander2019}. A comprehensive analysis of 3D NLTE abundance corrections and their relevance in studies of Galactic Chemical Evolution was recently presented by \citet{Storm2025} and \citet{Koutsouridou2025}. It is essential that the next step in observational stellar physics involves 3D NLTE spectroscopic grids, and ideally includes magnetism, for instance by employing 3D SSD (or MHD) model atmospheres \citep{Bhatia2022, Ludwig2023, Witzke2023, Kostogryz2024} in spectroscopic RT calculations. 

The prime advantage of 3D NLTE atmospheric and spectroscopic models is that they do not include ad-hoc free parameters (micro- or macroturbulence, mixing length), and thereby allow for analyses of fundamental stellar parameters, chemical abundances, but also studies of spatially- and time-dependent phenomena, from first principles.

Beyond chemical composition, model stellar spectra from 3D time-dependent RHD and MHD models for stars are  relevant in the context of other applications. These include: diagnostics of stellar oscillations from multi-epoch spectra and light-curves for  asteroseismology, variability and surface structure of stars in the context of studies of exoplanets, as well as studies of stellar multiplicity and pulsations. These models are also particularly important for science questions for which tiny effects on spectral line shifts or asymmetries may have drastic implications for conclusions. One interesting example is constraining the variability of the fundamental fine structure constant \citep[e.g.][]{Murphy2022}.

In the domain of observational studies of exoplanets, neither strong evidence for NLTE effects nor sub-surface convection or mass motions exists yet. It is difficult to probe these effects observationally in the transmission spectra, even despite theoretical models forecasting NLTE effects of up to a factor or 2 to 5 in the line depths, or in the mean atmospheric structure, such as severe (up to several $\sim 1000$ K) NLTE-driven heating in the outer atmospheric layers of gas planets, such as hot Jupiters. Standard externally irradiated 1D LTE spectral models \citep{Molliere2019} provide the first-order estimates of the atmospheric structure of exoplanets, and hence of abundance ratios of important molecular tracers, such as water, CO, CH$_4$, and CO$_2$. However, first attempts to compute models with NLTE radiative transfer highlight the substantial effects on atomic opacities and, hence, on the detectability of features in absorption or emission \citep[e.g.][]{Young2020, Fossati2021, Wright2022}. Interest in 3D convective planet atmosphere models is rapidly growing. These models include global atmospheric circulation, and thus yield fascinating observational predictions in the presence of stellar irradiation on the day-night side of tidally locked planets.

Methods of RT in the context of kilonova research \citep{Jerkstrand2011, Kerzendorf2014, Pognan2023, Shingles2023} have mostly been inspired by NLTE simulations for SN Ia \citep{Noebauer2019} and numerical methods have been mostly taken over. However, adapting the micro-physics, such as the chemical composition and the equation of state to the chemical mixtures prevalent in the outcomes of compact binary mergers of NS and NS-BH mergers. Here, shift to progressively heavy species, actinides and lanthanides produced in r-process during the merger, has imposed a need for calculations of radiative data for NLTE modelling of opacities and SEDs of kilonovae. Detailed calculations  confirmed strong pumping of critical transitions (e.g., He I and Sr II) by the radiation field emerging due to non-thermal NLTE effects \citep{Tarumi2023}. Also the NLTE effects on kilonovae opacities can be tremendous, leading to up to a factor of 1000 lower radiative opacities  at late epochs compared to LTE conditions \citep{Pognan2022a}. More detailed and extensive calculations for different chemical species, but also coupled 3D NLTE modelling \citep{vanBaal2023} are imperative to consolidate the diagnostic of these enigmatic astrophysical systems.

In summary, the past decade has seen tremendous progress in the numerical aspects of atmospheric radiation transfer --- moving towards multi-D geometry, convection, expansion, and non-local thermodynamic equilibrium ---  for vast classes of astronomical objects, ranging from exoplanets, to stars, and their post-explosion or post-merger remnants. These efforts, although tedious, technically (numerically and physically) complex, and barely accessible to the broader community of observers and phenomenologists, are exceptionally valuable and fundamental to progress in astrophysics. This is primarily because they are our primary (or even the only!) source of evidence that astrophysical inference, and hence all kinds of discoveries in astronomy, is reliable and trustworthy. To give an example, the importance of NLTE abundances and 3D stellar atmosphere grids was acknowledged quickly and unanimously in the studies of first stars \citep[][]{Beers2005, Bonifacio2025}, nucleosynthesis and nuclear physics \citep[][]{Kappeler2011, Sbordone2010}, origins of elements and Galactic chemical evolution \citep{Nomoto2006, Frebel2015}, formation and evolution of massive stars \citep{Woosley2007}, as well as in asteroseismology \citep{Aerts2010}, and exoplanet science \citep{Rauer2014}. However, even today, a decade later, those fields are still heavily reliant on rapid synthesis of LTE 1D spectra \citep{Sharma2018, Nelson2019, Gilmore2022}. Therefore, we consider it essential to develop NLTE and 3D spectrum synthesis codes that surpass established 1D LTE codes in their capability of employing 3D radiation-hydrodynamical inputs. 

We anticipate that the progress towards 3D NLTE will accelerate as RT codes become public, and this will enable a fundamentally new quality of observational astrophysics: the ability to understand, predict, and explain astronomical objects in space and in time. Unfortunately, is not possible to do this from first principles with any current 1D parameterized quasi-equilibrium (LTE) models.

\bmhead{Acknowledgements}
We extend our heartfelt thanks to all our colleagues who have made significant contributions to the material presented in this review article. The list is too long to be included here. Their passion for scientific exploration and discovery, along with many insightful conversations, has been a continual source of inspiration for us.

We are also grateful for the valuable discussions with colleagues during various meetings and workshops, including the RT24 workshop in Heidelberg, the Horses2 meeting in Berlin and the MIAPbP workshop in Munich.

A very small fraction of this review (Sect. \ref{sec:timing}) on computer time scaling was written with the help of ChatGPT. The content was produced by the authors (the text and numerical relationships), but subsequently ran through ChatGPT  in order to improve the structure and readability.

\phantomsection
\addcontentsline{toc}{section}{References}
\bibliography{refs}

\end{document}